\documentclass{aa}
\usepackage[varg]{txfonts}
\usepackage{amsmath}
\usepackage{graphicx}
\usepackage{natbib}
\usepackage{hyperref}
\usepackage{multirow}
\usepackage{arydshln}
\usepackage[normalem]{ulem}
\usepackage{color}
\bibpunct{(}{)}{;}{a}{}{,} 
\newcommand{\mps}{m\,s$^{-1}$}

\def\zwidth{\mbox{$\underline{\overline{z}}$}}
\begin{document}

\title{Comparison of time--distance inversion methods applied to SDO/HMI Dopplergrams}

\author{David Korda\inst{1}
  \and Michal {\v S}vanda\inst{1,}\inst{2}
  \and Junwei Zhao\inst{3}} 

\offprints{David Korda, \\ \email{korda@sirrah.troja.mff.cuni.cz}}

\institute{Astronomical Institute of Charles University, Faculty of Mathematics and Physics, V~Hole\v{s}ovi\v{c}k\'ach 2, Praha 8, CZ-180~00, Czech Republic
  \and Astronomical Institute of Czech Academy of Sciences, Fri\v{c}ova 298, Ond\v{r}ejov, CZ-25165, Czech Republic
  \and W. W. Hansen Experimental Physics Laboratory, Stanford University, Stanford, CA 94305-4085, USA} 

\date{Received 9 July 2019 / Accepted 5 August 2019}

\abstract
{The Helioseismic
and Magnetic Imager (HMI) onboard the Solar Dynamics Observatory (SDO) satellite has been observing the Sun since 2010. The uninterrupted series of Dopplergrams are ideal for studying the dynamics of the upper solar convection zone. Within the Joint Science Operations Center (JSOC) the time--distance inversions for flows and sound-speed perturbations were introduced. The automatic pipeline has produced flow and sound-speed maps every 8 hours. We verify the results of JSOC inversions by comparing the data products to equivalent results from inverse modelling obtained by an independent inversion pipeline.}
{We compared the results from the JSOC pipeline for horizontal flow components and the perturbations of the speed of sound at set of depths with equivalent results from an independently implemented pipeline using a different time--distance inversion scheme. Our inversion pipeline allows inversion for all quantities at once while allowing minimisation of the crosstalk between them. This gives us an opportunity to discuss the possible biases present in the JSOC data products.}
{For the tests we used the subtractive optimally
localised averaging (SOLA) method with a minimisation of the cross-talk. We compared three test inversions for each quantity at each target depth. At first, we used the JSOC setup to reproduce the JSOC results. Subsequently, we used the extended pipeline to improve these results by incorporating more independent travel-time measurements but keeping the JSOC-indicated localisation in the Sun. Finally, we inverted for flow components and sound-speed perturbations using a localisation kernel with properties advertised in the JSOC metadata.}
{We successfully reproduced the horizontal flow components. The sound-speed perturbations are strongly affected by the high level of the cross-talk in JSOC products. This leads to larger amplitudes in the inversions for the sound-speed perturbations. Different results were obtained when a target function localised around the target depth was used. This is a consequence of non-localised JSOC averaging kernels. We add that our methodology also allows inversion for the vertical flow.}
{}

\maketitle

\section{Introduction}
There are two basic methods available to study the solar interior. The first one is numerical modelling, which has been used for the description of the convection zone \citep[e.g.][]{Rempel_2014,Passos2017,Nelson2018} but these models must be confronted with observations \citep[see e.g. a review by][]{Hanasoge2015} or use observations as constraints \citep[e.g.][]{Hazra2018}. The other method is helioseismology, which is the only method that connects direct observations of the Sun and the conditions in its interior. Therefore, helioseismology is an important tool in studies of the solar interior and its results may serve as constraints for numerical models. 

Helioseismology has become a standard method of solar research. It is being used to study the internal structure of the Sun in a  global
helioseismology approach \citep[see a review by][]{Basu2016}. Thanks to global inversions, the profiles of speed-of-sound, density, and the adiabatic exponent were determined \citep[e.g.][and many more]{JCD1989,Drappen1991,Gough1996,Basu2009}. In contrast to global helioseismology, local helioseismology focuses on local perturbations of the Sun with respect to the reference model \citep{Lindsey1993}. Furthermore, local helioseismology has been successfully used in many research studies \citep[see a review by][]{2010ARA&A..48..289G}. 

\subsection{Time--distance helioseismology}
One of the local-helioseismic approaches is the time--distance method \citep{duvall_1993} which is a set of tools devoted to measuring, studying, and interpreting travel times of the waves travelling through the sub-surface layers of the solar convection zone. These waves are randomly excited via vigorous convection near the surface and as they propagate throughout the solar interior, their propagation speeds and paths are affected by deeper anomalies, for example anomalies in density, sound speed, temperature, and plasma flows. 

The signatures of travelling waves are observed as time-domain oscillations of solar observables. These ocillations are best observed in Doppler shifts of photospheric lines. The travel times of the waves are measured from the cross-covariance of the signals (e.g. the aforementioned Doppler velocity) at two different points on the solar surface. The travel time maximises the cross-covariance. From the difference between the measured travel time and the modelled one obtained from the background model, we may retrieve some information about the nature of the anomalies affecting the travel time. 

Time-distance heliosesmology has been used for inversions of many quantities, for example meridional circulation \citep{1997Natur.390...52G}, rotation \citep{1998ESASP.418..775G}, sound-speed perturbation \citep{Couvidat_2006}, and plasma flows \citep{Svanda_2013b}.

\subsection{Motivation of the comparative study}
Long-term studies require the routine availability of high-duty-cycle, homogeneous data products, which is also the case for time--distance helioseismology studies. Long-term observing campaigns producing the homogeneous, relatively high-cadence series capturing the solar oscillations are necessary. This approach was successfully tested with a Dynamics-Campaign series of Dopplergrams and intensitygrams computed from the NiI 676.8~nm photospheric line from the observations by Michelson Doppler Imager \citep[MDI;][]{MDI} aboard the Solar and Heliospheric Observatory (SOHO) satellite. Also, long-term ground series by the Global Oscillations Network Group \citep[GONG;][]{GONG} have produced suitable datasets. In 2010 a technological improvement to MDI was launched aboard the Solar Dynamics Observatory (SDO) satellite. The Helioseismic and Magnetic Imager \citep[HMI;][]{HMI1,HMI2} provides an almost uninterrupted series of Dopplergrams computed from the photospheric FeI 617.3~nm line with a cadence of 45~s. 

The measurements are stored within the Joint Science Operations Center (JSOC\footnote{\url{http://jsoc.stanford.edu}}) operated by Stanford University. Among products available within JSOC, processed products are available, including the helioseismic travel-time measurements \citep{JSOC_TT} and automatically processed time--distance inversion maps \citep{JSOC_pipeline}. A series of maps of the horizontal plasma flow and perturbations of sound speed at a set of depths are available. 

These results are often used in ongoing studies to represent the plasma properties in the shallow near-surface layers of the convection zone of the Sun. The products were carefully tested, but some studies \citep[such as][]{DeGrave_2015} suggested that there might be biases present. Recently, we developed and validated an independent implementation of the time--distance inversion pipeline based on a multichannel subtractive optimally localised averaging approach \citep[MC-SOLA;][]{Jackiewicz_2012,svanda_2011}. The pipeline was 
recently improved \citep{Korda_Svanda_2019}. It allows us to combine inversions of plasma flows and sound-speed perturbations in one inversion. Additionally, it combines difference and mean geometries and ridge and phase-speed filters. Our pipeline provides much more information about the inversion necessary for a proper interpretation of the results.

In this study, we compare the results of our pipeline (referred to as `our inversion' in the following) with the standard inverse-modelling data products from JSOC (referred to as JSOC from now on). 

\section{Time--distance helioseismology in a nutshell}
\subsection{Forward problem}
\label{subsec:forward}
Time--distance helioseismology relies on measurements and interpretation of the travel times $\tau$ of solar waves. The inhomogeneities or other anomalies in the solar structure introduce the perturbations $\delta \tau$ to the travel times, which may be directly related to the perturbed quantities $\delta q_{\beta}$, where $q_{\beta}$ indicates for example components of the plasma flow vector, speed of sound, density, and others. Index $\beta$ enumerates these quantities. 

The travel times and their perturbations are measured by cross-correlating the signals (in our case the Doppler shifts of the FeI 617.3~nm line) at two different spatial points on the surface of the Sun. The waves are excited by vigorous convection, therefore the travel times have a large random component -- realisation noise. The level of the realisation noise is usually so large that it is not possible to properly interpret the travel time computed between two points. In order to reduce the level of random noise, four averaging geometries are used. These are based on averaging of travel times between a point $\boldsymbol{r}$ and an annulus with the centre in $\boldsymbol{r}$ and radius $\Delta$. 

The first geometry, \emph{outflow-inflow,} denoted by (o-i), computes the difference between a travel time of the wave packets travelling from the point $\boldsymbol{r}$ to the rim of the annulus and a travel time of the wave packets travelling from the rim of the annulus to $\boldsymbol{r}$.

The \emph{east-west} and the \emph{north-south} geometries denoted by (e-w) and (n-s) are similar to (o-i) geometry but travel times on the annulus are weighted with a cosine (e-w) or a sine (n-s) of polar angle about $\boldsymbol{r}$. Because of the annulus weights, the travel times are sensitive to waves travelling in prominent directions. Geometries (o-i), (e-w), and (n-s) are referred to as the difference geometries because they are based on the difference of the travel times of the waves travelling in the opposite directions.

The \emph{mean} geometry is computed as an average of travel times in the direction from $\boldsymbol{r}$ to the rim of the annulus and travel times in the opposite direction.

The perturbed travel time $\delta \tau^a$ can be computed from the model

\begin{equation}
\delta \tau^a \left(\boldsymbol{r}\right) = \int \limits_{\odot} \mathrm{d}^2 \boldsymbol{r}'\, \mathrm{d} z \sum \limits_{\beta = 1}^P K^a_{\beta}\left(\boldsymbol{r}' - \boldsymbol{r}, z\right) \delta q_{\beta}\left(\boldsymbol{r}', z\right) + n^a\left(\boldsymbol{r}\right)\mathrm{,}
\label{eq:dtau}
\end{equation}
where $\boldsymbol{r}$ and $\boldsymbol{r}'$ are the horizontal positions and $z$ is the vertical position. $P$ is the number of quantities considered in the model and the term $n^a$ represents a realisation of the random noise. The functions $K^a_{\beta}$ are called sensitivity kernels. These are results of the forward modelling and depend on the reference background model. The upper index $a$ contains individual travel-time measurements (e.g. different averaging geometries, different values of $\Delta$, and different spatio-temporal filters).

\subsection{Inverse problem}
In the inverse modelling, we attempt to invert for $\delta q_{\alpha}$ from Eq. (\ref{eq:dtau}) when knowing the travel-time perturbations $\delta\tau^a$. Because of the noise term, we cannot invert this equation easily. We can compute an estimate of $\delta q_{\alpha}$ denoted by $\delta q_{\alpha}^{{\rm inv}}$ ($\alpha$ is an index of the quantity in the direction of the inversion) by invoking mathematical methods of inversions. 

There are two basic methods commonly used on how to construct this estimate: the regularised least squares (RLS) method and the subtractive optimally localised averaging (SOLA) method. 

\subsubsection{The RLS method}
\label{subsec:RLS}
The RLS method minimises the difference between measured travel times and theoretical travel times given by Eq. (\ref{eq:dtau}). The cost function is in the form

\begin{align}
\chi_{\rm RLS}^2 &= \sum \limits_{a = 1}^{M} \frac{1}{\sigma_a^2}\left[\delta \tau^a - \int \limits_{\odot} \mathrm{d}^2 \boldsymbol{r}'\, \mathrm{d} z \sum \limits_{\beta = 1}^P K^a_{\beta} \left(\boldsymbol{r}' - \boldsymbol{r}, z\right) \delta q_{\beta}\left(\boldsymbol{r}', z\right)\right]^2 + \nonumber \\
&+ \mu L\left(\delta q_{\beta}\right) \mathrm{,}
\label{eq:chiRLS}
\end{align}
where $L$ is regularisation operator \citep[see][for details]{JCD_1990}, $\mu$ is a trade-off parameter, $M$ is the number of independent travel-time geometries, and $\sigma^2$ is the variance of the observed travel times.

In Eq. (\ref{eq:chiRLS}) the first term is a misfit term between the measured and theoretical travel times. The second term is regularisation that is meant to ensure a smooth solution. Very often smooth second derivatives of the solution are required. The cost function is minimised according to $\delta q_{\beta}$. For more details about the implementation of this method in the JSOC pipeline, please see \citet{JSOC_pipeline}.

\subsubsection{The SOLA method}
\label{subsec:SOLA}
The SOLA method is principally different from RLS. It assumes that the inverted estimate of a specific quantity $\delta q_{\alpha}^{\mathrm{inv}}$ can be expressed in the form

\begin{equation}
\delta q_{\alpha}^{\mathrm{inv}} \left( \boldsymbol{r}_0; z_0\right) = \sum \limits_{i = 1}^{N} \sum \limits_{a = 1}^{M} w^{\alpha}_{a} \left(\boldsymbol{r}_i - \boldsymbol{r}_0;z_0\right)\delta \tau^a\left(\boldsymbol{r}_i\right)\mathrm{,}
\end{equation}
where $N$ is the number of horizontal positions, $z_0$ is the target depth, $r_i$ and $r_0$ are horizontal positions, and $w^{\alpha}_{a}$ are unknown weight functions. SOLA attempts to construct the localisation kernels peaking at the target depth by linearly combining the sensitivity kernels. In the end, the cost function, which is minimised, is in the form

\begin{align}
\chi_{\rm SOLA}^2 &= \int \limits_{\odot} \mathrm{d}^2 \boldsymbol{r}'\, \mathrm{d}z \sum\limits_\beta  \left[\mathcal{K}^{\alpha}_{\beta} \left(\boldsymbol{r}', z; z_0\right) - \mathcal{T}^{\alpha}_{\beta}\left(\boldsymbol{r}', z; z_0\right)\right]^2 + \nonumber\\
&+ \mu \sum \limits_{i,\,j,\,a,\,b} w^{\alpha}_a \left(\boldsymbol{r}_i; z_0\right) \Lambda^{ab} \left(\boldsymbol{r}_i - \boldsymbol{r}_j\right)w^{\alpha}_b \left(\boldsymbol{r}_j; z_0\right) + \nonumber\\
&+ \nu \sum \limits_{\beta \neq \alpha} \int \limits_{\odot} \mathrm{d}^2 \boldsymbol{r}'\, \mathrm{d}z \left[\mathcal{K}^{\alpha}_{\beta} \left(\boldsymbol{r}', z; z_0\right)\right]^2 +\epsilon \sum \limits_{a,\,i} \left[w^{\alpha}_a \left(\boldsymbol{r}_i; z_0\right)\right]^2 + \nonumber \\
&+ \sum \limits_{\beta = 1}^{P} \lambda^{\beta} \left[\int \limits_{\odot} \mathrm{d}^2 \boldsymbol{r}'\, \mathrm{d}z \mathcal{K}^{\alpha}_{\beta} \left(\boldsymbol{r}', z; z_0\right) - \delta^{\alpha}_{\beta}\right] \mathrm{,}
\label{eq:chiSOLA}
\end{align}
where $\mathcal{T}^{\alpha}_{\beta}$ is a user-selected target function localised around the area of interest. The inversion attempts to find a solution, which is sensitive in the considered region. The $\mu$, $\nu,$ and $\epsilon$ are trade-off parameters. The contribution of the random noise is assessed by the use of noise covariance matrix $\Lambda^{ab}$, which may be either modelled or measured from a large set of travel-time measurements. The $\lambda^{\beta}$ are Lagrange multipliers and $\delta^{\alpha}_{\beta}$ are the Kronecker delta functions. 

The cost function is minimised with respect to $w^{\alpha}_{a}$ and $\lambda^{\beta}$. The functions $\mathcal{K}^{\alpha}_{\beta}$ are averaging kernels and quantify the level of smearing of the inverted quantities. The connection between $\delta q_{\beta}$ and $\delta q_{\alpha}^{{\rm inv}}$ can be written using the averaging kernels as

\begin{equation}
\delta q_{\alpha}^{\mathrm{inv}} \left( \boldsymbol{r}_0; z_0\right)=\sum\limits_{\beta} \int \limits_{\odot} \mathrm{d}^2 \boldsymbol{r}'\, \mathrm{d}z\, \mathcal{K}^{\alpha}_{\beta} \left(\boldsymbol{r}' - \boldsymbol{r}_0, z; z_0\right) \delta q_{\beta} \left(\boldsymbol{r}',z\right)+\mathrm{noise}\ .
\label{eq:inv_by_akern}
\end{equation}
In other words, the averaging kernels describe an area where the inversion is sensitive to the considered quantities. The implementation of this method is described in detail by \citet{Korda_Svanda_2019}. It can be seen from Eq. (\ref{eq:inv_by_akern}) that all the considered quantities (different $\beta$s) in principle contribute to the inverted estimate for the $\alpha$-perturbation. It is only the shape of the vector averaging kernel $\mathcal{K}^{\alpha}_{\beta}$ for $\alpha \ne \beta$ that describes the contamination of the inversion by other-than-wanted quantities. Such a leakage of the components not in the direction of the inversion is called \emph{cross-talk}. 

The last important output from the inversion is an estimate for the variance $\sigma^2_{\alpha}$ of the inverted quantity $\delta q_{\alpha}^{\mathrm{inv}}$, 

\begin{equation}
\sigma^2_{\alpha} = \sum \limits_{i,\,j,\,a,\,b} w^{\alpha}_a \left(\boldsymbol{r}_i; z_0\right) \Lambda^{ab} \left(\boldsymbol{r}_i - \boldsymbol{r}_
j\right) w^{\alpha}_b (\boldsymbol{r}_j; z_0)\mathrm{.}
\label{eq:noise}
\end{equation}
This value is an estimate of the level of random noise present in the inverted maps. A complete description of the MC-SOLA method can be found in \cite{Jackiewicz_2012} or \cite{svanda_2011}.

\vspace{\baselineskip}
\noindent The principal difference between the RLS and SOLA methods is the misfit term. In RLS, the solution minimises differences between the theoretical (forward-modelled) and observed travel times. The SOLA method minimises the difference between the targeted localisation and the localisation returned from the inversion. Compared with RLS, SOLA allows for the cross-talk minimisation, because the averaging kernel is part of the cost function. In the RLS approach, the averaging kernel may be computed, but its shape does not enter the solution of the inverse problem. 

\section{Comparison of the results of inverse modelling}

\subsection{The JSOC inversions}
The inverted maps from JSOC were downloaded using the web-based interface from the series {\tt hmi.tdVinvrt\_synopHC}. In the following, we demonstrate the comparisons by using the maps near the centre of the solar disc on 16 Aug 2010. We focus on the inversions performed by using the Born-approximation sensitivity kernels \citep{BG07,GB04} and the linearised travel times measured by the Gizon-Birch method \citep{GB02}. The other methods available were found to produce similar results \citep{JSOC_TT,JSOC_pipeline}. For comparison, we downloaded two horizontal components of the vector flows and also inversions for the perturbations of sound speed. The maps from JSOC have a sample size of 0.12 heliographic degrees per pixel. The sampling in the vertical direction is irregular with indicated depths of sensitivity. In the data products, there are maps for the depths of 0--1, 1--3, 3--5, 5--7, 7--10, 10--13, 13--17, 17--21, 21--26, 26--30, and 30--35 Mm \citep{JSOC_pipeline}. 

The standard JSOC inversion products have an averaging over 8 hours, and consecutive maps follow the same sampling. Our standard inversions, on the other hand, are averaged over 24 hours. Therefore, we averaged three consecutive JSOC maps to mimic our time resolution and averaging. Our inversions consist of Postel-projected maps near the centre of the solar disc with a pixel size of 1.4~Mm, which is roughly comparable to the resolution of JSOC data products. In the spatial domain, we carefully remap JSOC inversions to our respective coordinate system, using a cubic interpolation so that a pixel-to-pixel comparison is possible. We also do not consider about 50~Mm from the edge of each frame, where an edge effect is very strong, especially in the case of inversions for the sound speed. 

The JSOC inversions of the sound speed are computed as a fractional perturbation $\delta c_s/c_s$, whereas our inversion returns directly $\delta c_s$ in velocity units. We therefore multiplied the JSOC inverted maps for the sound speed by a model sound-speed profile from model S \citep{Model_S}, which was weighted by the inversion averaging kernel to account for depth sensitivity of the results. The depth weighting is explained below in a mathematical context. 

Except for the inverted maps of the physical quantities, we also considered the noise levels \citep[given by][in Table 3]{JSOC_pipeline} and we also managed to obtain the averaging kernels for the horizontal flows for the first four depths. Therefore, in our comparison we focus on depths no larger than only a few Mm below the photosphere. Such a selection is also motivated by the findings of \cite{svanda_2011} and \cite{Jackiewicz_2008}, who concluded that inversions principally similar to ours are dominated by random noise already at depths of about 5~Mm for 24h averaging for the horizontal components of the flow and even at shallower depths for the vertical component. 

\begin{figure*}
        \sidecaption
        \includegraphics[height=8cm]{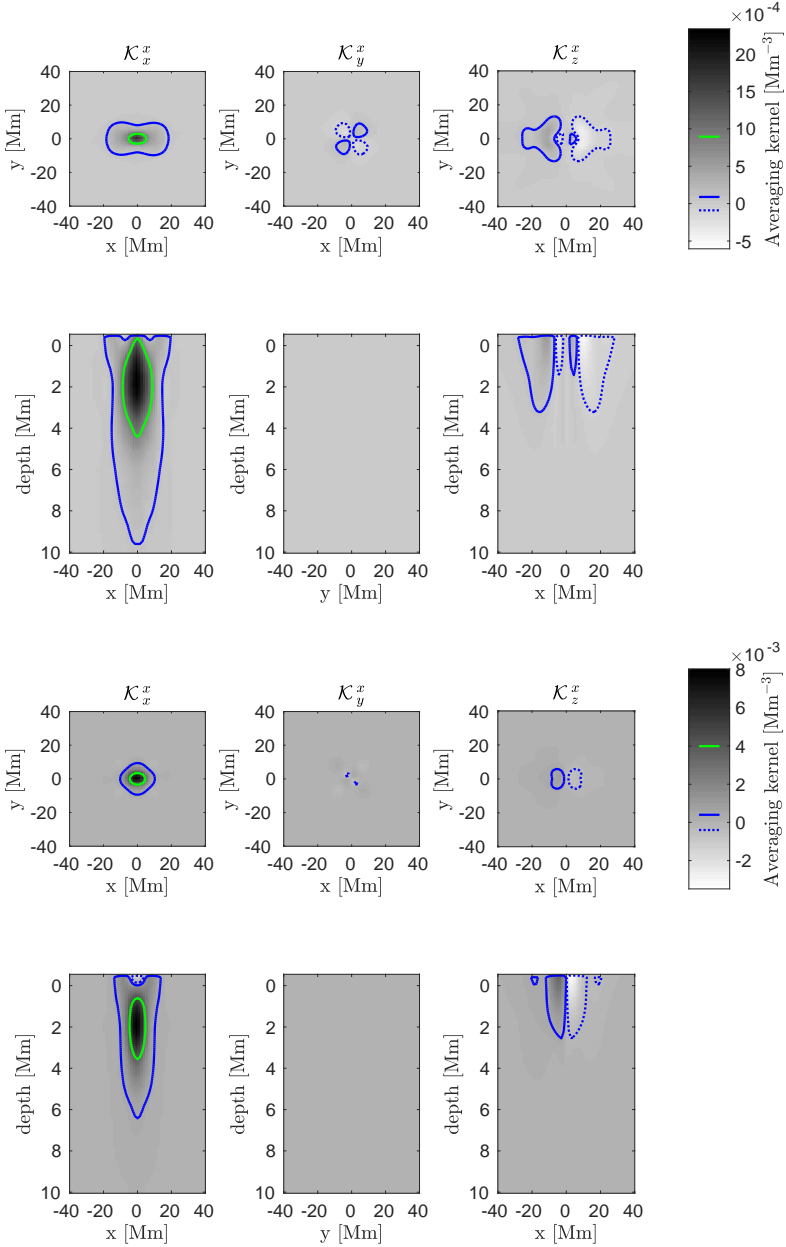}
        \caption{JSOC averaging kernel for $v_x$ inversion at the depth of 0--1 Mm. The columns indicate the contributions from the individual quantities considered in the inversion, the terms not in the direction of the inversion indicate the leakage of the other quantities -- the cross-talk. The solid green curve corresponds to the half-maximum of the averaging kernel at the target depth, the blue solid and blue dotted lines correspond to $+5\%$ and $-5\%$ of the maximum of the averaging kernel at the target depth, respectively. In some plots of this kind the green dotted and the red solid lines are also present, which correspond to minus half-maximum of the averaging kernel at the target depth and to the half-maximum of the target function at the target depth, respectively. In the top row, there are horizontal slices of the averaging kernel at the target depth and in the bottom row, there are vertical slices perpendicular to the symmetries. }
        \label{pic:RLS_rakern_x_0.5}
\end{figure*}

\begin{figure*}
        \sidecaption
        \includegraphics[height=8cm]{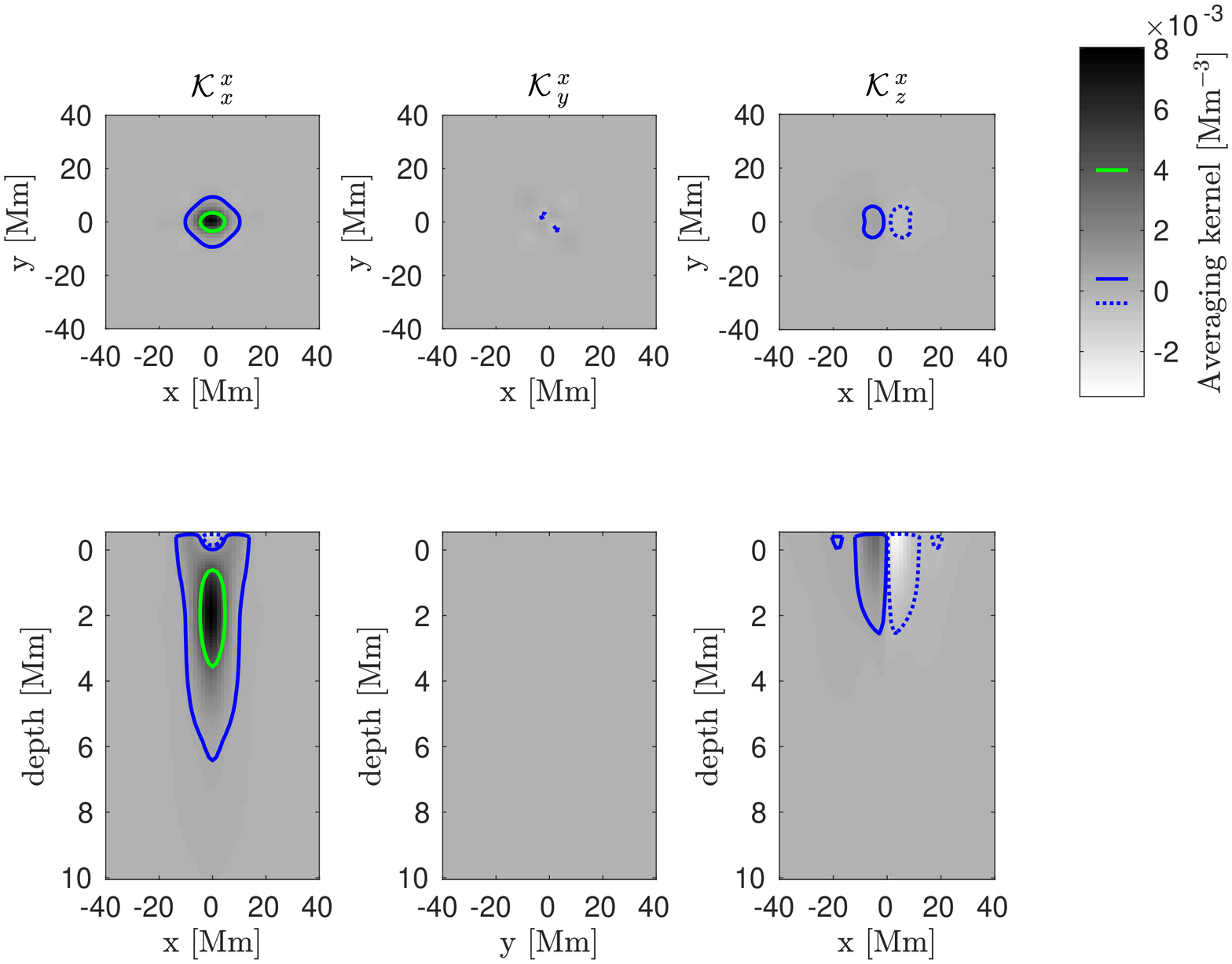}
        \caption{JSOC averaging kernel for $v_x$ inversion at the depth of 1--3 Mm. See Fig. \ref{pic:RLS_rakern_x_0.5} for details.}
        \label{pic:RLS_rakern_x_2.0}
\end{figure*}

Figures \ref{pic:RLS_rakern_x_0.5} and \ref{pic:RLS_rakern_x_2.0} show the averaging kernels for the  $v_x$ inversion at 0--1 and 1--3 Mm depths. One can see that the depth sensitivity of both inversions is nearly the same. Another problem is the level of cross-talk. It is not clear how to interpret data in terms of individual quantities without cross-talk minimisation. Other JSOC averaging kernels can be found in Appendix \ref{app:jsoc}.

It is very difficult to objectively assess the depth sensitivity of the inversion from figures like Fig.~\ref{pic:RLS_rakern_x_0.5}. Therefore, we defined two numerical indicators to describe the characteristic depth $\langle z \rangle$ and a vertical width $\zwidth$ of the averaging kernel:
\begin{equation}
\langle z \rangle = \int \limits_{\odot} \mathrm{d}^2 \boldsymbol{r}'\, \mathrm{d}z\, \mathcal{K}^{\alpha}_{\alpha} z \mathrm{,} \quad \quad \zwidth = \sqrt{\langle z^2 \rangle - \langle z \rangle^2}. 
\end{equation}
The angle brackets signify the vertical average with the weighting by the averaging kernel. 

In Table \ref{tab:RLS_stat} we give $\langle z \rangle$ and $\zwidth$ for the first four considered depths of the JSOC inversions. For completeness, we also give the root-mean-square (RMS) values of the inverted maps. It can be seen that the depth sensitivity for the first three depths is nearly the same, as demonstrated also by \citet{DeGrave_2015} in their Fig.~10. The averaging kernels for all these inversions have a similar extension between the surface to around 4 Mm of depth with a peak at 2 Mm. The inversion for the last depth 5--7~Mm has a larger sensitivity to depth of about 5.5 Mm depth. In all cases, the averaging kernel is not localised around the indicated target depth. 

A similar sensitivity for inversions at different indicated depths should in principle lead to a larger correlation between inversions at different depths. The matrix of the correlation coefficients can be found in Table \ref{tab:RLS_corr}. The correlation coefficients of inverted flows at different depths are large, as expected from the shape of the averaging kernels. On the other hand, something unexpected happens between the depths of 1--3 and 3--5 in the case of inversion for the speed of sound, where the correlation coefficient with shallower depths tends to zero or even becomes negative. We do not have an explanation for such behaviour. Inversions dominated by noise or a strong but varying contribution of the cross-talk might explain such an issue. 


\begin{table}
\caption{Statistical properties of the JSOC models.}
\label{tab:RLS_stat}
\centering
\begin{tabular}{l c c c c}
\hline\hline
Target depth [Mm] & 0--1 & 1--3 & 3--5 & 5--7\\
$\langle z \rangle$ [Mm] & 3.4 & 2.8 & 3.5 & 4.9\\
$\zwidth$ [Mm] & 2.9 & 2.4 & 2.8 & 3.4\\
\hline
RMS ($v_x^{{\rm inv}}$) [\mps] & 38 & 58 & 42 & 23\\
RMS ($v_y^{{\rm inv}}$) [\mps] & 38 & 58 & 42 & 24\\
RMS ($\delta c_s^{{\rm inv}}$) [\mps] & 21 & 18 & 13 & 14\\
\hline
\end{tabular}
\end{table}

\begin{table}
\caption{Mutual correlation coefficients of the JSOC inverted maps}
\label{tab:RLS_corr}
\centering
\begin{tabular}{r r r r r}
\hline\hline
& 0--1 & 1--3 & 3--5 & 5--7\\
\hline
\multicolumn{5}{c}{$v_x^{{\rm inv}}$} \\
\hline
0--1 & 1.00 & 0.96 & 0.99 & 0.88\\
1--3 & 0.96 & 1.00 & 0.98 & 0.78\\
3--5 & 0.99 & 0.98 & 1.00 & 0.88\\
5--7 & 0.88 & 0.78 & 0.88 & 1.00\\
\hline
\multicolumn{5}{c}{$v_y^{{\rm inv}}$} \\
\hline
0--1 & 1.00 & 0.96 & 0.99 & 0.85\\
1--3 & 0.96 & 1.00 & 0.98 & 0.74\\
3--5 & 0.99 & 0.98 & 1.00 & 0.85\\
5--7 & 0.85 & 0.74 & 0.85 & 1.00\\
\hline
\multicolumn{5}{c}{$\delta c_s^{{\rm inv}}$} \\
\hline
0--1 & 1.00 & 0.89 & $-0.00$ & $-0.13$\\
1--3 & 0.89 & 1.00 & 0.02 & $-0.10$\\
3--5 & $-0.00$ & 0.02 & 1.00 & 0.58\\
5--7 & $-0.13$ & $-0.10$ & 0.58 & 1.00\\
\hline
\end{tabular}
\end{table}

\subsection{Our inversions}
Our time--distance inversion method is principally different from the JSOC inversion and is also very variable. Therefore, we have many options to obtain inverted estimates that we could use for comparisons. To demonstrate the variability, we select three approaches:

\begin{description}
        \item[\bf JSOC-like inversion] First, we use the JSOC averaging kernels as target functions for our inversions. JSOC results were obtained by separate inversions, when flows were inverted using the difference travel-time geometries and the perturbations of the sound speed were inverted using the mean travel-time geometry only. Additionally, in agreement with JSOC inversions, we focused on phase-speed filtered travel times \citep{Duvall_1997}, that is, we use TD1 to TD11 filters with five annulus radii for each filter. In the figures, this case is labelled OUR1.
        
        \item[\bf JSOC-like target] Second, we use the JSOC averaging kernels as the target for our inversion as in the previous case. However, the considered travel-time measurements fit our usual approach. That is, we use not only the phase-speed filtered travel times as in the previous step but also the ridge-filtered travel times for the $f$-mode ridge and the first four acoustic ridges. For each ridge, we consider 16 annulus radii. We combine both the difference and the mean travel-time geometries in one inversion. In the captions of the figures, this case is labelled OUR2. Except for the selection of the target, the inversion is principally the same as described by \cite{Korda_Svanda_2019}. 
        
        \item[\bf JSOC-indicated target] Finally, we combine all the travel-time geometries given in the previous paragraph but use a different target function. We set the target to have a Gaussian shape in both the horizontal and the vertical directions with their respective full widths at half maximum (FWHM) to represent the depth-sensitivity indicated by the JSOC data description. In the vertical direction, the Gaussian is further multiplied by a smooth function so that it reaches zero before the surface at $z = 0$. The target function is localised around the selected target depth. The details of the selected target functions are summarised in Table \ref{tab:target}. Inversions with this setup are labelled OUR3 in the figures and tables.

\end{description}

\begin{table}
\caption{Parameters and properties of the target functions. Target depths and $\mathrm{FWHM_z}$ are chosen to correspond to indicated JSOC depths}
\label{tab:target}
\centering
\begin{tabular}{l r r r r}
\hline\hline
\multicolumn{5}{c}{Target functions}\\
\hline
$\mathrm{FWHM_h}$ [Mm] & 10 & 10 & 10 & 10\\
$\mathrm{FWHM_z}$ [Mm] & 0.5 & 1 & 1 & 1\\
Target depth [Mm] & 0.5 & 2 & 4 & 6\\
$\langle z \rangle$ [Mm] & 0.5 & 2.1 & 4.1 & 6.1\\
$\zwidth$ [Mm] & 0.2 & 0.4 & 0.4 & 0.4\\
\hline
JSOC corresponding depth [Mm] & 0--1 & 1--3 & 3--5 & 5--7\\
\hline
\end{tabular}
\end{table}

For each approach, we compare the shapes of the averaging kernels and give examples for the depth of 2 Mm. For the same depth, we also give maps of inverted quantities, we select $v_x$ to represent the horizontal flow and $\delta c_s$ for the speed of sound. The figures for the remaining depths can be found in Appendices. We note that the colour scales are intentionally not the same for all plots to allow for better visibility of the structure of the inverted maps. 

For all four considered depths, we provide statistical properties of the inversions at each target depth, and also correlation coefficients with the JSOC inversions for each quantity. 


\begin{figure*}
\sidecaption
\includegraphics[height=8cm]{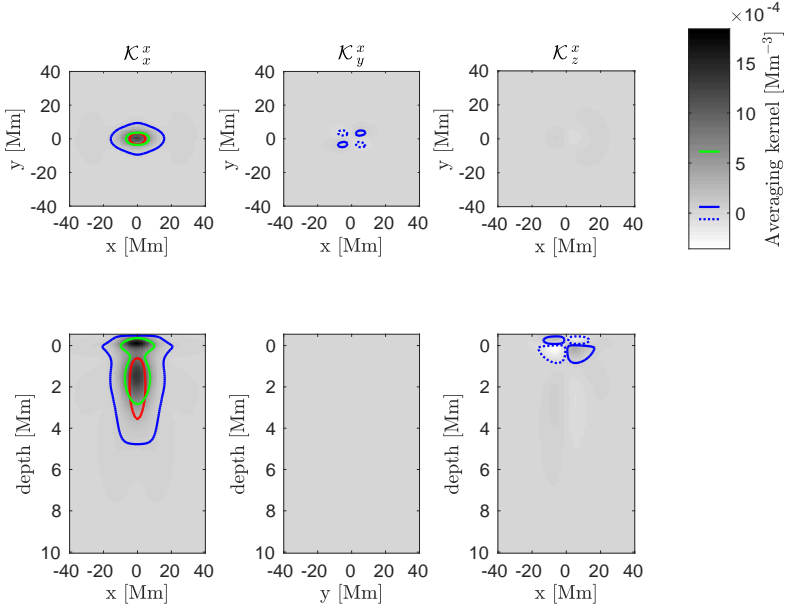}
\caption{Averaging kernels for $v_x$ inversion at the depth of 2.0 Mm, JSOC-like inversion. See Fig. \ref{pic:RLS_rakern_x_0.5} for details.}
\label{pic:SOLA_rakern_x_2.0_Adiff}
\end{figure*}

\begin{figure*}
\sidecaption
\includegraphics[height=8cm]{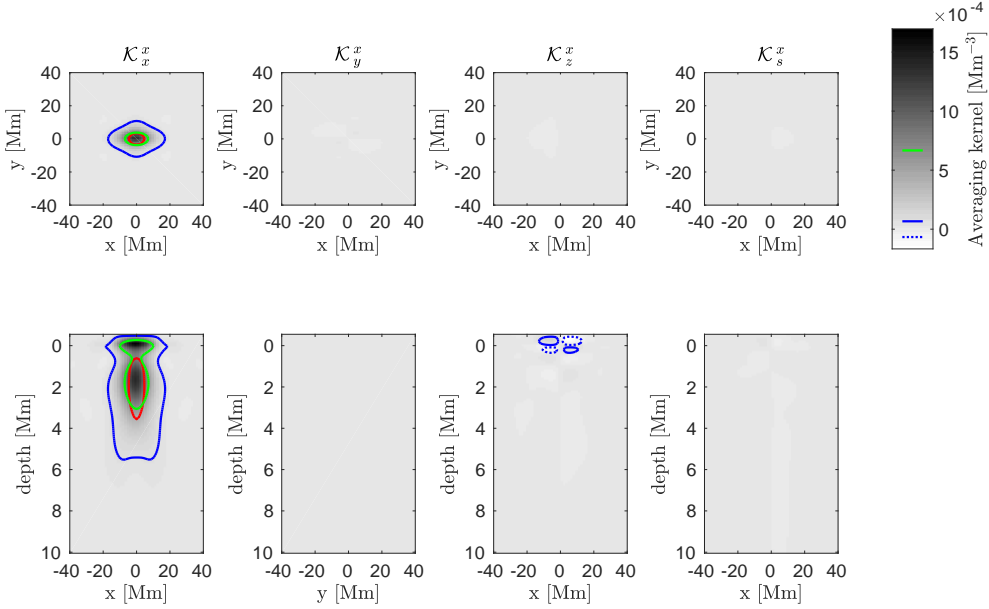}
\caption{Averaging kernels for $v_x$ inversion at the depth of 2.0 Mm, JSOC-like target. See Fig. \ref{pic:RLS_rakern_x_0.5} for details. We note that compared to Fig.~\ref{pic:SOLA_rakern_x_2.0_Adiff} an additional column appears as the sound-speed perturbations are also considered in the inversion. }
\label{pic:SOLA_rakern_x_2.0_A}
\end{figure*}

\begin{figure*}
\sidecaption
\includegraphics[height=8cm]{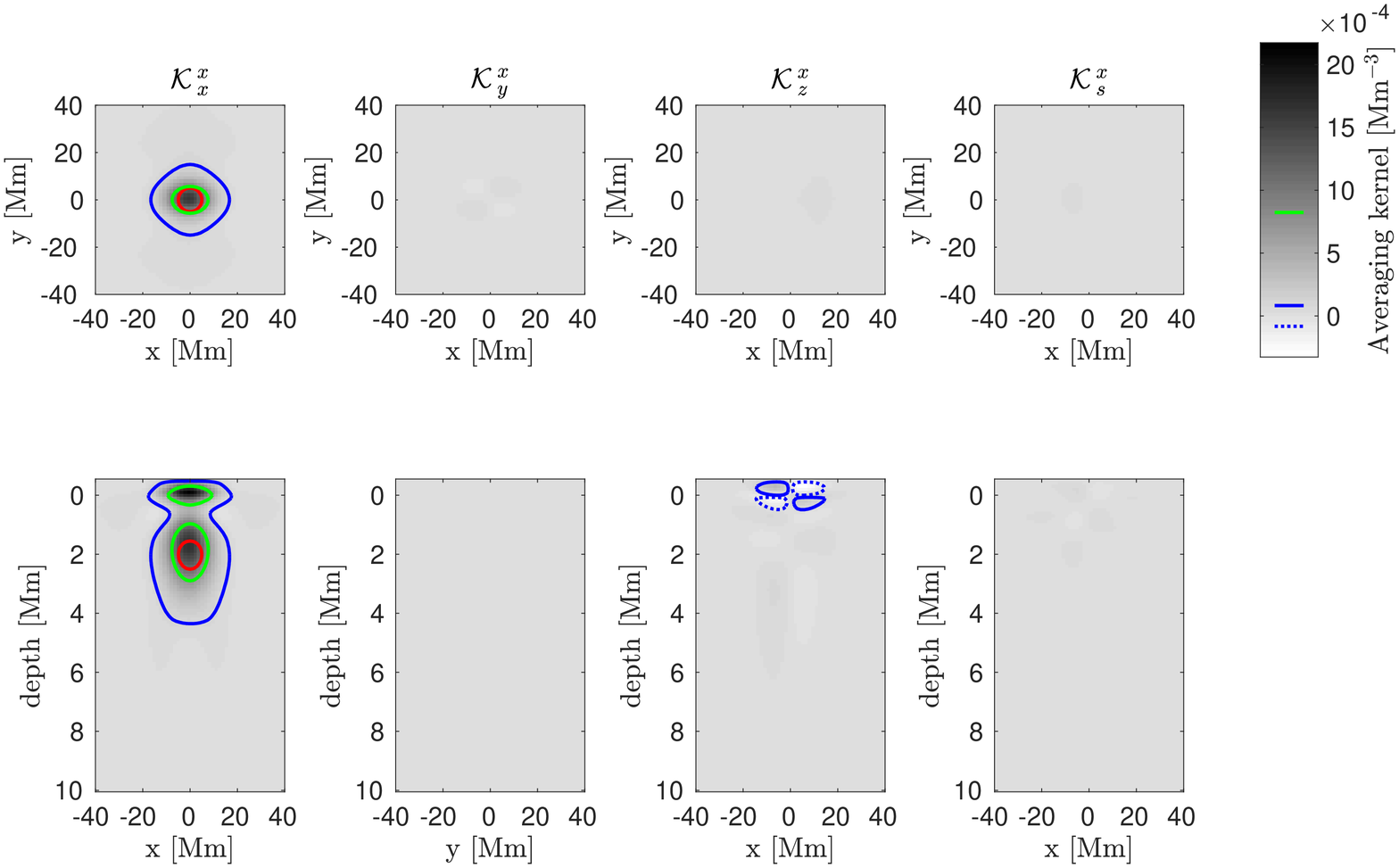}
\caption{Averaging kernels for $v_x$ inversion at the depth of 2.0 Mm, JSOC-indicated target. See Fig. \ref{pic:RLS_rakern_x_0.5} for details.}
\label{pic:SOLA_rakern_x_2.0_GdG}
\end{figure*}

\begin{figure*}
\sidecaption
\includegraphics[width=0.7\textwidth]{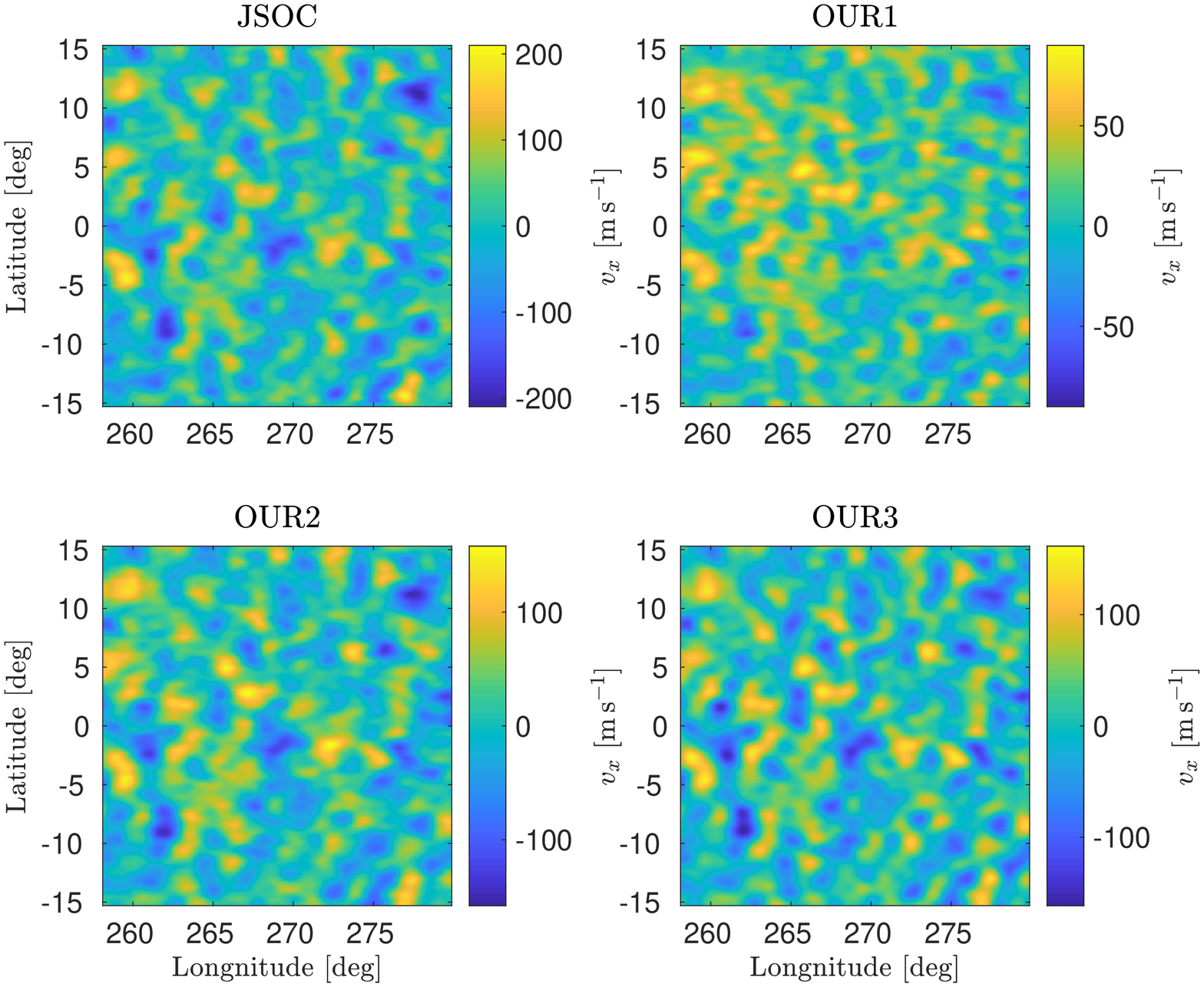}
\caption{Inversions for $v_x^{{\rm inv}}$ at 0.5 Mm depth.} 
\label{pic:vx_2.0}
\end{figure*}

\section{Results}
For a detailed description of the similarities and differences in the two inversion pipelines we picked up a depth of 2.0~Mm, which corresponds to the indicated depth sensitivity of 1--3~Mm in JSOC products. This depth is not so large as to be strongly influenced by random noise, but it also is not so shallow as to be strictly limited to the surface only. The results are generalised for the remaining three depths further on in the paper. 

\subsection{Horizontal flows}
The horizontal components of the flow vector are considered to be standard products of almost any time--distance inversion pipelines. The near-surface maps are usually dominated by flows within supergranular cells \citep[e.g.][]{Duvall2000}. These supergranular flows are nearly horizontal, almost radial within a cell and have an amplitude of a few hundred meters per second. Clear signatures of the supergranules make the comparison between various time--distance results easier. 

\subsubsection{JSOC-like inversion}
\label{sect:vx1}

Figure \ref{pic:SOLA_rakern_x_2.0_Adiff} shows our averaging kernel for $v_x$ inversion at 2.0 Mm depth, to be compared with Fig.~\ref{pic:RLS_rakern_x_2.0} from JSOC, which served as the target for our inversion. The green solid curve corresponds to the half-maximum of the averaging kernel at the target depth and the red solid curve corresponds to the half-maximum of the target function at the target depth (which again is the JSOC averaging kernel). One can see that the inversion fits the target function well but that there is an additional sensitivity at the surface in our averaging kernel. This leads to a lower $\langle z \rangle = 2.3$ Mm, when compared with the corresponding JSOC inversion. The vertical width $\zwidth = 2.4$~Mm is slightly larger than the case of the target function. The panels in the middle and right columns indicate the leakage of the other-than-wanted flow components, which translates to the cross-talk. The level of cross-talk is much lower than that of the corresponding JSOC inversion. Because we do not know the real quantity, we cannot compute the level of the cross-talk. We can estimate magnitudes of the cross-talk components of the averaging kernel as an integral of absolute value of $\mathcal{K}^x_y$ and $\mathcal{K}^x_z$. In the case of $\mathcal{K}^x_y$, the ratio of JSOC to JSOC-like inversion is 4.95. In the case of $\mathcal{K}^x_z$, the ratio of JSOC to JSOC-like inversion is 6.21. That is because our SOLA-based approach allows for cross-talk minimisation, whereas JSOC RLS-based inversion does not. The level of random noise for our inversion is about 4~\mps, which is about half of the JSOC-indicated noise level. Here we have to note that our random-noise level estimates are based on a proper estimate from the travel-time noise covariance matrix (see Eq.~\ref{eq:noise}), while the noise-level estimate for JSOC is based on empirical reproducibility of the inverted flow patterns \citep{JSOC_pipeline}. 

Figure \ref{pic:vx_2.0} compares the maps of $v_x^{\rm inv}$ of both pipelines at depths of 2.0~Mm (OUR) and 1--3 Mm (JSOC). The top left panel shows a result of the JSOC pipeline and the top right panel shows our result obtained by mimicking JSOC inversion using our inversion pipeline. The RMS of the JSOC result is 58~\mps\ and the RMS of our result is 23~\mps. The correlation coefficient of these two maps is equal to 0.89. The flow structure is comparable; the lower magnitude in the case of our inversion is caused by a larger-than-expected smoothing in the horizontal direction, where our averaging kernel is more extended than that of JSOC. 

\subsubsection{JSOC-like target}
In the case of our second approach, the inversion has more degrees of freedom to fit the target in a form of the JSOC averaging kernel. A new averaging kernel is shown in Fig. \ref{pic:SOLA_rakern_x_2.0_A}. Despite the fact that the misfit between the averaging kernel and the target function is much smaller than was the case in Section~\ref{sect:vx1}, the inversion is still partially sensitive at shallower depths. Together with a low sensitivity at greater depths, the mean value of inverted depth $\langle z \rangle = 2.6$ Mm is about 8\% lower than expected. The vertical extend $\zwidth = 2.4$~Mm is equal to $\zwidth$ of the target function. The fit of the target function by the averaging kernel is very reasonable and also the level of the cross-talk is sufficiently small. The estimated random-noise level for this inversion is 7~\mps, and again the JSOC random-noise estimate is about the same value. 
In Fig. \ref{pic:vx_2.0} (bottom left) one can see a map of the inverted $v_x^{\rm inv}$ using our pipeline. The result is nearly the same as the JSOC result in both amplitude and structure. The RMS of our inversion is 48~\mps and the RMS of the JSOC inversion is 58~\mps. The correlation coefficient between these inversions is 0.93. 

\subsubsection{JSOC-indicated target}
Lastly, for the horizontal flow we used JSOC-indicated target depth and ran a full inversion. The averaging kernel is displayed in Fig. \ref{pic:SOLA_rakern_x_2.0_GdG}. Because of the noise minimisation, there is a higher-than-expected sensitivity at surface regions. Otherwise, the fit of the target function (a Gaussian placed at the depth of 2.0~Mm) is very good. At the same time, the level of the cross-talk is sufficiently low. The mean inverted depth, $\langle z \rangle = 1.6$ Mm is in good agreement with the mean target depth. On the other hand, the vertical extent, $\zwidth=1.2$~Mm, is larger than expected. The averaging kernel must be more extended than the well-localised target function because of averaging of the random noise in a larger volume. The estimate of the random noise is 6~\mps{} for this inversion.

In Fig. \ref{pic:vx_2.0} one can compare maps of inverted $v_x^{\rm inv}$ of both pipelines (our result in the bottom right panel, the JSOC result in the top left panel) at depths of 2.0 Mm and 1--3 Mm. The flow patterns and the RMS are comparable in both inversions. The RMS of our inversion is 50~\mps and the RMS of the JSOC inversion is 58~\mps. The correlation coefficient at 2.0 Mm is 0.94. 

The correlation coefficient is much the same also at 0.5 Mm because of similar sensitivity of the JSOC inversion at these depths. At greater depths, the correlation decreases because of different localizations of ours and the JSOC averaging kernels. At 6.0 Mm the correlation coefficient is only 0.25. At other depths, the RMS of our results is higher because of the better localised averaging kernel which is connected with lower smearing of the real quantity.

\vspace{\baselineskip}
\noindent The observed similarity of the maps of the horizontal flow is a consequence of three facts. First, the sensitivity of the JSOC inversion expressed by the averaging kernels is not particularly localised in depth. The JSOC inversion takes an average of real quantity $\delta q^{\alpha}$ in a large range of depths, from the surface up to around 4 Mm. Second, there is a small vertical gradient of the inverted quantities in the Sun. Lastly, the large amplitude of the inverted quantity plays an important role, which is reflected by the large signal-to-noise ratio. Furthermore, a leakage from the other components is negligible. These features collectively lead to very similar results of the JSOC pipeline and our pipeline in the case of inversions for the horizontal flow components.

The statistical results for all target depths can be found in Table \ref{tab:stat_horizontal}. Figures similar to Figs. \ref{pic:SOLA_rakern_x_2.0_Adiff} and \ref{pic:vx_2.0} for other target depths can be found in Appendix \ref{app:horizontal}.

\begin{table}[!ht]
\caption{Statistical properties of $v_x^{{\rm inv}}$}
\label{tab:stat_horizontal}
\centering
\begin{tabular}{l r r r r}
\hline\hline
Target depth [Mm] & 0.5 & 2 & 4 & 6\\
\hline
\multicolumn{5}{c}{JSOC-like inversion} \\
\hline
corr(OUR1, JSOC) & 0.86 & 0.89 & 0.88 & 0.82\\
RMS (OUR1) [\mps] & 18 & 23 & 20 & 25\\
$\sigma_x$ [\mps] & 3 & 4 & 4 & 8\\
$\langle z \rangle$ [Mm] & 2.7 & 2.3 & 2.8 & 4.2\\
$\zwidth$ [Mm] & 2.7 & 2.4 & 2.7 & 3.3\\
\hline
\multicolumn{5}{c}{JSOC-like target} \\
\hline
corr(OUR2, JSOC) & 0.93 & 0.93 & 0.93 & 0.84\\
RMS (OUR2) [\mps] & 36 & 48 & 39 & 24\\
$\sigma_x$ [\mps] & 5 & 7 & 5 & 4\\
$\langle z \rangle$ [Mm] & 3.0  & 2.6 & 3.2 & 3.9\\
$\zwidth$ [Mm] & 2.6 & 2.4 & 2.7 & 3.0\\
\hline
\multicolumn{5}{c}{JSOC-indicated target} \\
\hline
corr(OUR3, JSOC) & 0.91 & 0.94 & 0.79 & 0.25\\
RMS (OUR3) [\mps] & 94 & 50 & 59 & 34\\
$\sigma_x$ [\mps] & 9 & 6 & 21 & 16\\
$\langle z \rangle$ [Mm] & 1.6 & 1.6 & 4.7 & 5.8\\
$\zwidth$ [Mm] & 2.0 & 1.2 & 1.9 & 2.3\\
\hline
\end{tabular}
\end{table}


\subsection{Sound-speed perturbations}

\begin{figure}
\sidecaption
\includegraphics[height=8cm]{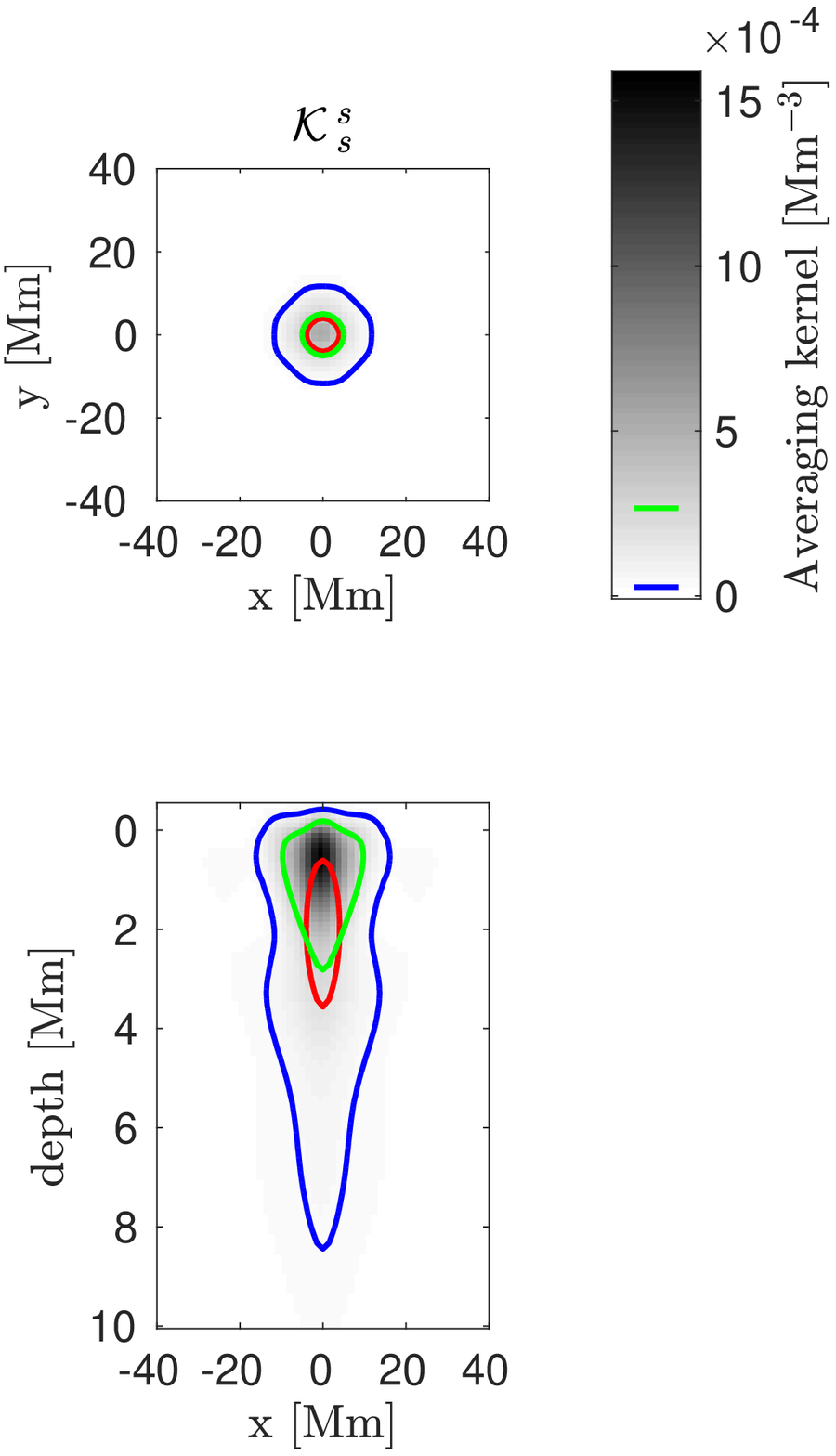}
\caption{Averaging kernels for $\delta c_s$ inversion at the depth of 2.0 Mm, JSOC-like inversion. See Fig. \ref{pic:RLS_rakern_x_0.5} for details.}
\label{pic:SOLA_rakern_s_2.0_Adiff}
\end{figure}

The inversions for the sound-speed perturbations (wave-speed perturbations in the JSOC terminology) are other typical data products of time--distance helioseismology; they are often connected with temperature variations (according to the ideal gas equation of state, the sound speed is proportional to the square root of the gas temperature). From the models \citep[e.g.][]{Rempel_2014,DeGrave_2014} it follows that the sound-speed perturbation amplitudes are about an order of magnitude smaller than the horizontal flow velocities in the quiet-Sun regions. 

JSOC inversions also use principally different travel-time measurements. Whereas for the horizontal flow the difference travel-time geometries are used, for the wave-speed the mean travel-time geometry is employed. This needs to be taken into account. 

\begin{figure*}
\sidecaption
\includegraphics[height=8cm]{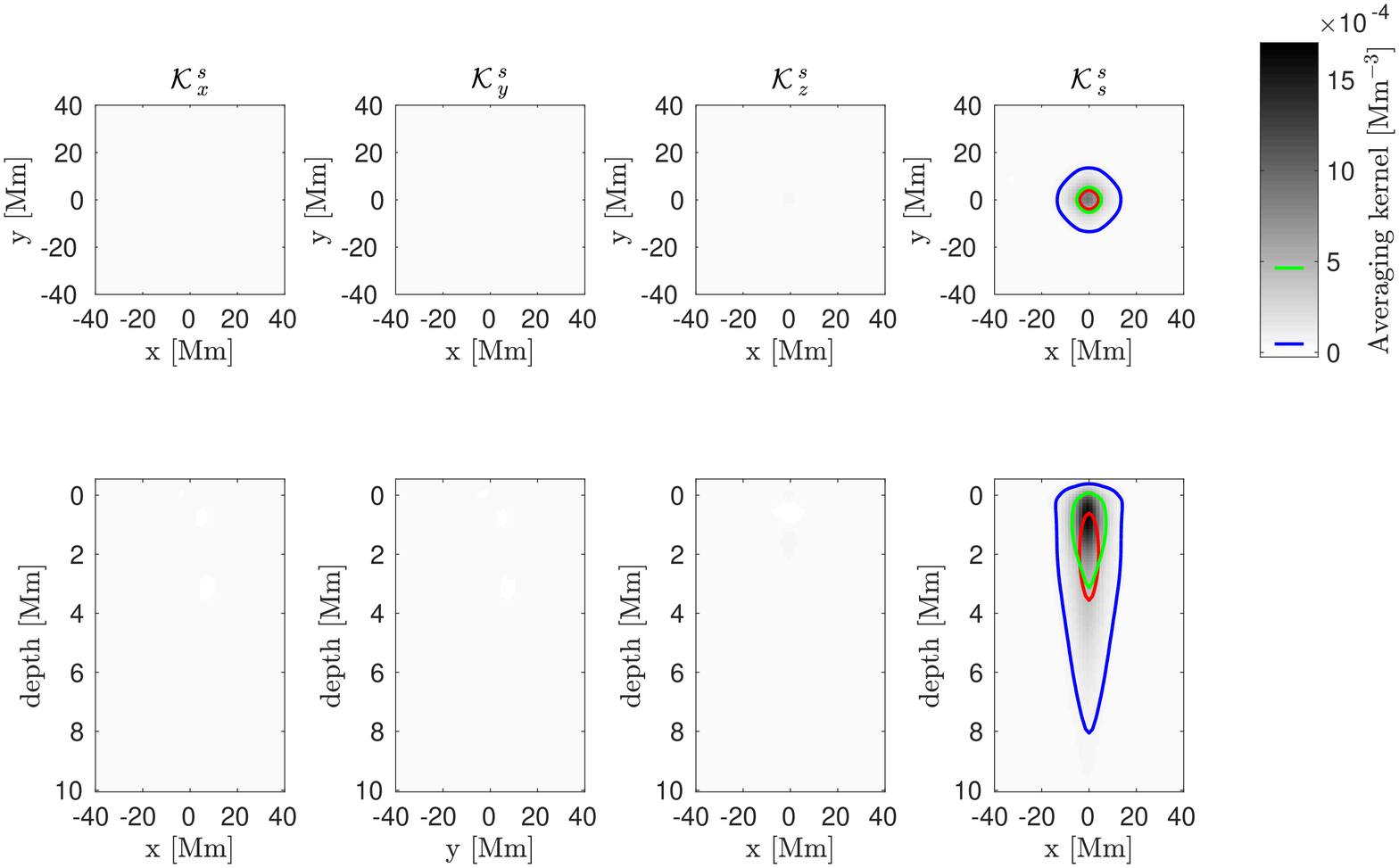}
\caption{Averaging kernels for $\delta c_s$ inversion at the depth of 2.0 Mm, JSOC-like target. See Fig. \ref{pic:RLS_rakern_x_0.5} for details. We note that compared to Fig.~\ref{pic:SOLA_rakern_s_2.0_Adiff} there are three additional columns as all three components of the flow are also considered in this inversion. }
\label{pic:SOLA_rakern_s_2.0_A}
\end{figure*}

\begin{figure*}
\sidecaption
\includegraphics[height=8cm]{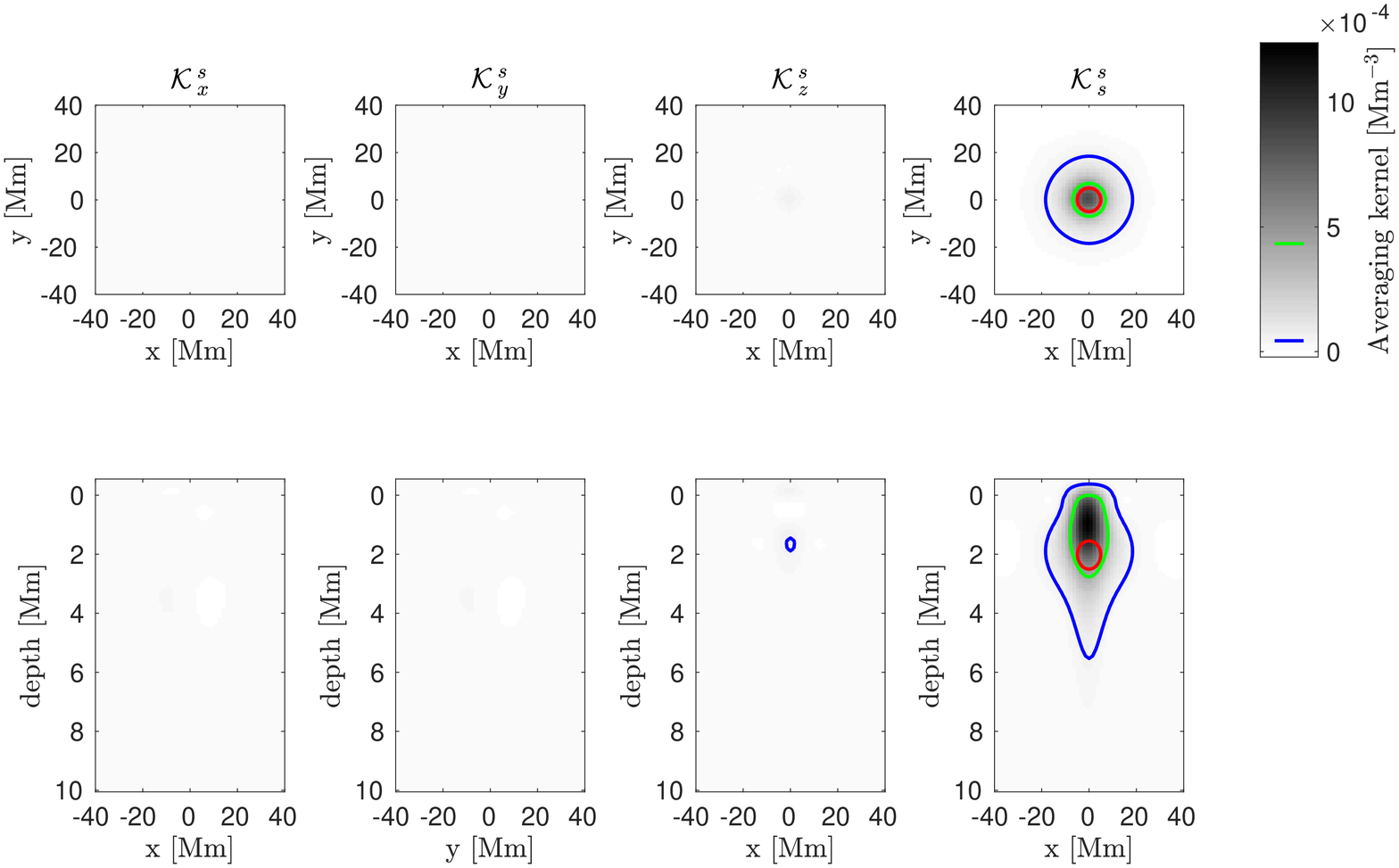}
\caption{Averaging kernels for $\delta c_s$ inversion at the depth of 2.0 Mm, JSOC-indicated target. See Fig. \ref{pic:RLS_rakern_x_0.5} for details.}
\label{pic:SOLA_rakern_s_2.0_GdG}
\end{figure*}

\subsubsection{JSOC-like inversion}
For the JSOC-like inversions we also combined only mean travel-time sensitivity kernels in our inversion. The corresponding averaging kernel is shown in Fig.~\ref{pic:SOLA_rakern_s_2.0_Adiff}. The inversion is more sensitive around a depth of 1 Mm (see red line in Fig. \ref{pic:rakern_h_cs_all}). The mean depth, $\langle z \rangle = 3.1$ Mm is more than targeted which is also a consequence of sensitivity of the inversion at greater depths. The sensitivity at first 3 Mm is only $1.44$ times larger than sensitivity of deeper layers. The variations of the depth, $\zwidth = 2.7$ Mm, is also higher. The fit of the target function is not perfect because the noise minimisation is stronger. The noise estimate of this inversion is 3 \mps. The cross-talk cannot be evaluated because the inversion setup does not consider contributions from other quantities.

In Fig. \ref{pic:cs_2.0} one can see the maps of the sound-speed perturbations. 
The RMSs of ours and the JSOC inversion are 10~\mps and 18~\mps, respectively. The correlation between these two inversions is 0.35. The difference may be caused by different cross-talk contributions. It is not possible to minimise the cross-talk in this inversion setup, but its contribution might be important. The signal-to-noise ratios for the two inversions are also smaller than the case of the inversion for horizontal flows, which also decrease the correlation naturally. 

\subsubsection{JSOC-like target}
\begin{figure*}
    \sidecaption
\includegraphics[width=0.7\textwidth]{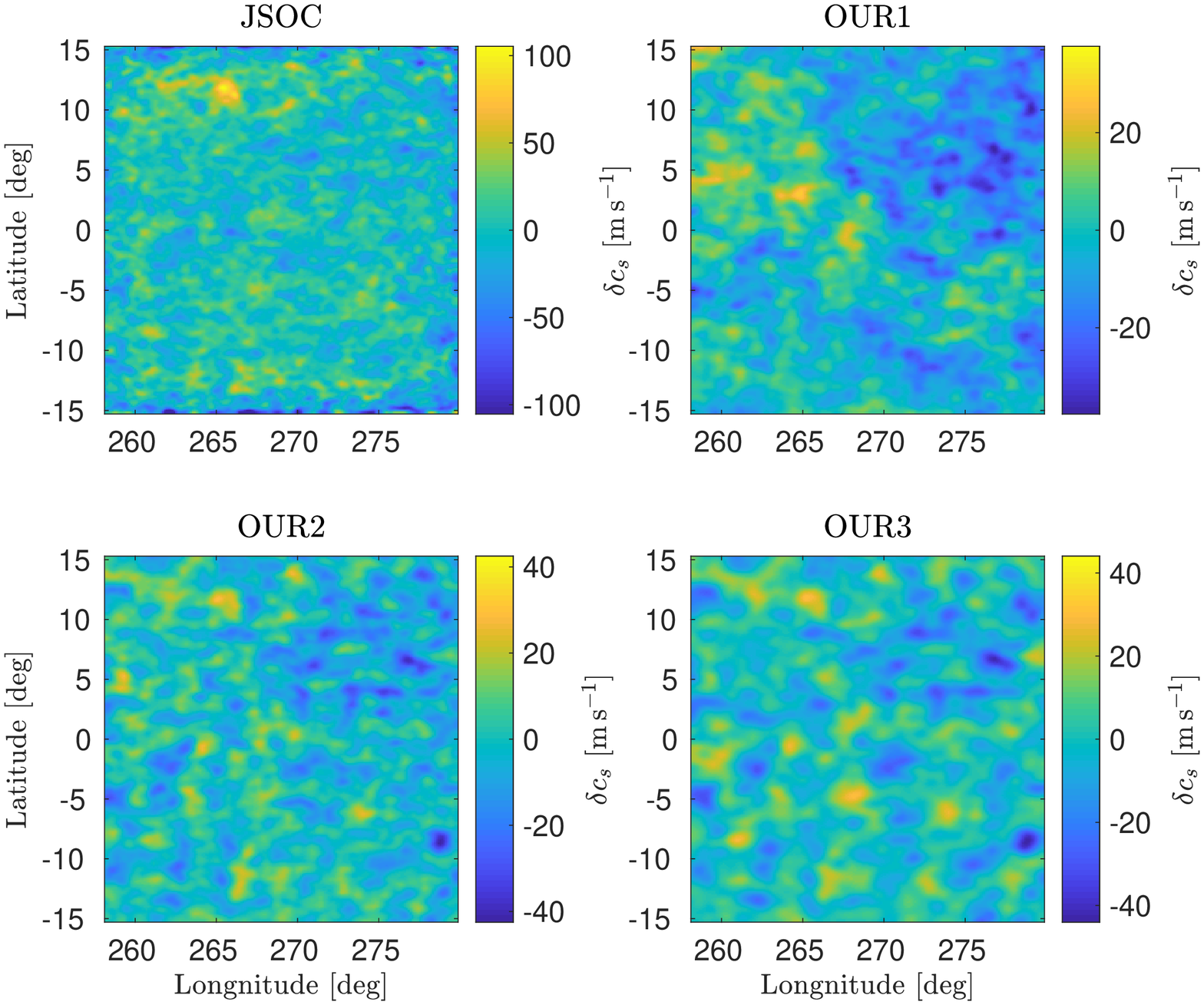}
\caption{Inversions for $\delta c_s^{{\rm inv}}$ at 2.0 Mm depth.}
\label{pic:cs_2.0}
\end{figure*}

Combined inversion of flows and sound-speed perturbations provides more information about the inversion. In the case of the sound-speed perturbations, especially information about the cross-talk between the flows and the sound-speed perturbations is given. Figure \ref{pic:SOLA_rakern_s_2.0_A} shows the averaging kernels for our $\delta c_s$ inversion. The fit of the target function by the averaging kernels is quite convincing and the level of the cross-talk is negligible. The mean inverted depth $\langle z \rangle = 2.7$ Mm is nearly the same as the mean target depth. The variation of the averaging kernel in depth, $\zwidth = 2.7$ Mm, is about 13\% larger than the variation of the target function. The RMS of the random noise is 3 \mps{} for this inversion.

As in previous sections, the JSOC result at 1–3 Mm depth can be found in the top left corner and JSOC-like target inversion at 2.0 Mm depth in the bottom left corner of Fig. \ref{pic:cs_2.0}. The RMSs of our map and of the JSOC map are 9 \mps{} and 18 \mps, respectively. The correlation between the two maps is large with a correlation coefficient of 0.64. The significant correlation and larger RMS of the JSOC inversion than of our inversion together with a very similar localisation in the Sun is consistent with the JSOC inversions being contaminated by the positively correlated cross-talk from the other quantities. \citet{Korda_Svanda_2019} found that the cross-talk may consist of about one half of the inverted estimate in the case of the inversions for the sound speed. The differences in RMS might be explained by such cross-talk contamination. 

At a shallower depth where random noise is not so important, the correlation is high. At greater depths, the correlation decreases mostly because of random noise.

\subsubsection{JSOC-indicated target}
The inversion for the sound-speed perturbations using a Gaussian target function peaking at the depth of 2.0~Mm is the last comparison we performed. Figure~\ref{pic:SOLA_rakern_s_2.0_GdG} shows the plots of the corresponding averaging kernel. The fit of the target function is poor, and the resulting averaging kernel is significantly more sensitive to the surface layers. The cross-talk is negligible. The mean inverted depth, $\langle z \rangle = 2.4$ Mm, is about 14 \% larger than the targeted. This is caused by the fact that the averaging kernel does not decrease quickly enough with depth, which is also expressed by the large variation of the kernel at depth, $\zwidth = 2.3$~Mm. The noise contribution to this inversion is estimated to be 5 \mps. This is a very strong constraint, which is largely responsible for the poorer-than-expected fit of the target function and the averaging kernel. 

The map of the sound-speed perturbations for the JSOC-indicated target inversion at 2.0 Mm depth is shown in Fig. \ref{pic:cs_2.0} (bottom right). The JSOC result at 1--3~Mm depth is shown in the top left panel. The RMS of our result is 10~\mps, while the RMS of the JSOC result is 18~\mps. This is in agreement with the positively correlated contribution of the cross-talk. The JSOC RMS is higher at all depths. The correlation coefficient of inversion at 2.0 Mm is 0.64. At other depths, the correlation coefficients are lower but still high and positive.

The statistical properties for other depths are listed in Table~\ref{tab:stat_sound}. Figures similar to Figs. \ref{pic:SOLA_rakern_s_2.0_Adiff} and \ref{pic:cs_2.0} for other target depths can be found in Appendix \ref{app:sound}. One should note that the sound-speed perturbations in the quiet-Sun regions have small amplitudes, much smaller than the horizontal flow. As a consequence one needs to target a very low level of random noise. The inversion then leads to a very large misfit between the target function and the averaging kernels, especially for the larger depths. Virtually all the sound-speed inversions for all four investigated depths are sensitive towards the surface. Deep snapshot inversions for the sound-speed perturbations with averaging times of a few hours and signal-to-noise ratios larger than unity are not possible.  

\begin{table}[!ht]
\caption{Statistical properties of $\delta c_s^{{\rm inv}}$}
\label{tab:stat_sound}
\centering
\begin{tabular}{l r r r r}
\hline\hline
Target depth [Mm] & 0.5 & 2 & 4 & 6\\
\hline
\multicolumn{5}{c}{JSOC-like inversion} \\
\hline
corr(OUR1, JSOC) & 0.30 & 0.35 & 0.14 & 0.19\\
RMS (OUR1) [\mps] & 9 & 10 & 9 & 7\\
$\sigma_{s}$ [\mps] & 2 & 3 & 2 & 4\\
$\langle z \rangle$ [Mm] & 3.3 & 3.1 & 3.4 & 3.9\\
$\zwidth$ [Mm] & 2.8 & 2.7 & 2.8 & 2.9\\
\hline
\multicolumn{5}{c}{JSOC-like target} \\
\hline
corr(OUR2, JSOC) & 0.50 & 0.64 & 0.18 & 0.11\\
RMS (OUR2) [\mps] & 12 & 9 & 7 & 6\\
$\sigma_{s}$ [\mps] & 4 & 3 & 2 & 3\\
$\langle z \rangle$ [Mm] & 3.1 & 2.7 & 2.8 & 3.3\\
$\zwidth$ [Mm] & 2.7 & 2.7 & 2.8 & 3.0\\
\hline
\multicolumn{5}{c}{JSOC-indicated target} \\
\hline
corr(OUR3, JSOC) & 0.46 & 0.64 & 0.32 & 0.32\\
RMS (OUR3) [\mps] & 9 & 10 & 12 & 9\\
$\sigma_{s}$ [\mps] & 3 & 5 & 5 & 4\\
$\langle z \rangle$ [Mm] & 2.7 & 2.4 & 3.0 & 3.1\\
$\zwidth$ [Mm] & 2.8 & 2.3 & 2.8 & 3.0\\
\hline
\end{tabular}
\end{table}


\subsection{Vertical flows}
To complete the story, one piece of puzzle is missing. Our inversion scheme not only provides horizontal flow components and the wave speeds, as JSOC does, but it also allows us to reliably invert for the vertical flow $v_z$.  

We can follow the same approach as for the other physical quantities described above, but we cannot compare our results with JSOC data products. The methodology in principle also allows inversion for the $v_z$, however in the past this turned out to be unreliable \citep{Zhao2007}. 

The vertical flow has a similar issue as the sound-speed perturbations, which is a smaller magnitude in the near-surface layers. Since again the vertical flow is mostly composed of the vertical upflows in the centres of the supergranules and downflows at their edges, its amplitude is at least an order of magnitude smaller than the magnitude of the horizontal components \citep[e.g.][]{Duvall2010}. Consequently, there is significant danger of contamination by the cross-talk from the horizontal flow components with a much larger amplitude. The cross-talk is very likely the cause of the failed validation of the RLS inversion for the vertical flow by \cite{Zhao2007}. This issue was studied by \cite{svanda_2011} in detail. 

Therefore, the importance of the noise and cross-talk terms in the inversion cost function is much greater, leading naturally to a larger misfit between the target function and the averaging kernels. 

\begin{figure*}
\sidecaption
\includegraphics[height=8cm]{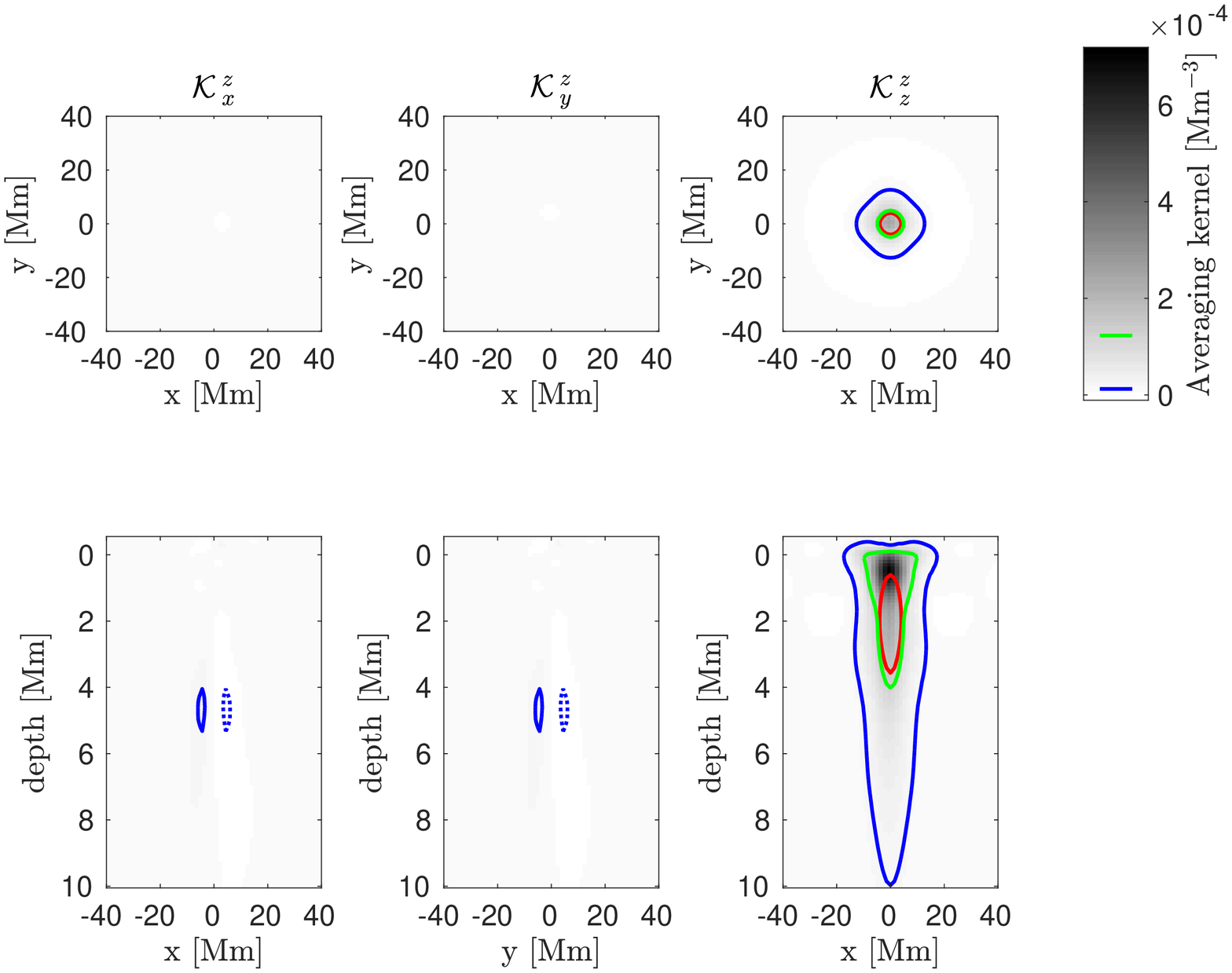}
\caption{Averaging kernels for $v_z$ inversion at the depth of 2.0 Mm, JSOC-like inversion. See Fig. \ref{pic:RLS_rakern_x_0.5} for details.}
\label{pic:SOLA_rakern_z_2.0_Adiff}
\end{figure*}

\begin{figure*}
\sidecaption
\includegraphics[height=8cm]{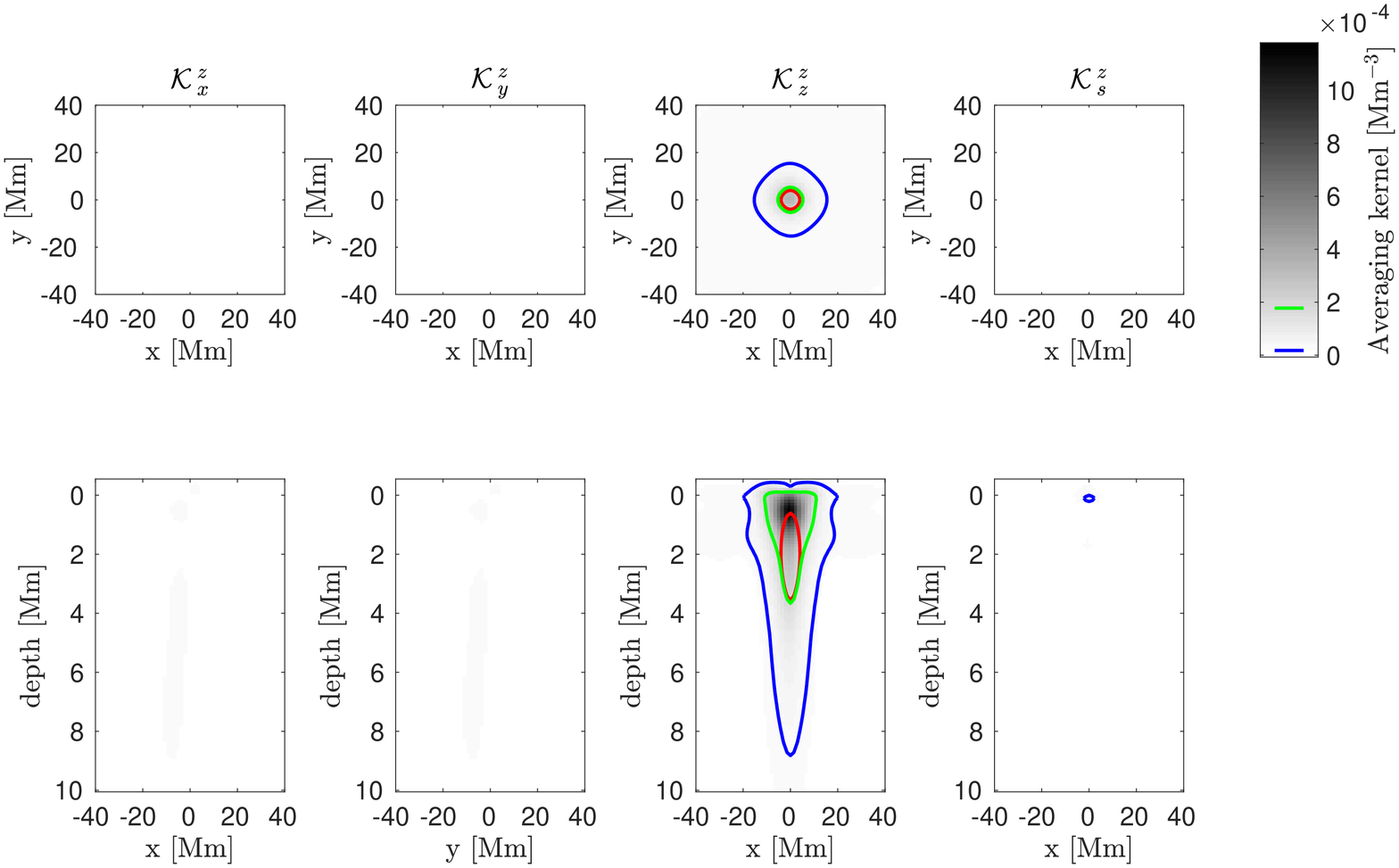}
\caption{Averaging kernels for $v_z$ inversion at the depth of 2.0 Mm, JSOC-like target. See Fig. \ref{pic:RLS_rakern_x_0.5} for details. We note that compared to Fig.~\ref{pic:SOLA_rakern_z_2.0_Adiff} there is an additional column because the sound-speed perturbations are also considered in this inversion. }
\label{pic:SOLA_rakern_z_2.0_A}
\end{figure*}

\begin{figure*}
\sidecaption
\includegraphics[height=8cm]{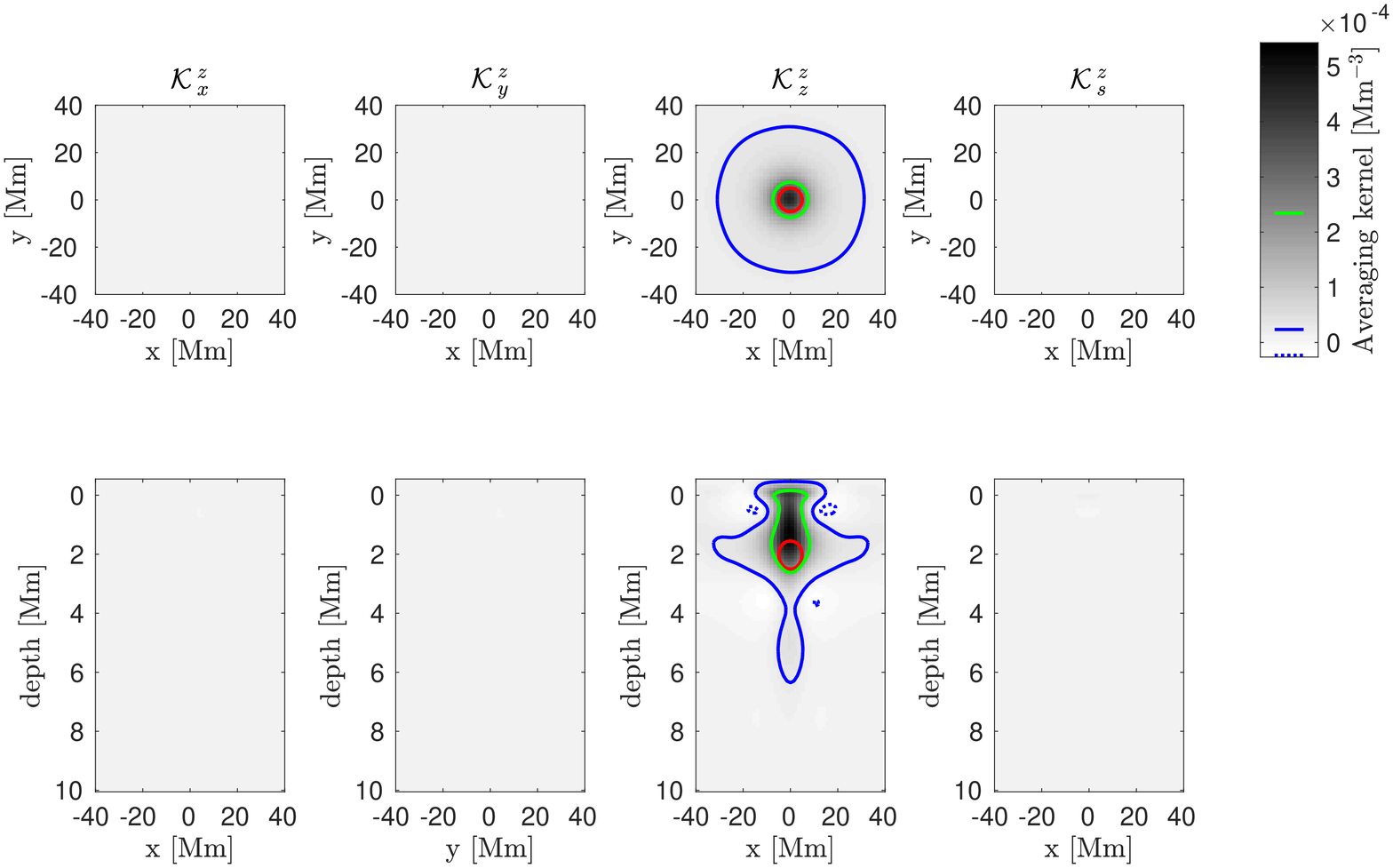}
\caption{Averaging kernels for $v_z$ inversion at the depth of 2.0 Mm, JSOC-indicated target. See Fig. \ref{pic:RLS_rakern_x_0.5} for details.}
\label{pic:SOLA_rakern_z_2.0_GdG}
\end{figure*}

That is already visible to the naked eye in Figs.~\ref{pic:SOLA_rakern_z_2.0_Adiff}, \ref{pic:SOLA_rakern_z_2.0_A}, and \ref{pic:SOLA_rakern_z_2.0_GdG}, where the averaging kernels for the $v_z$ inversion are plotted for the three approaches we use throughout this study. Obviously, the averaging kernels are strongly peaked in the near-surface layers and they slowly decrease toward greater depths. The mean calculated depths are $\langle z \rangle = 3.1$~Mm, $\langle z \rangle = 1.5$~Mm, and $\langle z \rangle = 1.4$~Mm respectively, with vertical extents of $\zwidth = 2.7$~Mm, $\zwidth = 1.6$~Mm, and $\zwidth = 1.3$~Mm, respectively. For all these inversions the cross-talk contribution is minimised. 

The above given numbers obtained for the target depth of 2.0~Mm may seem satisfactory. Unfortunately, as is evident from the summary Table~\ref{tab:vertical} for all the depths, the mean depths and the estimated vertical extents are similar for all the studied target depths; that is, the noise and cross-talk terms are so strongly regularised that the averaging kernels do not accurately represent the target function. The inversion is therefore naturally sensitive to the near-surface layers only, as evident from figures available in Appendix \ref{app:vertical}. 

This is in agreement with the validation study by \cite{svanda_2011}, where it was found that for travel times averaged over 24 hours, only flow inversions for up to 1~Mm depth yield inversion maps with the signal-to-noise ratio larger than unity. The inversions at larger depths are hopelessly buried in random noise.

\begin{figure*}
\includegraphics[width=\textwidth]{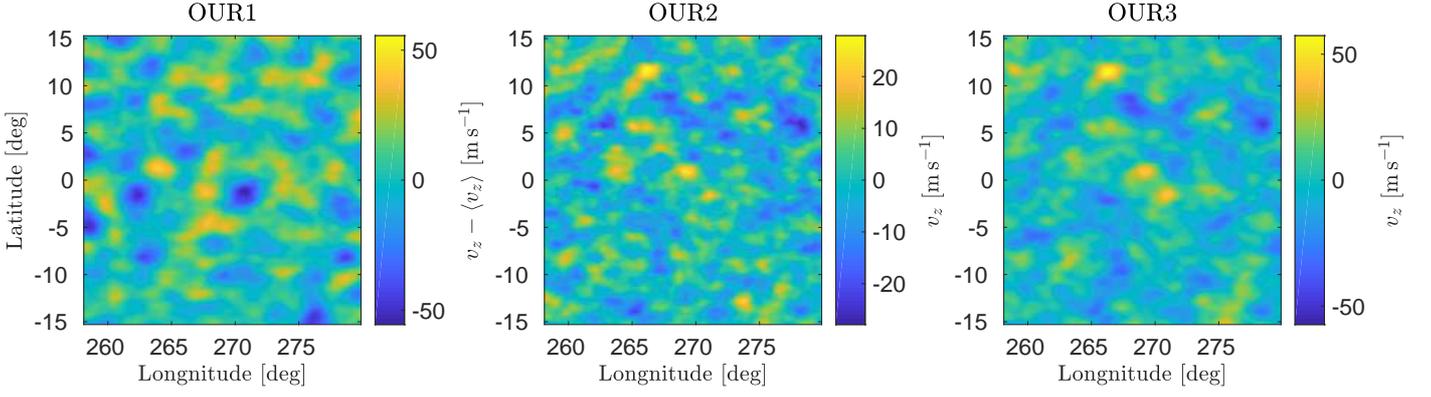}
\caption{Inversions for $v_z^{{\rm inv}}$ at 2.0 Mm depth.}
\label{pic:vz_2.0}
\end{figure*}


\begin{table}[!ht]
\caption{Statistical properties of $v_z^{{\rm inv}}$}
\label{tab:stat_vertical}
\centering
\begin{tabular}{l r r r r}
\hline\hline
Target depth [Mm] & 0.5 & 2 & 4 & 6\\
\hline
\multicolumn{5}{c}{JSOC-like inversion} \\
\hline
RMS (OUR1) [\mps] & 22 & 15 & 12 & 7\\
$\sigma_z$ [\mps] & 2 & 2 & 4 & 2\\
$\langle z \rangle$ [Mm] & 3.1 & 3.1 & 3.1 & 3.1\\
$\zwidth$ [Mm] & 2.7 & 2.7 & 2.7 & 2.7\\
\hline
\multicolumn{5}{c}{JSOC-like target} \\
\hline
RMS (OUR2) [\mps] & 6 & 7 & 5 & 8\\
$\sigma_z$ [\mps] & 2 & 3 & 2 & 3\\
$\langle z \rangle$ [Mm] & 1.5 & 1.5 & 1.5 & 2.2\\
$\zwidth$ [Mm] & 1.6 & 1.6 & 1.7 & 2.1\\
\hline
\multicolumn{5}{c}{JSOC-indicated target} \\
\hline
corr(OUR2, OUR3) & 0.75 & 0.79 & 0.33 & 0.43\\
RMS (OUR3) [\mps] & 16 & 12 & 7 & 3\\
$\sigma_z$[\mps] & 2 & 4 & 2 & 1\\
$\langle z \rangle$ [Mm] & 1.3 & 1.4 & 1.7 & 1.6\\
$\zwidth$ [Mm] & 1.6 & 1.3 & 1.9 & 1.9\\
\hline
\end{tabular}
\label{tab:vertical}
\end{table}

\section{Conclusions}

We compared inversion results from two independent travel-time time--distance inversion pipelines: RLS inversions implemented at the JSOC data centre and MC-SOLA inversions we developed recently. For mutual comparisons, we used three different setups for our inversion. First, we used the JSOC target function as an input and used the set of travel-time measurements comparable to those used in JSOC inversions, that is phase-speed filtered sensitivity kernels and separate inversions for flows and for the sound speed. Subsequently, we used the JSOC target function but implemented a much broader set of travel-time measurements in the inversion. Third, a full inversion was applied to the task when the target was constructed to agree to the parameters indicated by the JSOC metadata. 

Overall the JSOC measurements are accurately reproduced using our inversion pipeline. From the different comparisons we make the following conclusions.
\begin{itemize}
\item The agreement on the inverted estimates for the horizontal components of the flow is good in all cases. The correlation decreases with depth due to the increasing influence of random noise. Surprisingly, it appears that the correlation also decreases with an increasing number of degrees of freedom (the number of independent travel-time measurements combined in the inversion) and with a selection of the target function that does not resemble that of JSOC inversion. Both these effects are related to the increasing vertical extent of the averaging kernels. For instance, for the greatest considered depth 5--7~Mm the vertical extent of the averaging kernel is about 3.4~Mm for the JSOC inversion and about 2.3~Mm for our inversion. 
\item As for the sound-speed perturbations, the agreement is satisfactory. The RMS of the maps inverted by our method is smaller than that from JSOC. This discrepancy may be explained as an action of the cross-talk with flows, as demonstrated recently using synthetic data \citep{Korda_Svanda_2019}. Random-noise levels are also very important in the inversions for the sound-speed perturbations. The expected magnitudes are comparable to the magnitudes of the vertical component of the flow. Therefore, it is very difficult to target a localised inversion at a depth larger than some 1~Mm, again using the 24h averaged travel times. 
\end{itemize}

\begin{figure*}
        \includegraphics[width=0.30\textwidth]{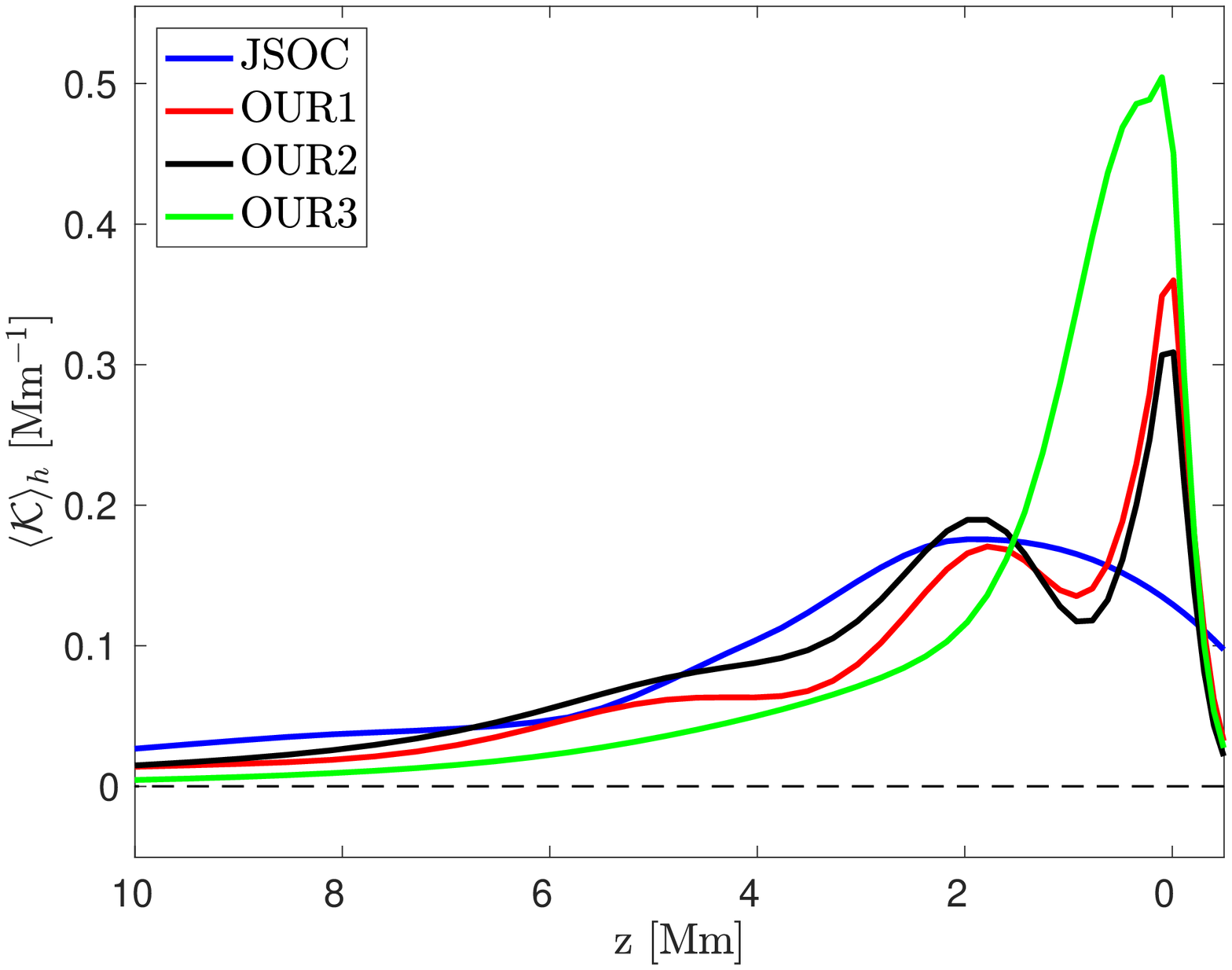}
        \includegraphics[width=0.30\textwidth]{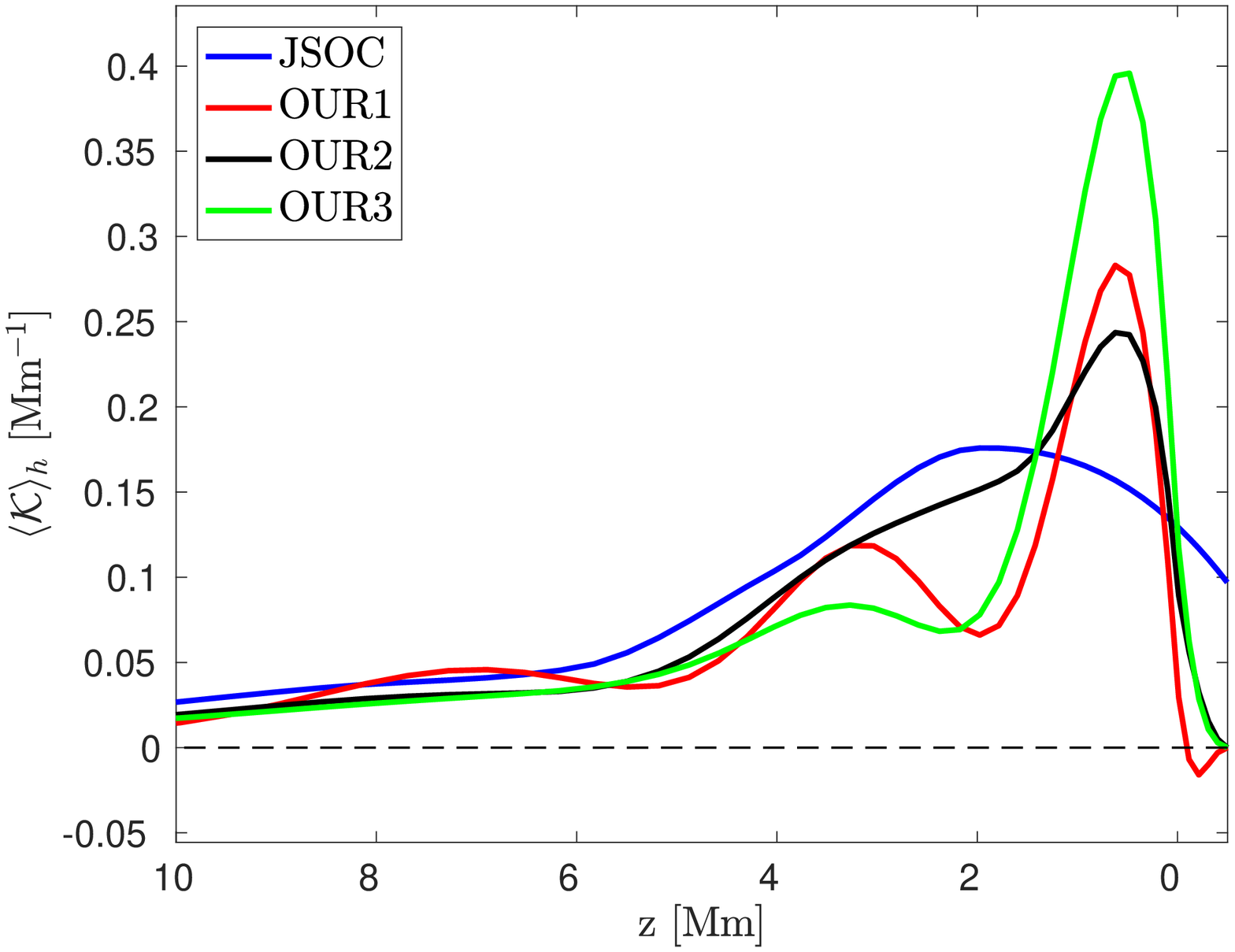}
        \includegraphics[width=0.30\textwidth]{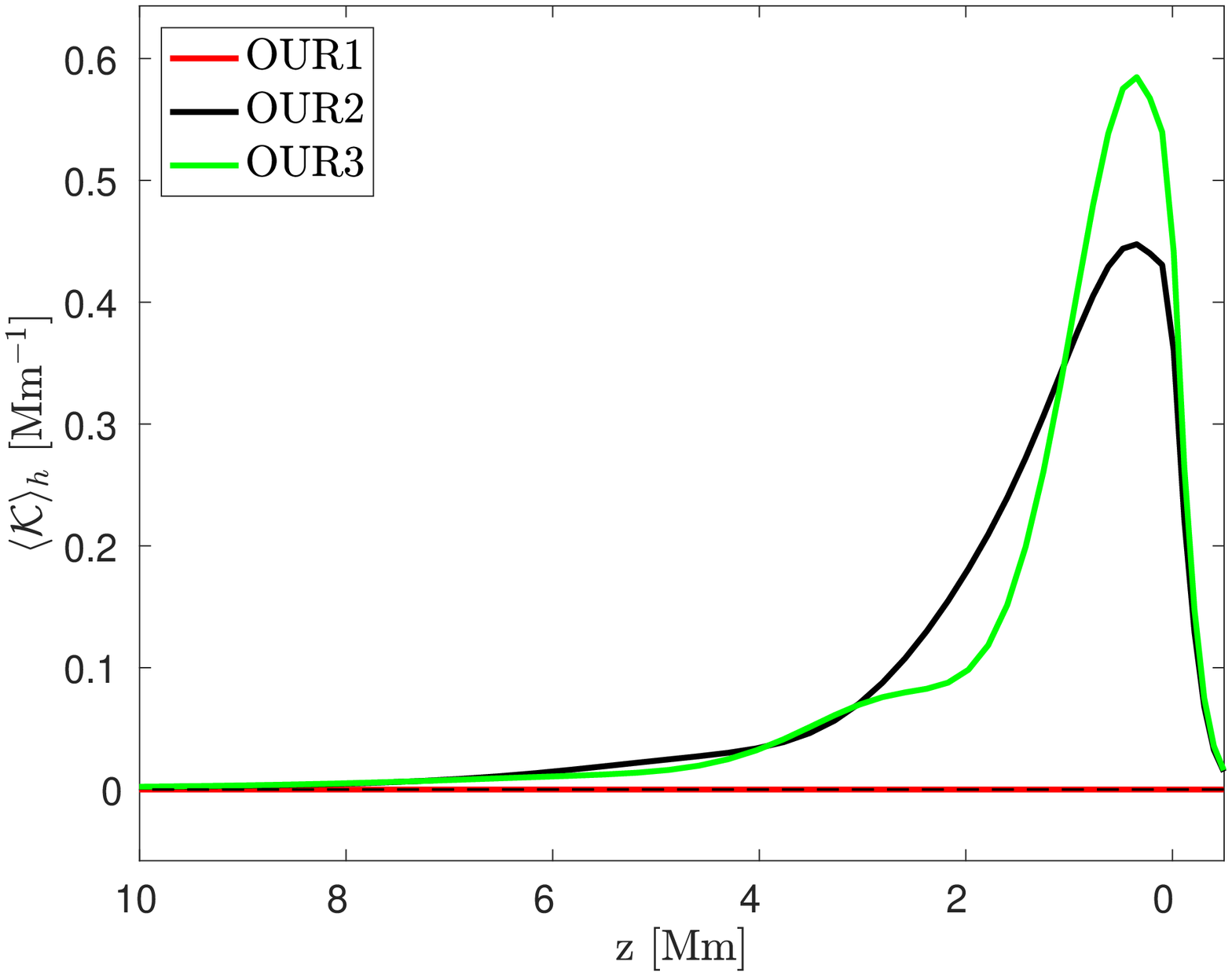}
        \\
        \includegraphics[width=0.30\textwidth]{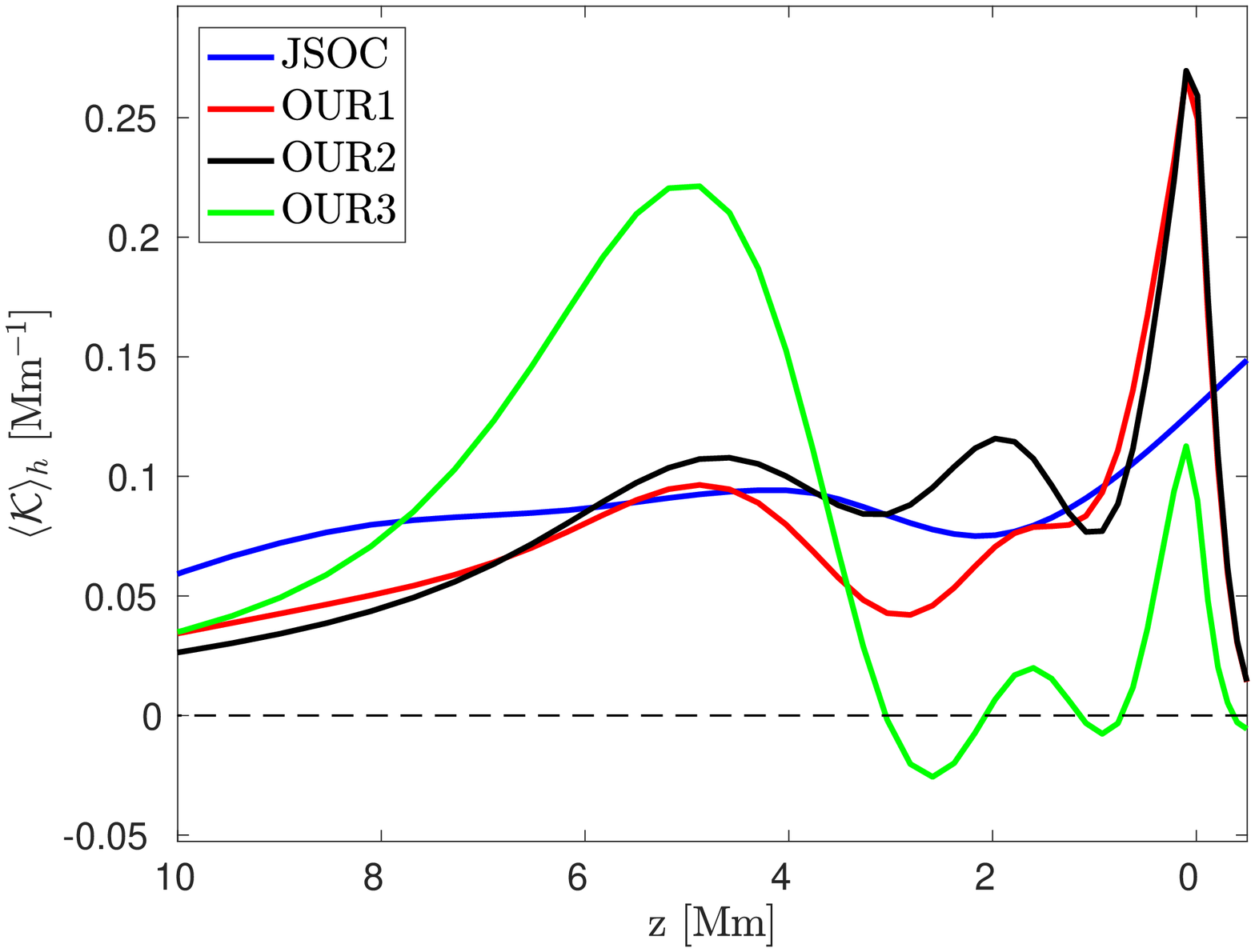}
        \includegraphics[width=0.30\textwidth]{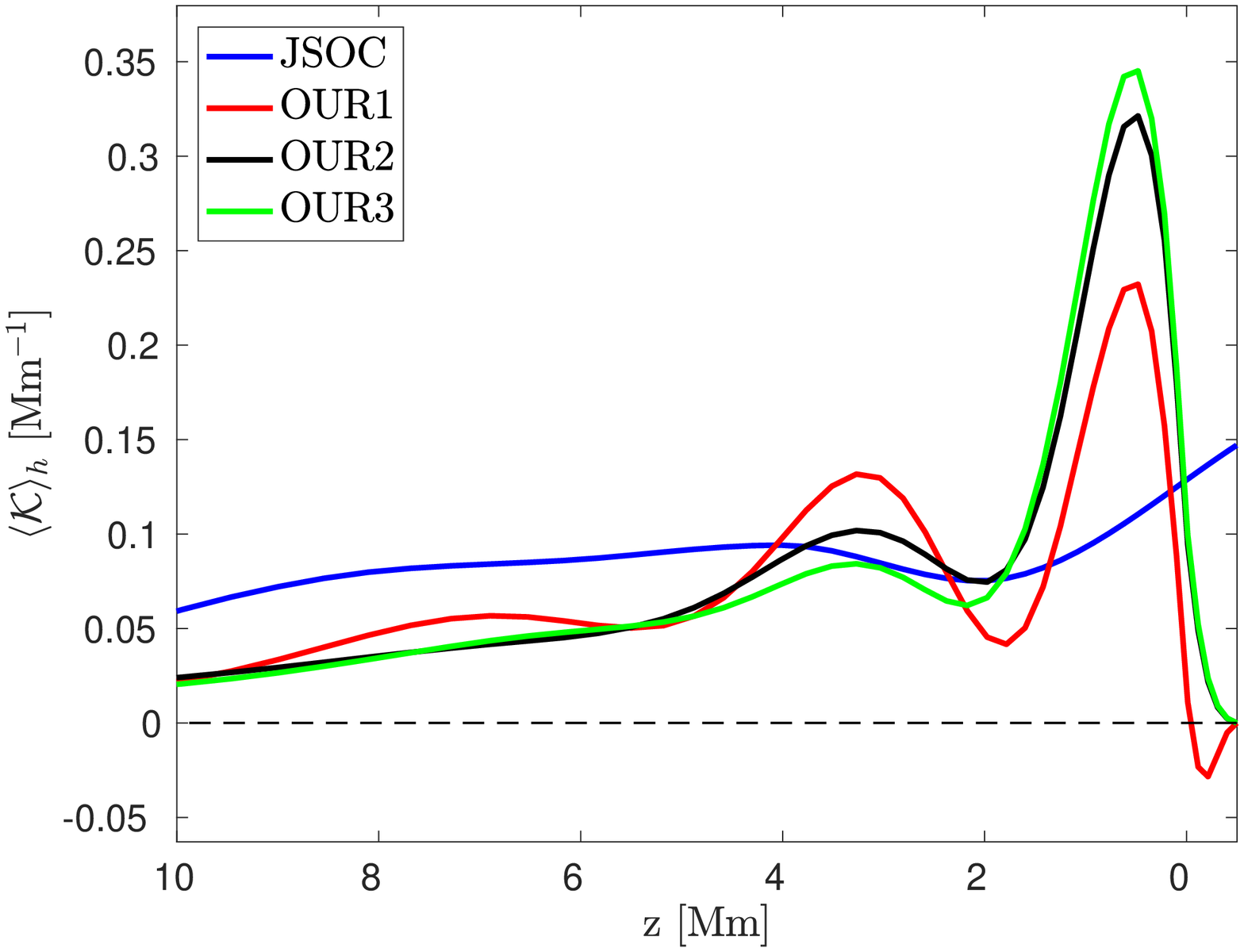}
        \includegraphics[width=0.30\textwidth]{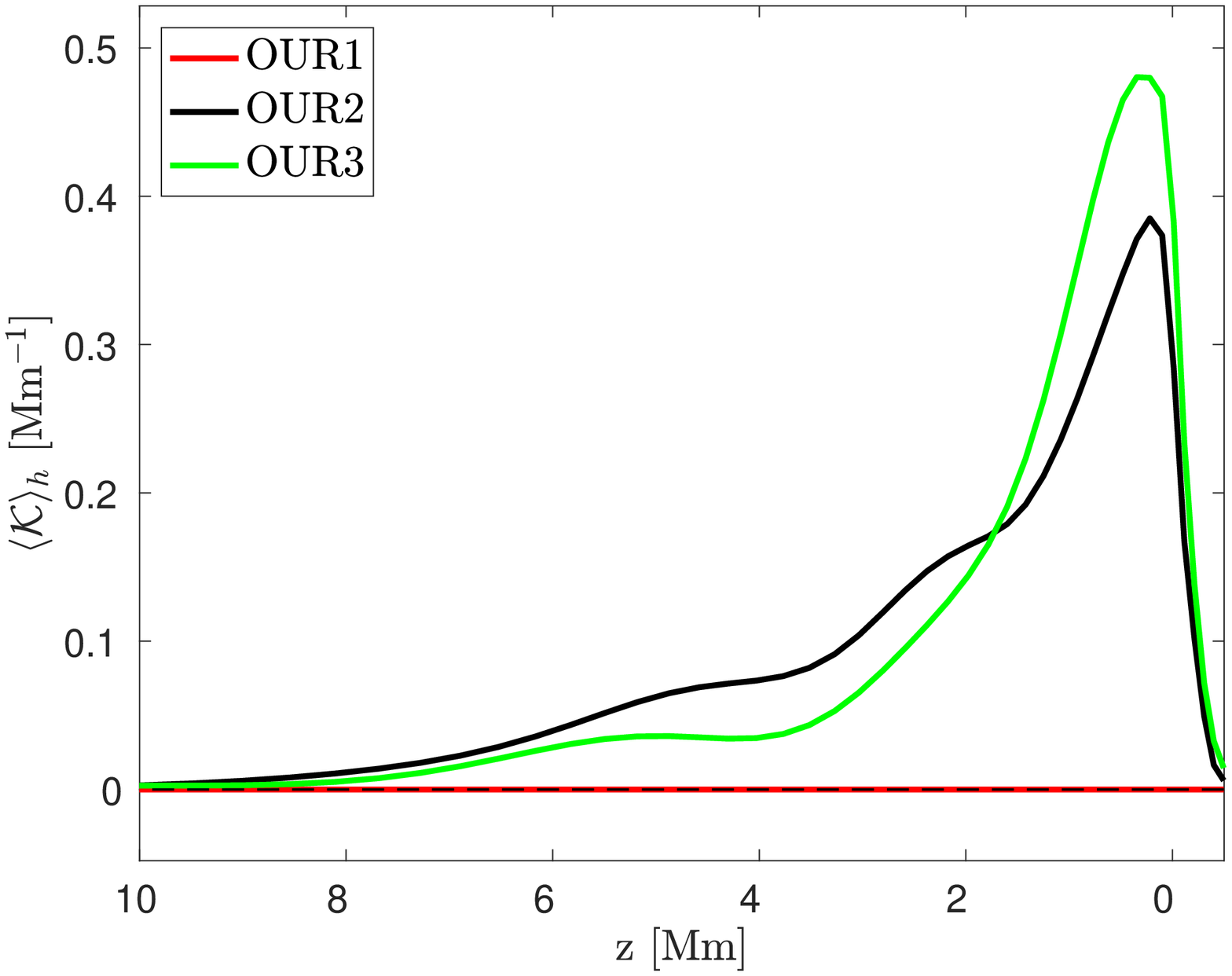}
        \caption{Examples of the horizontally averaged averaging kernels as a function of depth. The first and second rows show the kernels for the depths of 0--1~Mm (or 0.5~Mm for our inversion) and 5--7~Mm (6.0~Mm depth), respectively. Left-most column: kernel for $v_x.$  Middle: kernel for $\delta c_s.$  Right-most column: kernel for vertical $v_z$. We note that the averaging kernels for the $v_z$ inversion using a \emph{JSOC-like inversion} (denoted as OUR1) have vanishing horizontal intergals because only the difference travel-time geometries were employed \citep{Burston_2015}.}
        
        \label{pic:1Dakerns}
\end{figure*}

The different setups of our inversions lead to quite different localisations in the Sun, expressed by the averaging kernels. To indicate such an issue we plot examples of the horizontally averaged averaging kernels for individual inversion components in Fig.~\ref{pic:1Dakerns}. It is  obvious from these figures that the different localisations must lead to a difference in the inverted quantities. In general, the JSOC averaging kernels are broader, whereas kernels from our inversions do not have such a large extent. On the other hand, our inversion averaging kernels often have sidelobes outside of the expected depths. This is usually the case for the parasitic peak near the surface, which is a consequence of the noise minimisation, which affects essentially all the sensitivity kernels that have a strong sensitivity here \citep[see e.g. Fig.~2][]{Svanda_2013b}. 

It is also clearly visible that for the vertical flow $v_z$ and the sound-speed perturbations $\delta c_s$ it is impossible to have a meaningful (that is with the signal-to-noise ratio larger than unity) inversion at larger depths using travel times averaged over 24 hours. When the stress on the noise term in the cost function is too large,  the misfit term automatically increases and the resulting averaging kernel peaks near the surface in much shallower layers than expected. The indicated target depth is therefore misleading, giving an impression of the in-depth inversion, whereas the true localisation is near the surface. It is therefore virtually impossible to discuss the depth-dependence of the inverted quantities from such a set of inversions. 

We can safely conclude that the JSOC inversion maps are representative of the flows and wave speeds in the near-surface layers of the solar convection zone. One needs to bear in mind, however, that there still might be issues inherently present in the data that could possibly lead to an overinterpretation. The main weakness of the JSOC inversions is their poor localisation in depth of the solar convection zone, where the indicated depths are not well represented by the averaging kernels. As a consequence, a large positive correlation appears when comparing the inversions of the flows at various depths. Such a false correlation may for instance easily be misinterpreted as a depth coherence of flows on supergranular scales. 

The strengths of JSOC inversions are their general availability within the publicly accessible data centre, where everyone can access them and use them for research. Our inversions are not available to the public as we do not have resources to make them routinely available. Furthermore, JSOC inversions have a large spatial coverage by tiling up most of the solar disc. It is possible that our inversions are only reliable near the disc centre, where the assumption of the horizontally homogeneous background holds. Foreshortening effects farther from the disc centre cause our inversions to be less reliable.

\begin{acknowledgements}
D.K. is supported by the Grant Agency of Charles University under grant No. 532217. M.{\v S}. is supported by the project RVO:67985815. M.{\v S}. and D.K. are supported by the grant project 18-06319S awarded by the Czech Science Foundation. The sensitivity kernels were computed by the code {\sc Kc3} kindly provided by Aaron Birch. This research has made use of NASA Astrophysics Data System Bibliographic Services.
\end{acknowledgements}

\bibliographystyle{aa} 
\bibliography{BIBL}

\begin{appendix}
\onecolumn
\section{The JSOC results}
\label{app:jsoc}

\begin{figure*}[!h]
\sidecaption
\includegraphics[height=8cm]{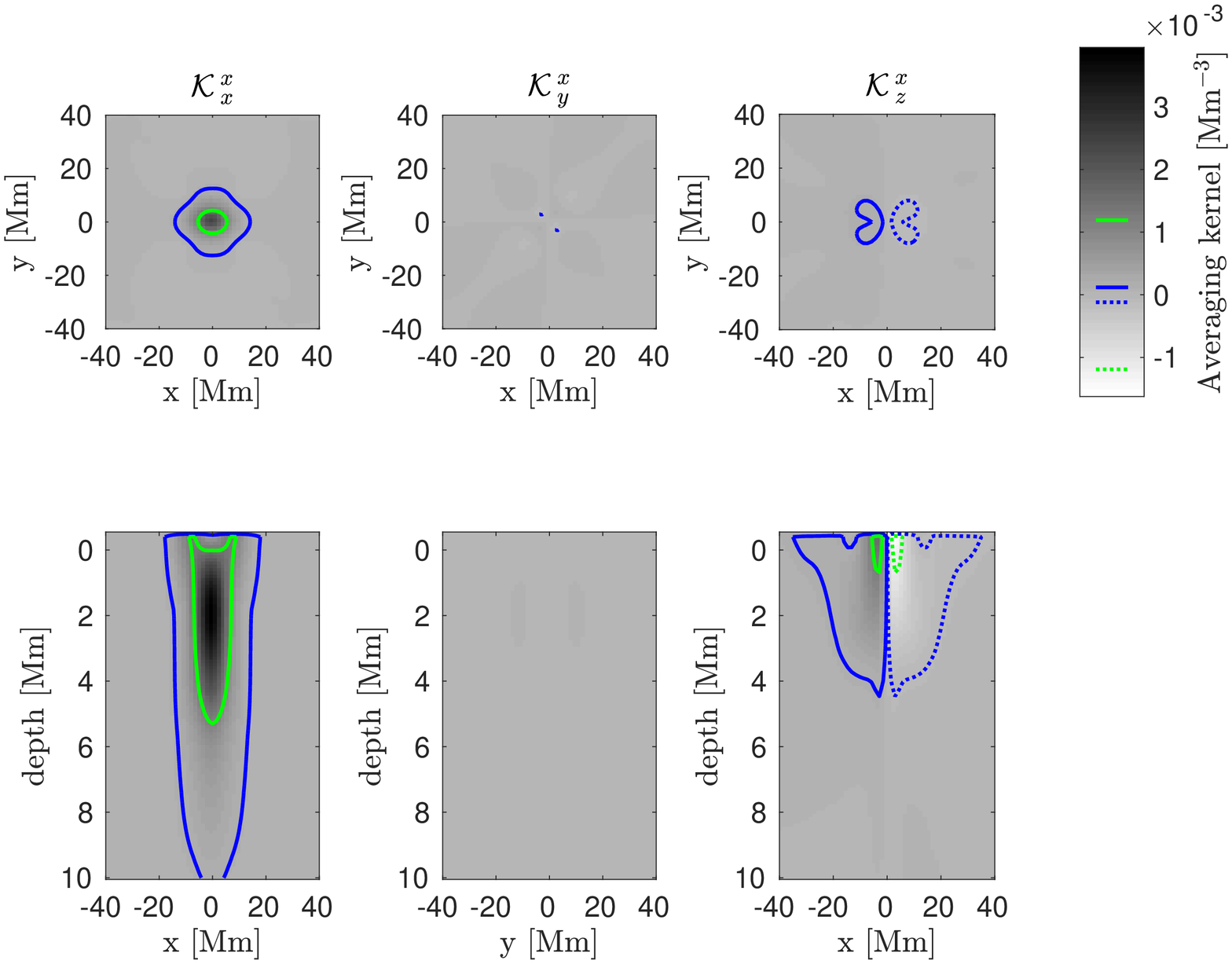}
\caption{JSOC averaging kernels for $v_x$ inversion at the depth of 3--5 Mm. See Fig. \ref{pic:RLS_rakern_x_0.5} for details.}
\end{figure*}

\begin{figure*}[!h]
\sidecaption
\includegraphics[height=8cm]{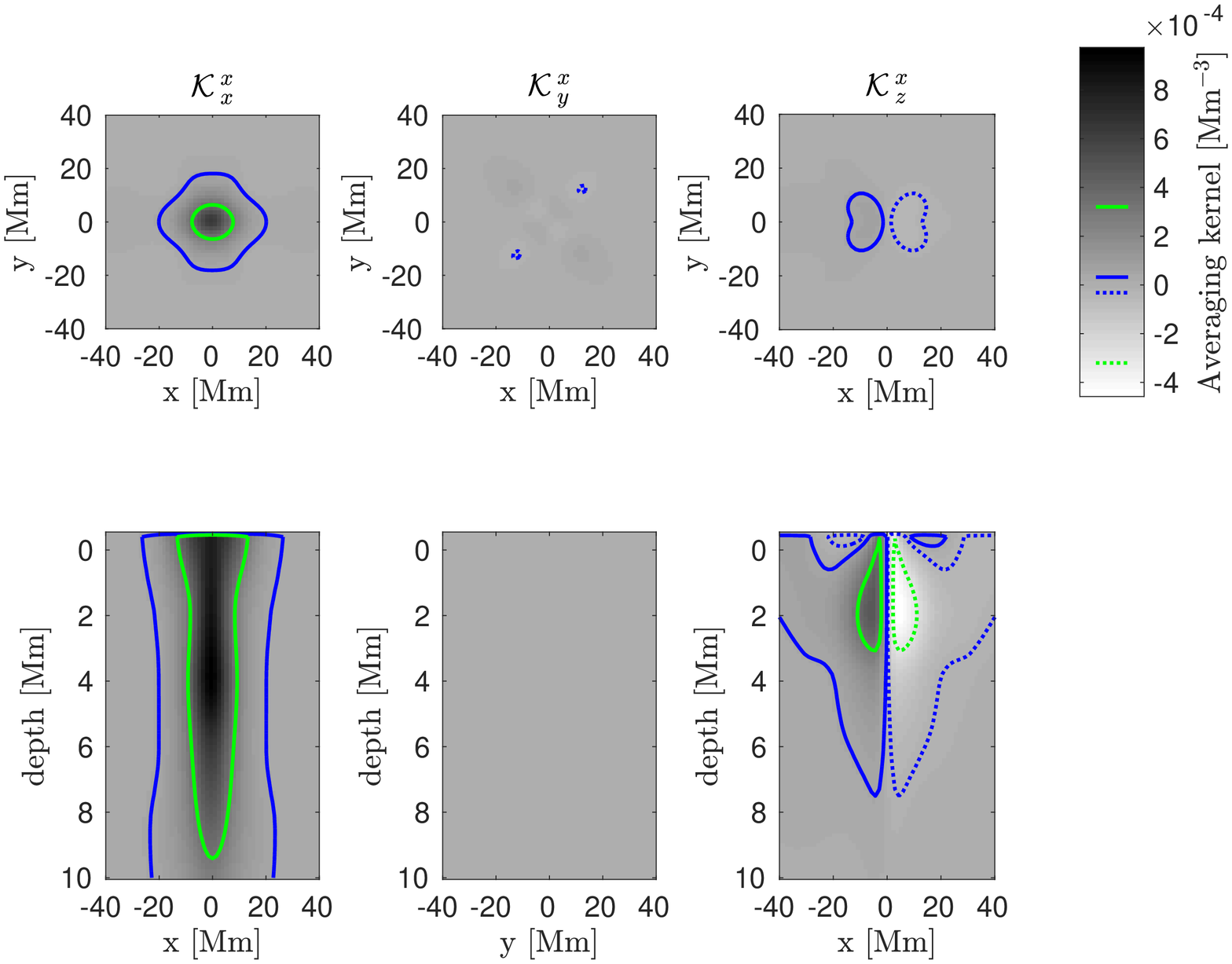}
\caption{JSOC averaging kernels for $v_x$ inversion at the depth of 5--7 Mm. See Fig. \ref{pic:RLS_rakern_x_0.5} for details.}
\end{figure*}

\clearpage

\section{Horizontal flows}
\label{app:horizontal}

\begin{figure*}[!h]
\sidecaption
\includegraphics[height=8cm]{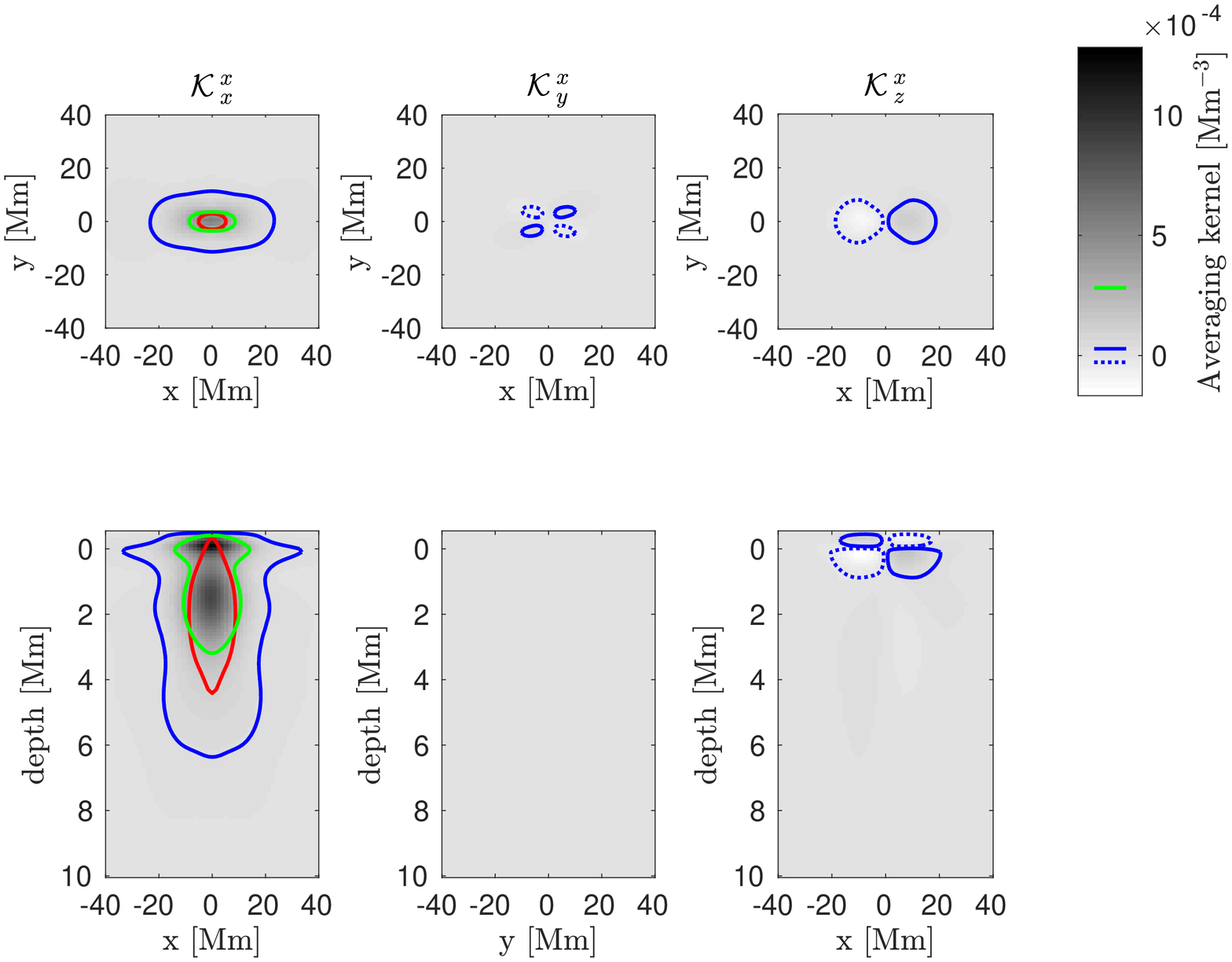}
\caption{Averaging kernels for $v_x$ inversion at the depth of 0.5 Mm, JSOC-like inversion. See Fig. \ref{pic:RLS_rakern_x_0.5} for details.}
\end{figure*}

\begin{figure*}[!h]
\sidecaption
\includegraphics[height=8cm]{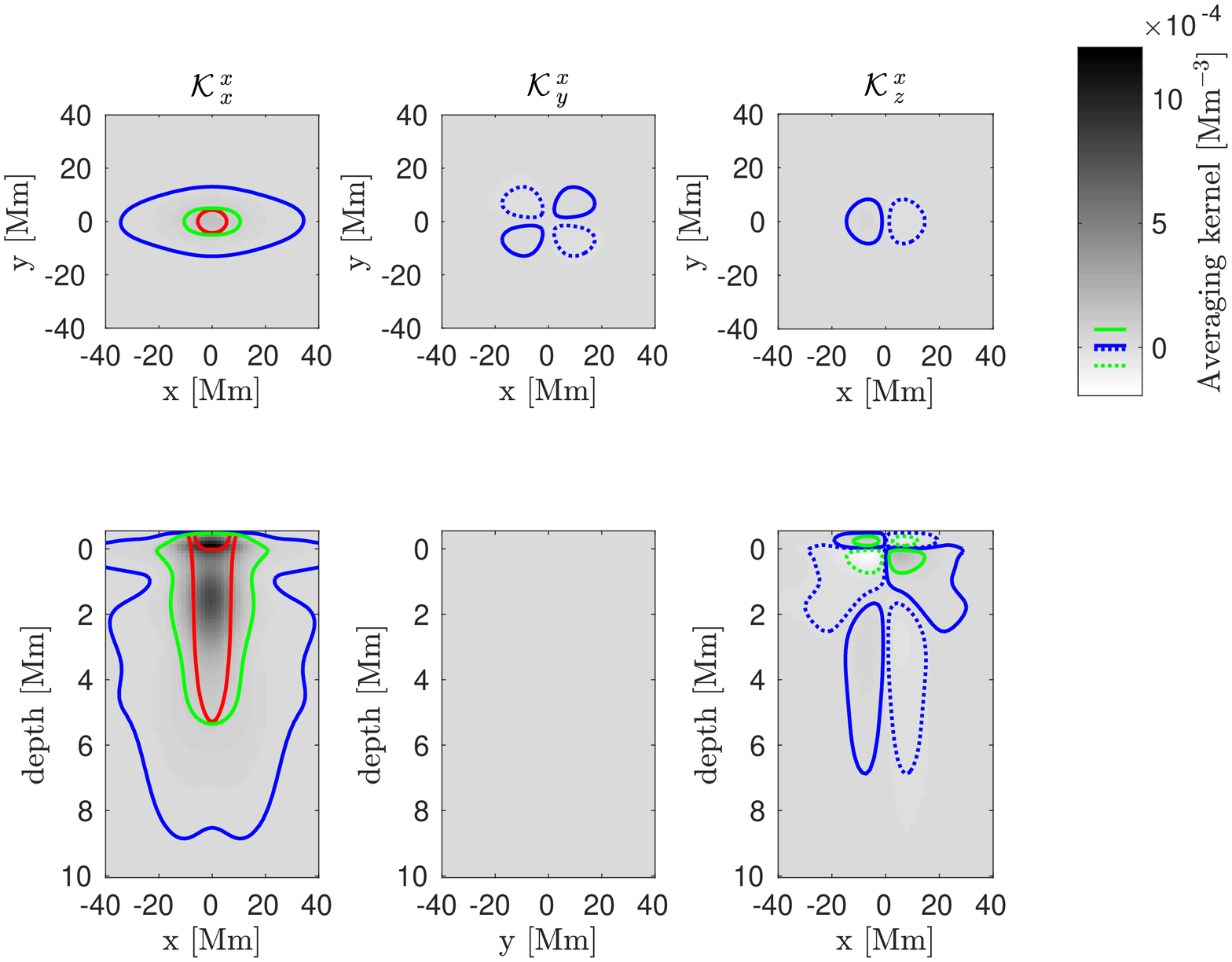}
\caption{Averaging kernels for $v_x$ inversion at the depth of 4.0 Mm, JSOC-like inversion. See Fig. \ref{pic:RLS_rakern_x_0.5} for details.}
\end{figure*}

\begin{figure*}[!h]
\sidecaption
\includegraphics[height=8cm]{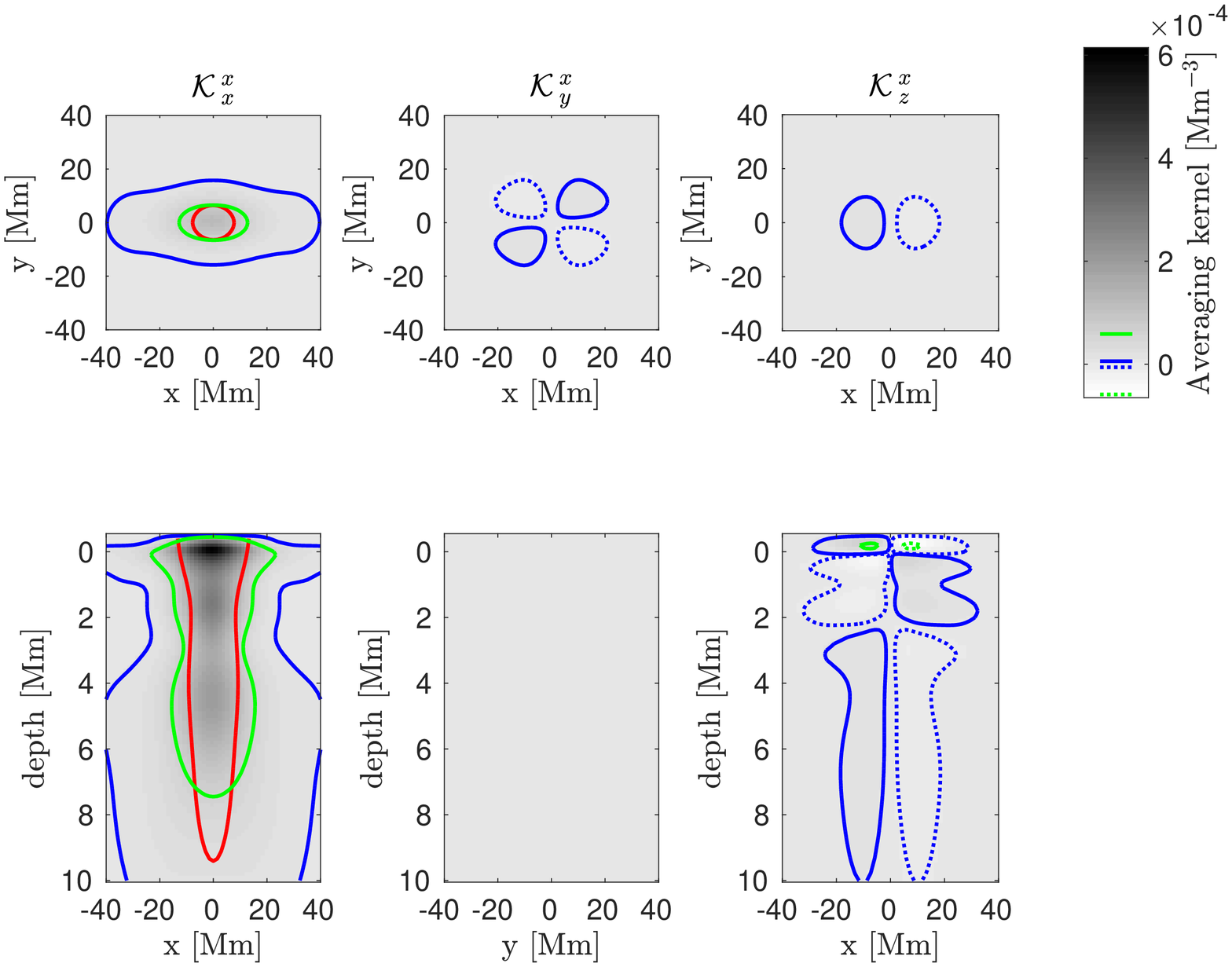}
\caption{Averaging kernels for $v_x$ inversion at the depth of 6.0 Mm, JSOC-like inversion. See Fig. \ref{pic:RLS_rakern_x_0.5} for details.}
\end{figure*}

\begin{figure*}[!h]
\sidecaption
\includegraphics[height=8cm]{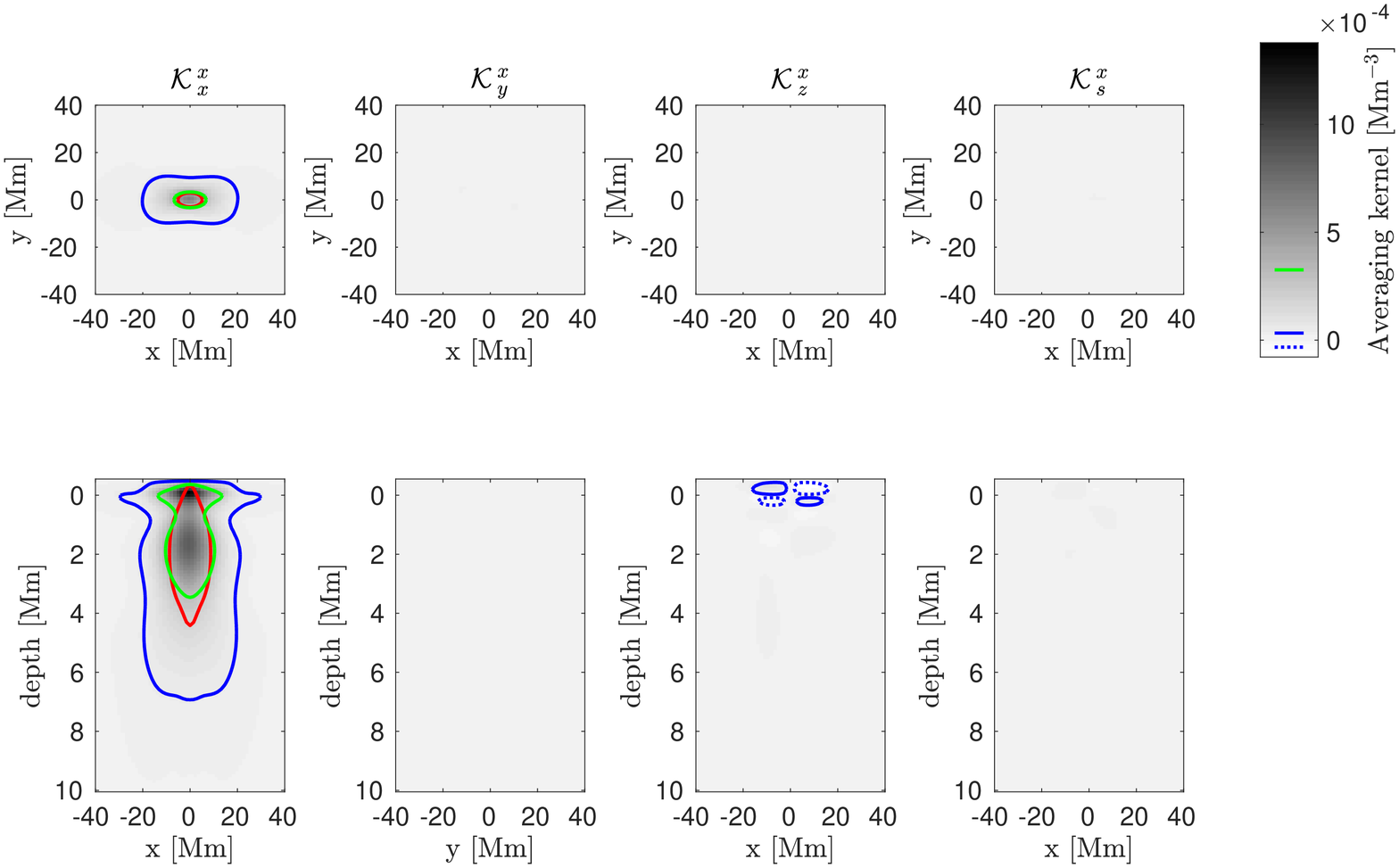}
\caption{Averaging kernels for $v_x$ inversion at the depth of 0.5 Mm, JSOC-like target. See Fig. \ref{pic:RLS_rakern_x_0.5} for details.}
\end{figure*}

\begin{figure*}[!h]
\sidecaption
\includegraphics[height=8cm]{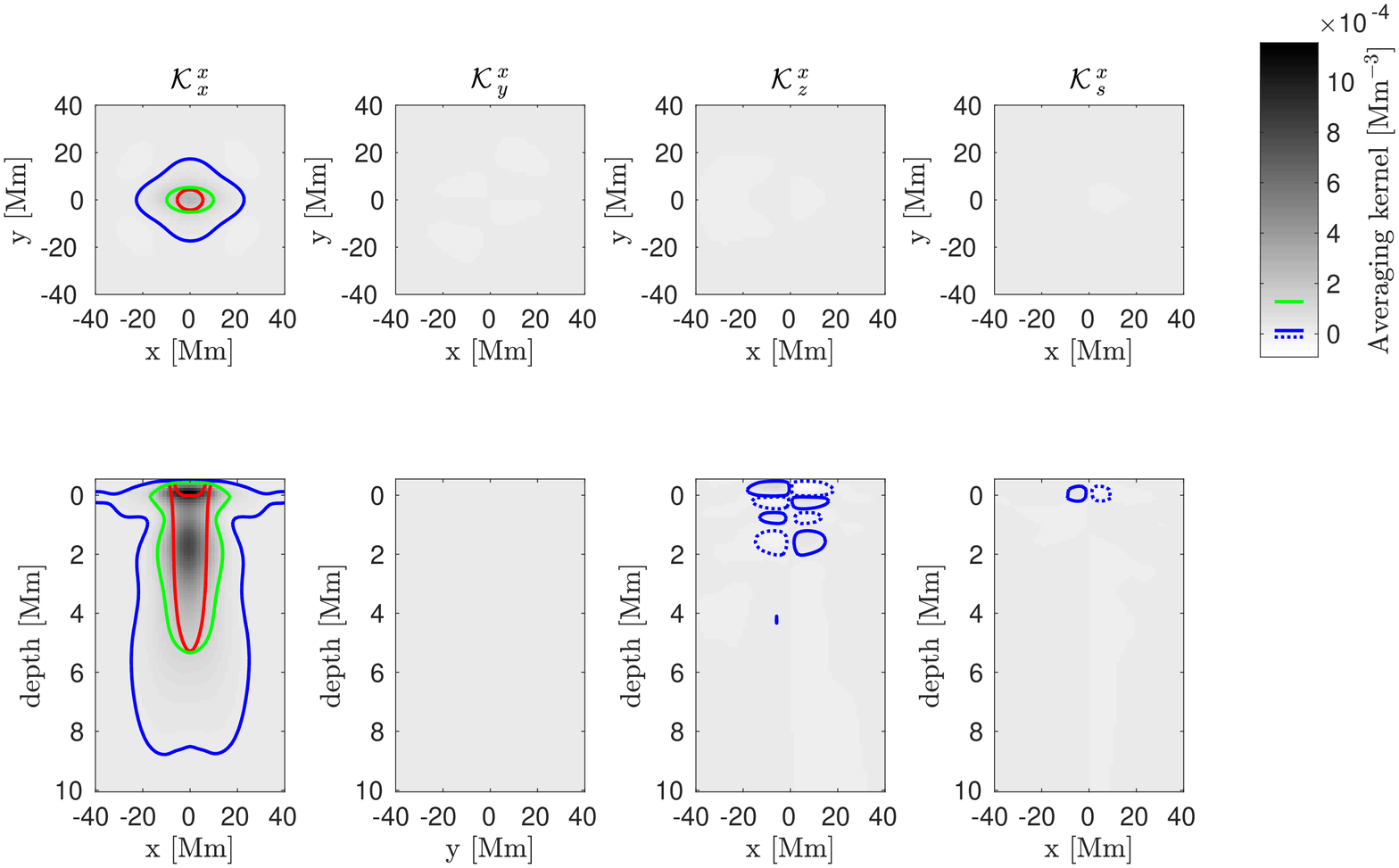}
\caption{Averaging kernels for $v_x$ inversion at the depth of 4.0 Mm, JSOC-like target. See Fig. \ref{pic:RLS_rakern_x_0.5} for details.}
\end{figure*}

\begin{figure*}[!h]
\sidecaption
\includegraphics[height=8cm]{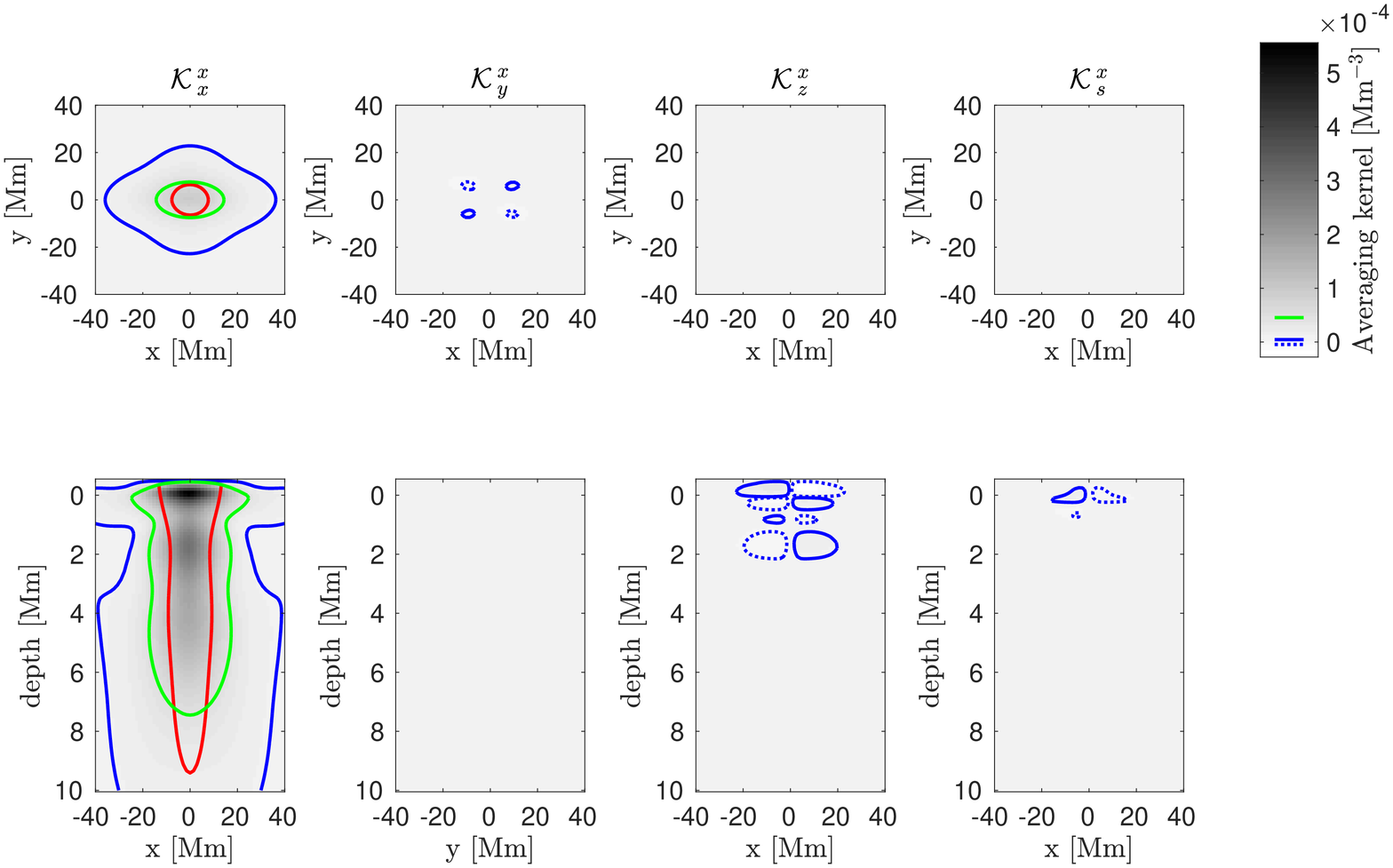}
\caption{Averaging kernels for $v_x$ inversion at the depth of 6.0 Mm, JSOC-like target. See Fig. \ref{pic:RLS_rakern_x_0.5} for details.}
\end{figure*}

\begin{figure*}[!h]
\sidecaption
\includegraphics[height=8cm]{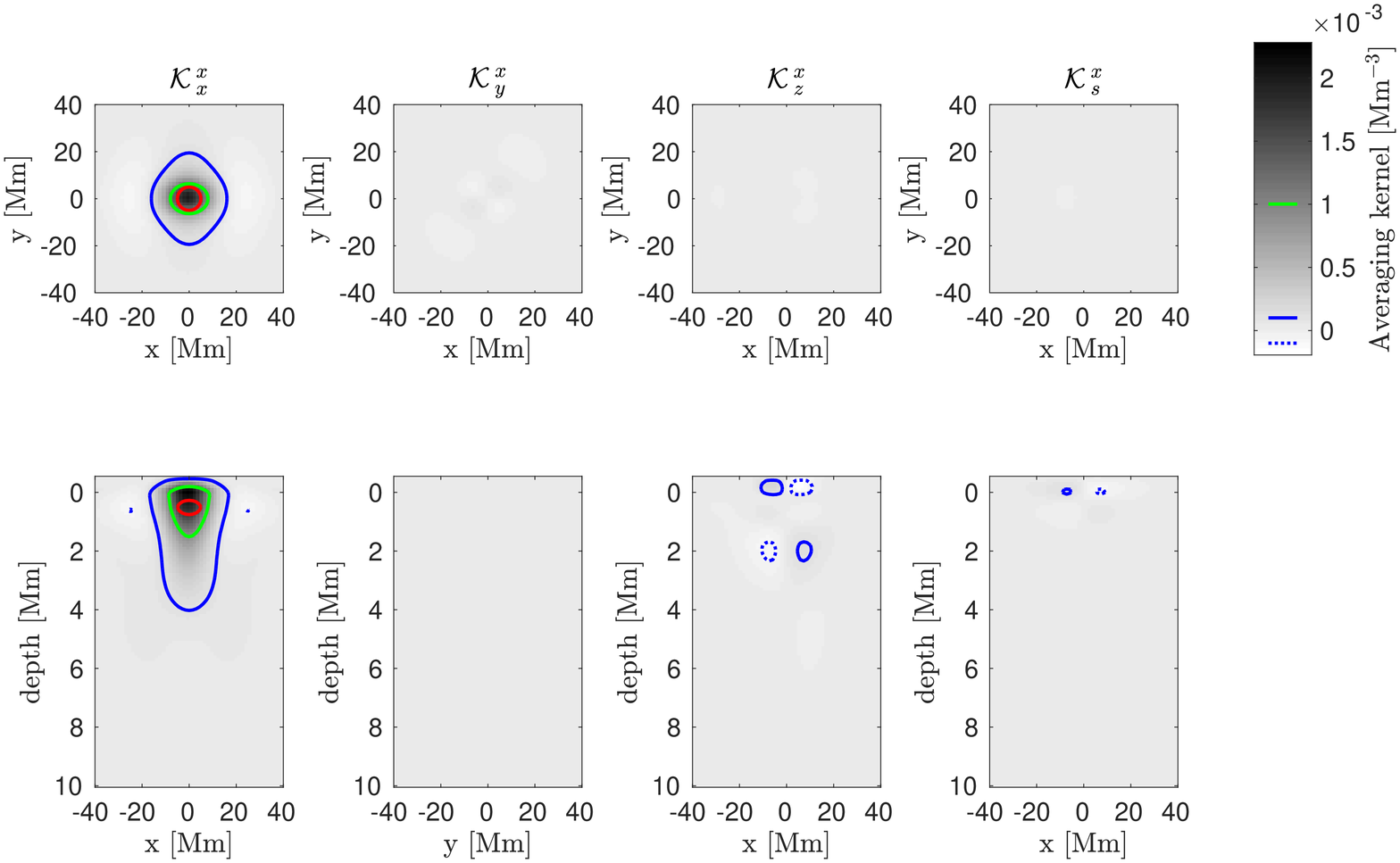}
\caption{Averaging kernels for $v_x$ inversion at the depth of 0.5 Mm, JSOC-indicated target. See Fig. \ref{pic:RLS_rakern_x_0.5} for details.}
\end{figure*}

\begin{figure*}[!h]
\sidecaption
\includegraphics[height=8cm]{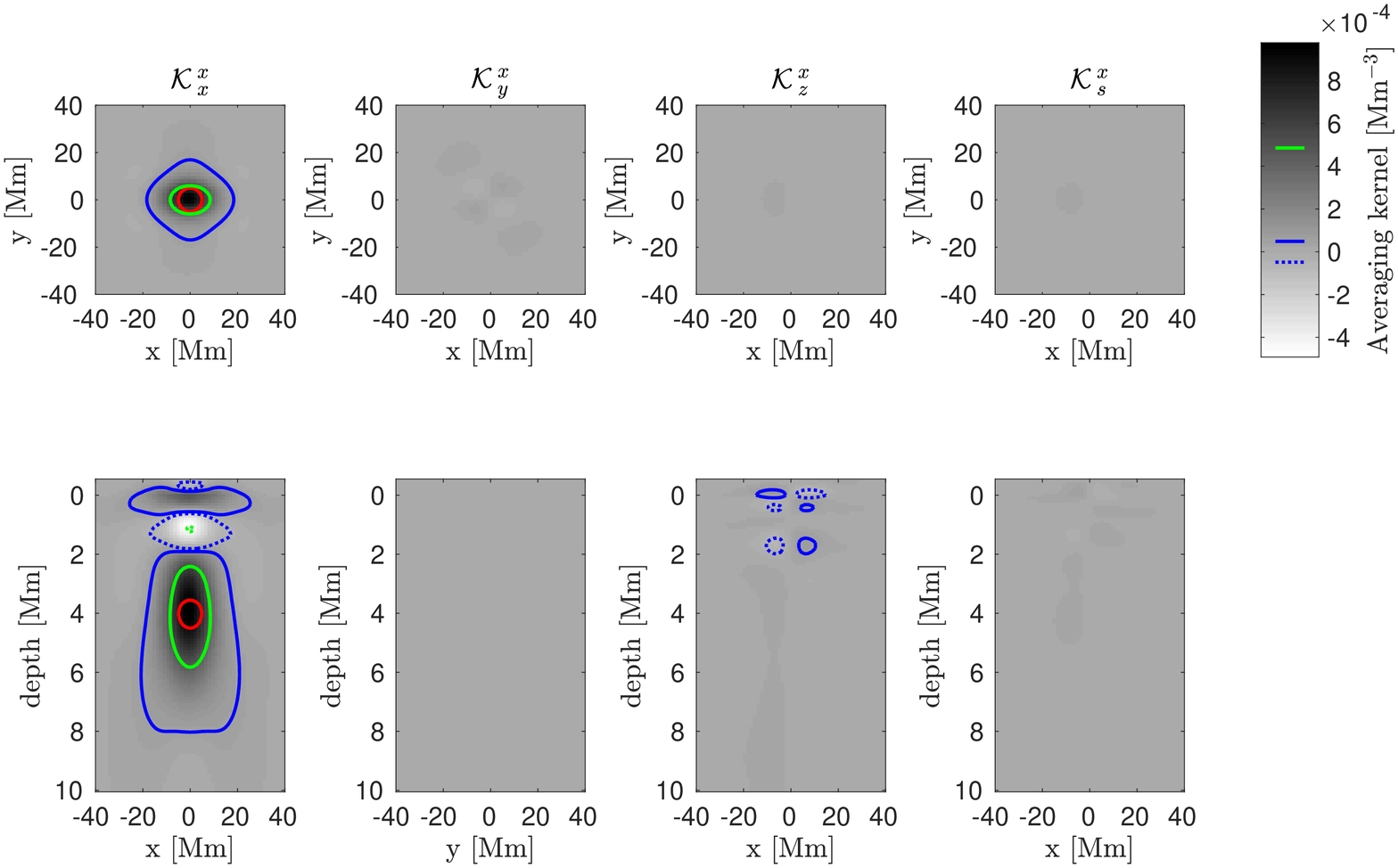}
\caption{Averaging kernels for $v_x$ inversion at the depth of 4.0 Mm, JSOC-indicated target. See Fig. \ref{pic:RLS_rakern_x_0.5} for details.}
\end{figure*}

\begin{figure*}[!h]
\sidecaption
\includegraphics[height=8cm]{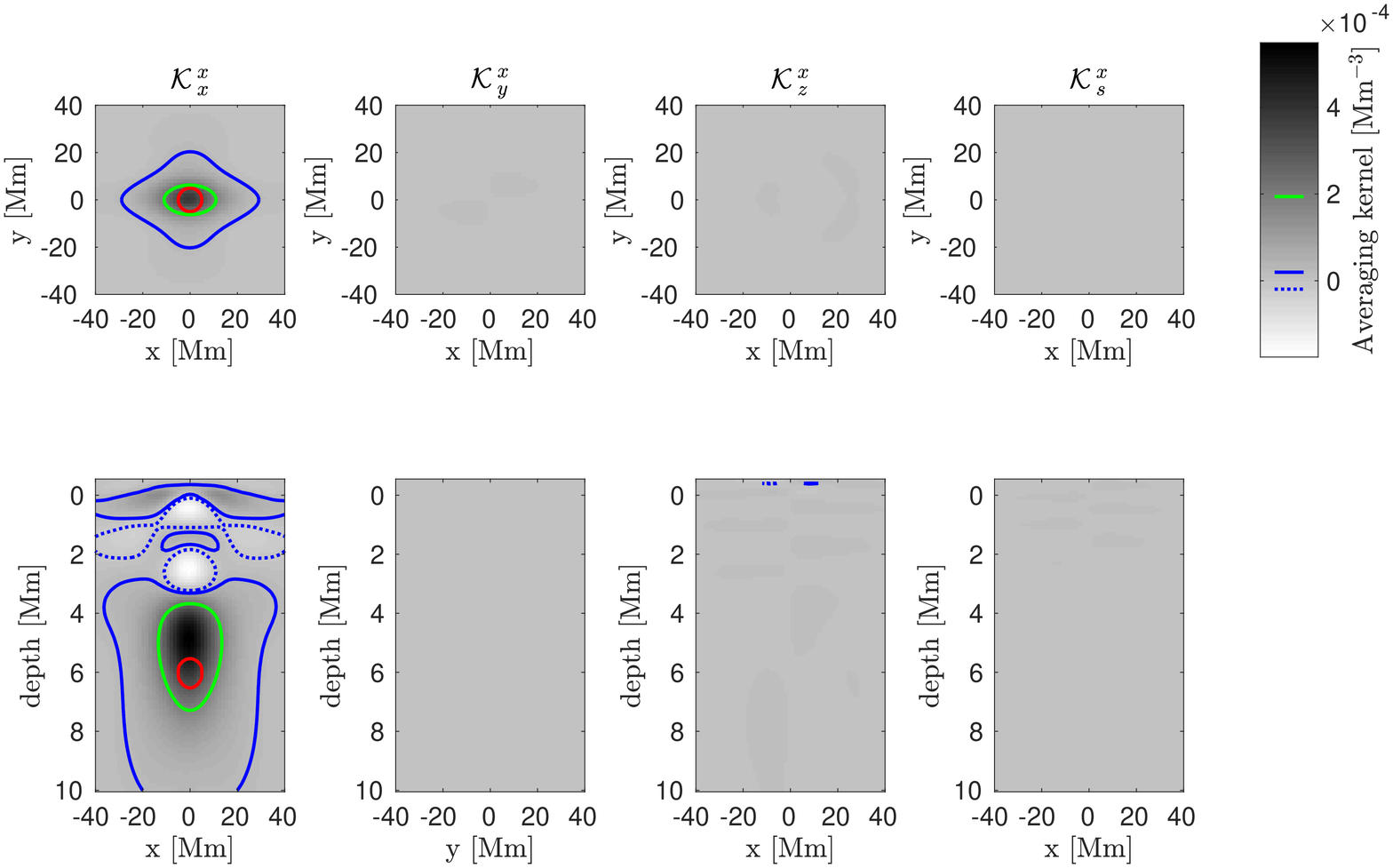}
\caption{Averaging kernels for $v_x$ inversion at the depth of 6.0 Mm, JSOC-indicated target. See Fig. \ref{pic:RLS_rakern_x_0.5} for details.}
\end{figure*}

\begin{figure*}[!h]
        \includegraphics[width=0.30\textwidth]{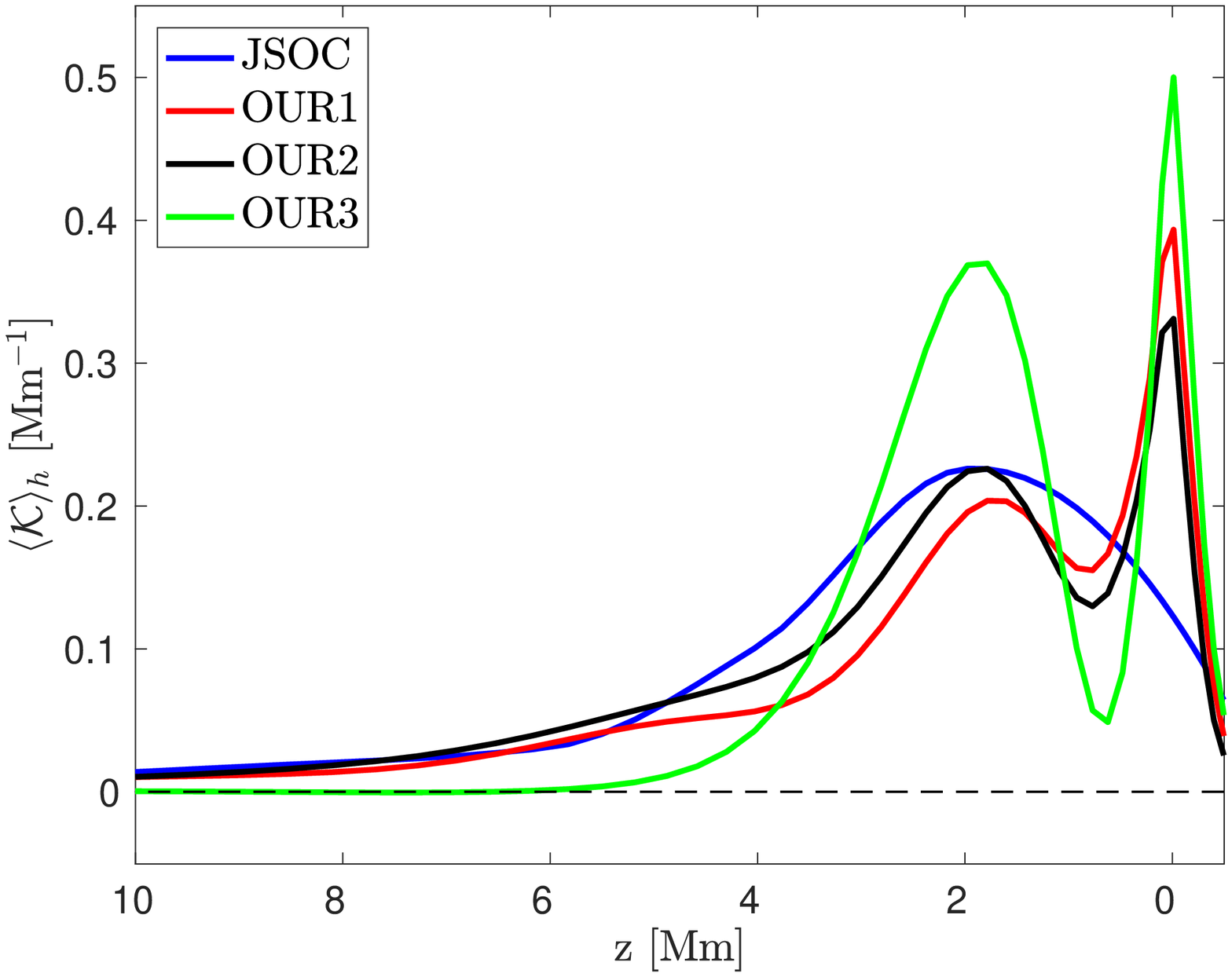}
        \includegraphics[width=0.30\textwidth]{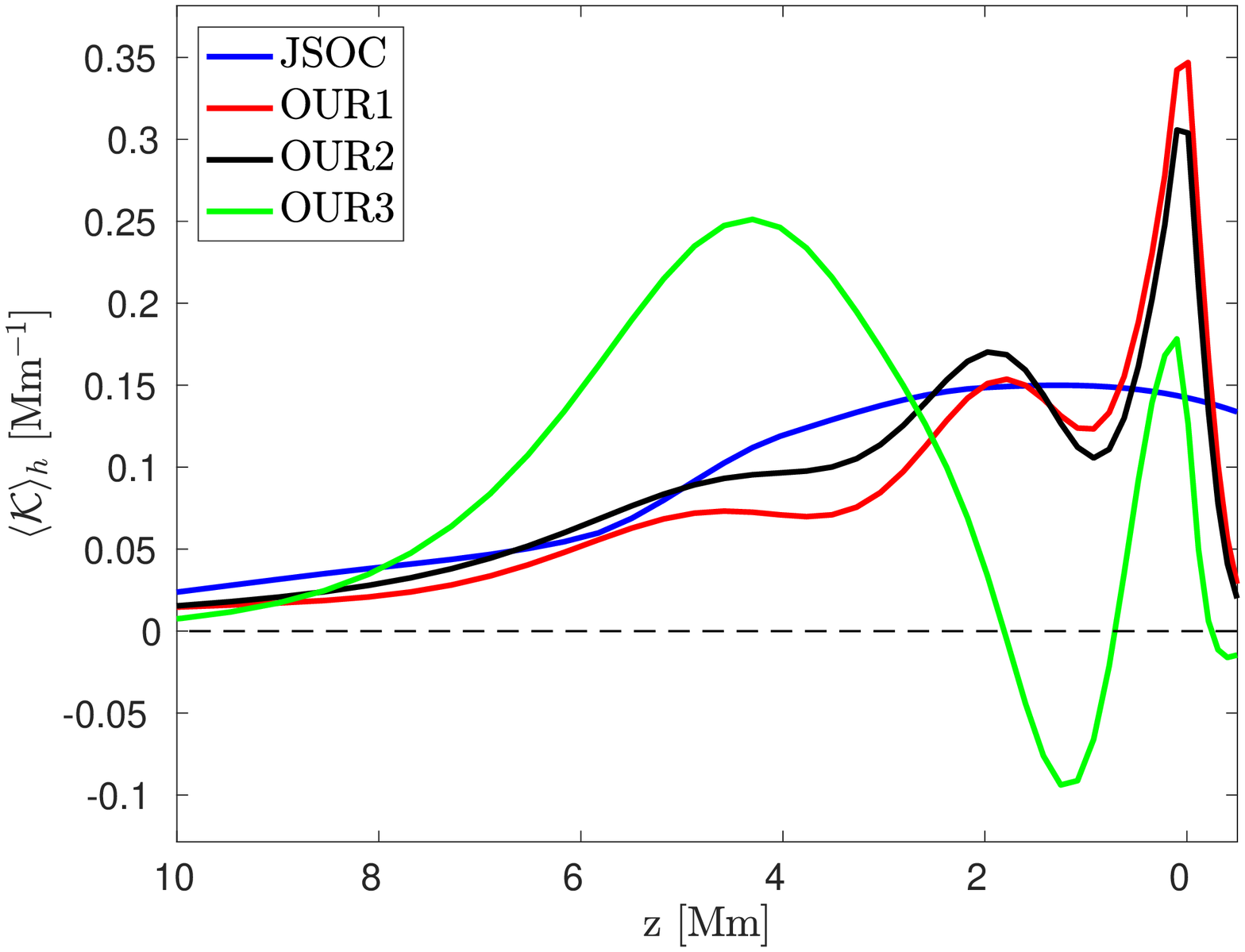}
        \caption{Horizontally averaged averaging kernels for $v_x$ inversions as a function of depth. Left: kernels for the depth of 1--3~Mm (or 2.0~Mm for our inversion), Right: kernels for the depth of 3--5~Mm (or 4.0~Mm depth).}
\end{figure*}

\begin{figure*}[!h]
\sidecaption
\includegraphics[height=8cm]{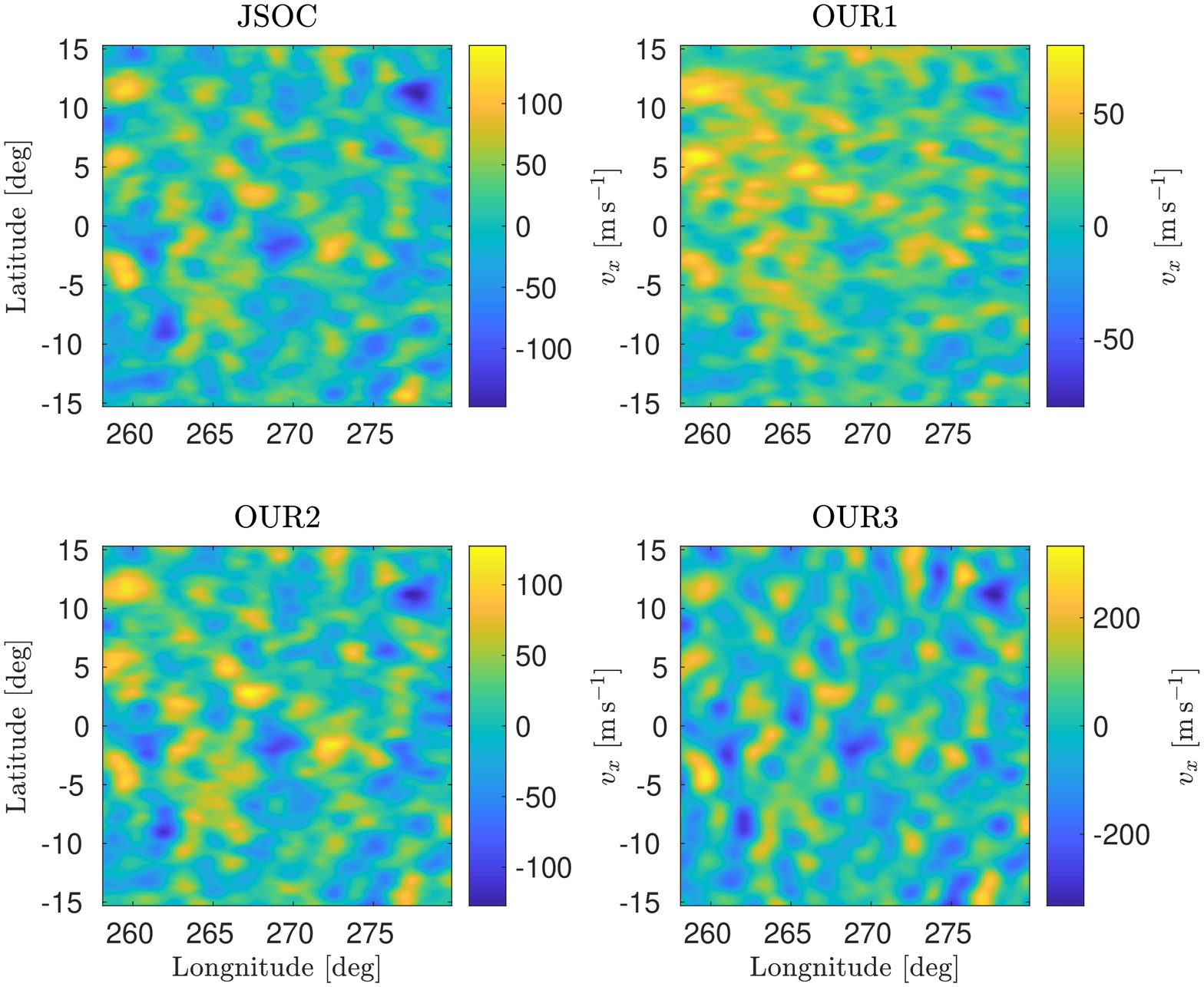}
\caption{Inversions for $v_x^{{\rm inv}}$ at 0.5 Mm depth.} 
\end{figure*}

\begin{figure*}[!h]
\sidecaption
\includegraphics[height=8cm]{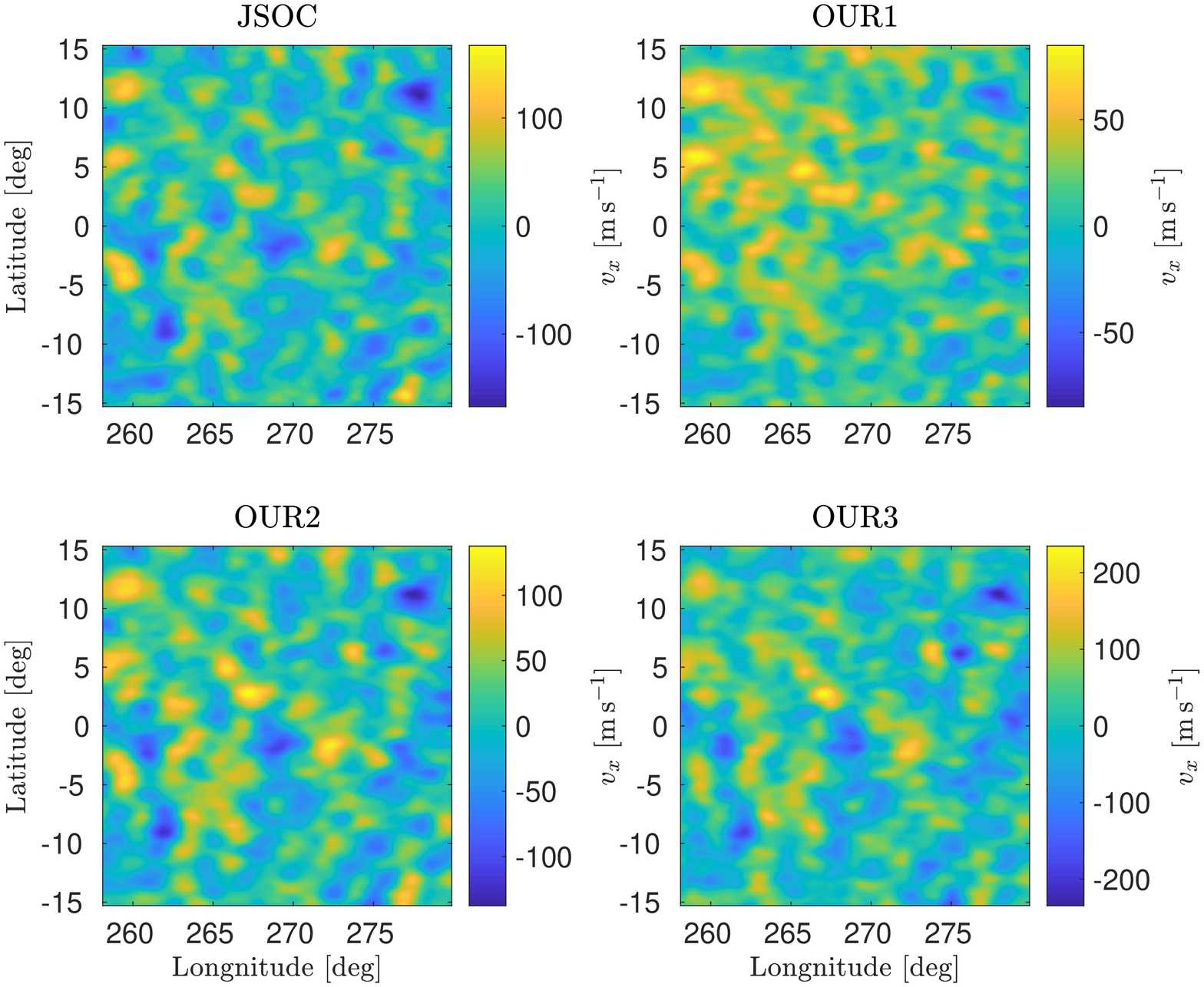}
\caption{Inversions for $v_x^{{\rm inv}}$ at 4.0 Mm depth.} 
\end{figure*}

\begin{figure*}[!h]
\sidecaption
\includegraphics[height=8cm]{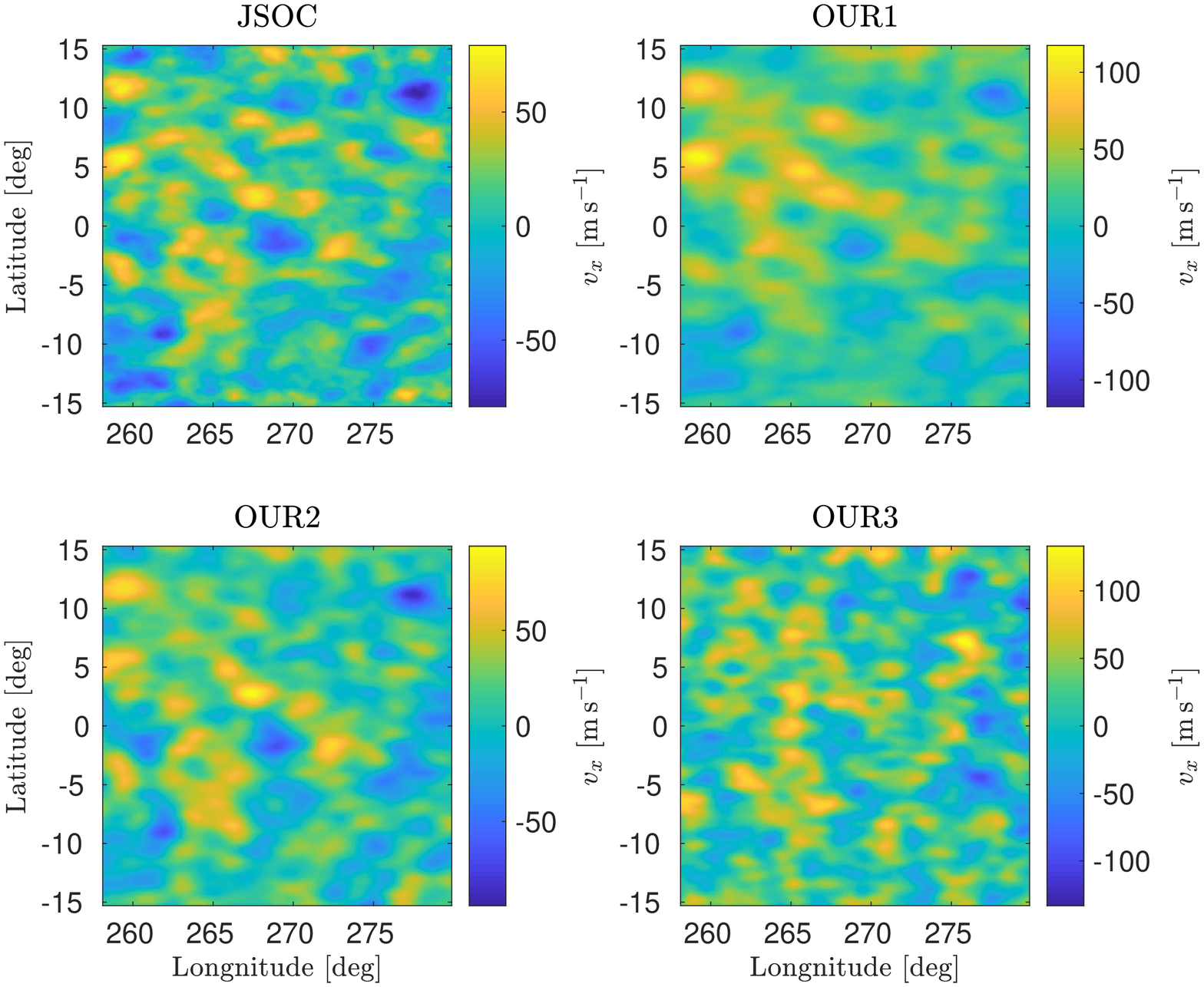}
\caption{Inversions for $v_x^{{\rm inv}}$ at 6.0 Mm depth.} 
\end{figure*}

\clearpage

\section{The sound-speed perturbations}
\label{app:sound}

\begin{figure*}[!h]
\sidecaption
\includegraphics[width=4cm]{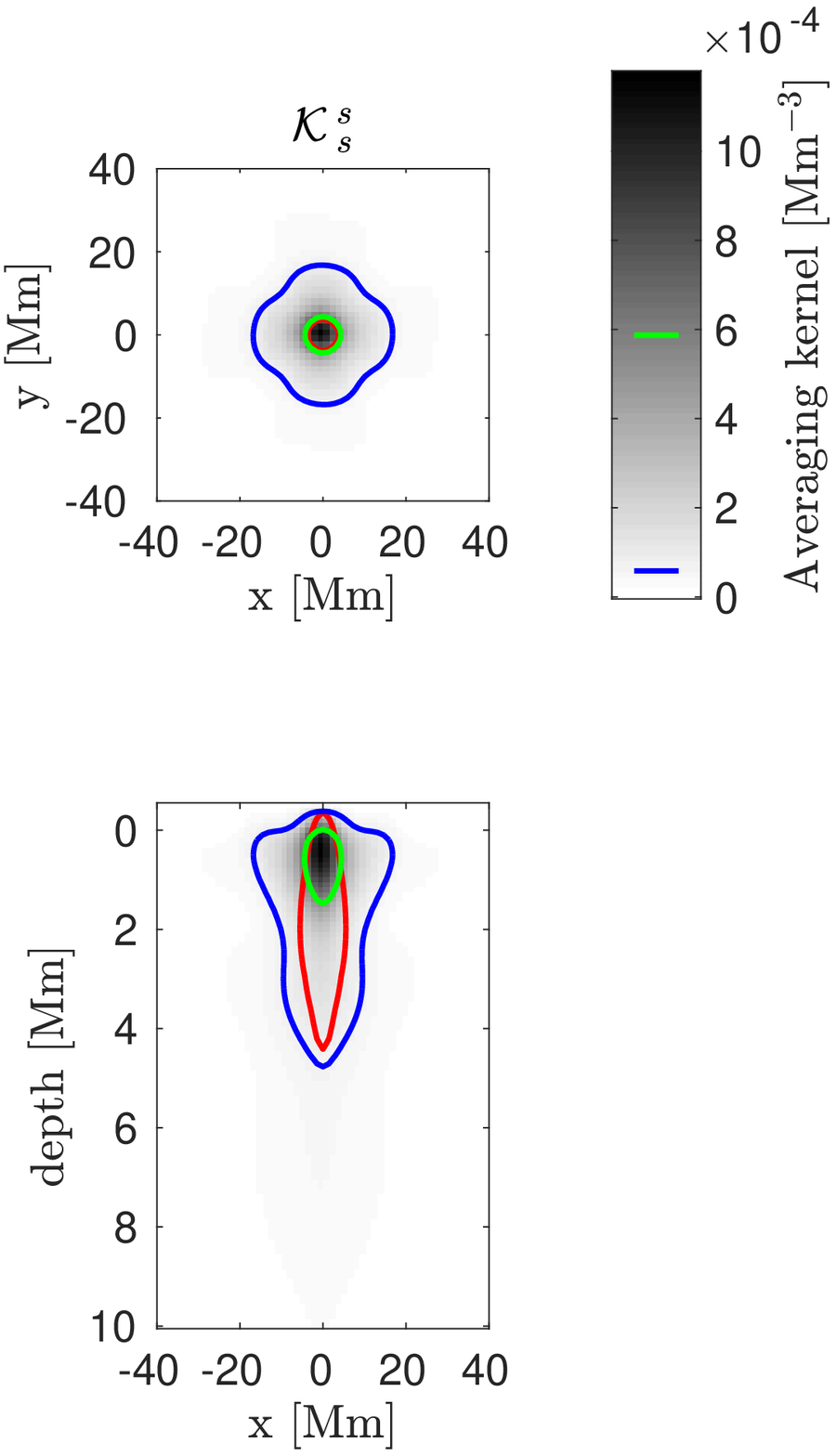}
\caption{Averaging kernels for $\delta c_s$ inversion at the depth of 0.5 Mm, JSOC-like inversion. See Fig. \ref{pic:RLS_rakern_x_0.5} for details.}
\end{figure*}

\begin{figure*}[!h]
\sidecaption
\includegraphics[width=4cm]{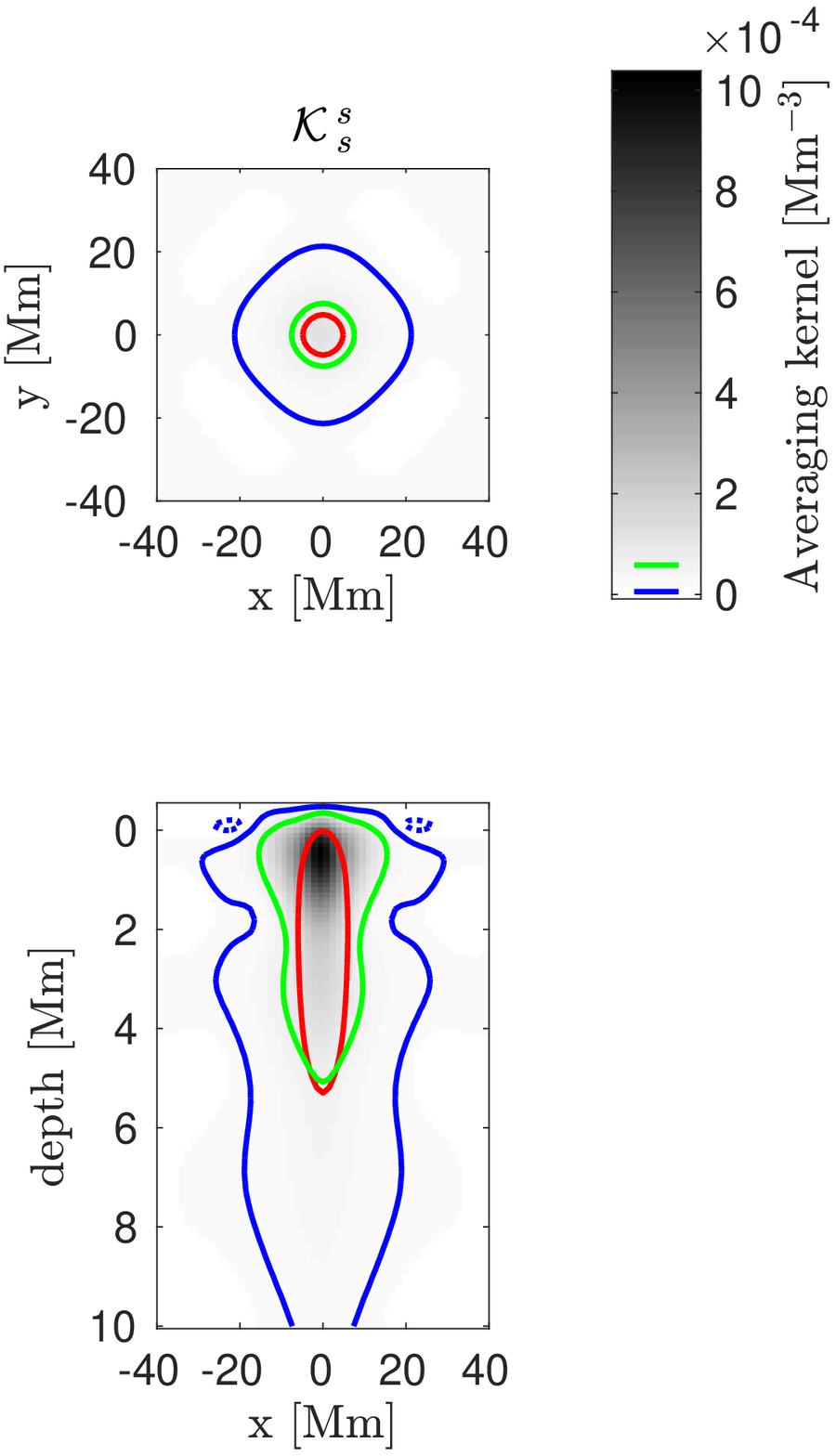}
\caption{Averaging kernels for $\delta c_s$ inversion at the depth of 4.0 Mm, JSOC-like inversion. See Fig. \ref{pic:RLS_rakern_x_0.5} for details.}
\end{figure*}

\begin{figure*}[!h]
\sidecaption
\includegraphics[width=4cm]{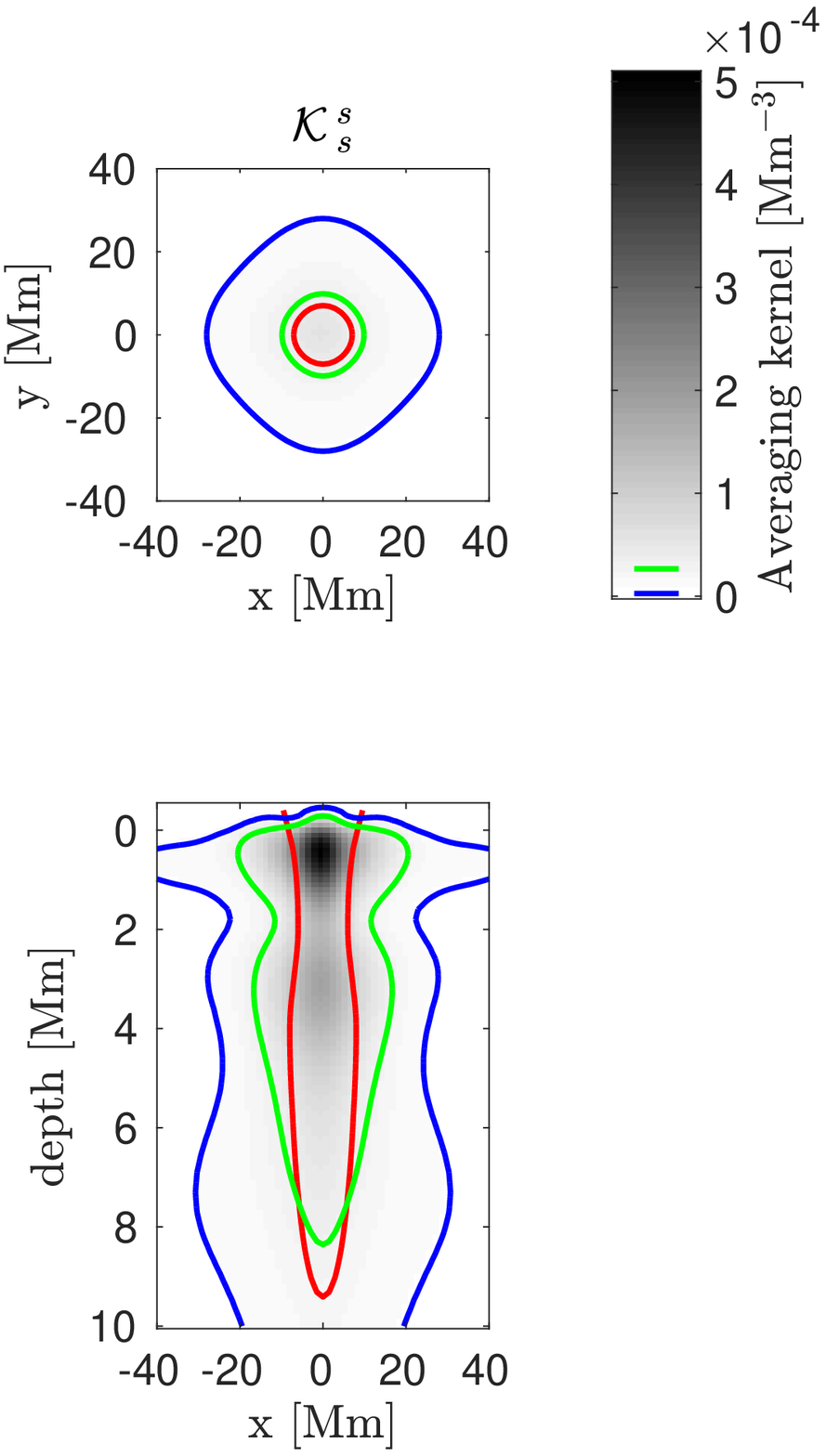}
\caption{Averaging kernels for $\delta c_s$ inversion at the depth of 6.0 Mm, JSOC-like inversion. See Fig. \ref{pic:RLS_rakern_x_0.5} for details.}
\end{figure*}

\begin{figure*}[!h]
\sidecaption
\includegraphics[height=8cm]{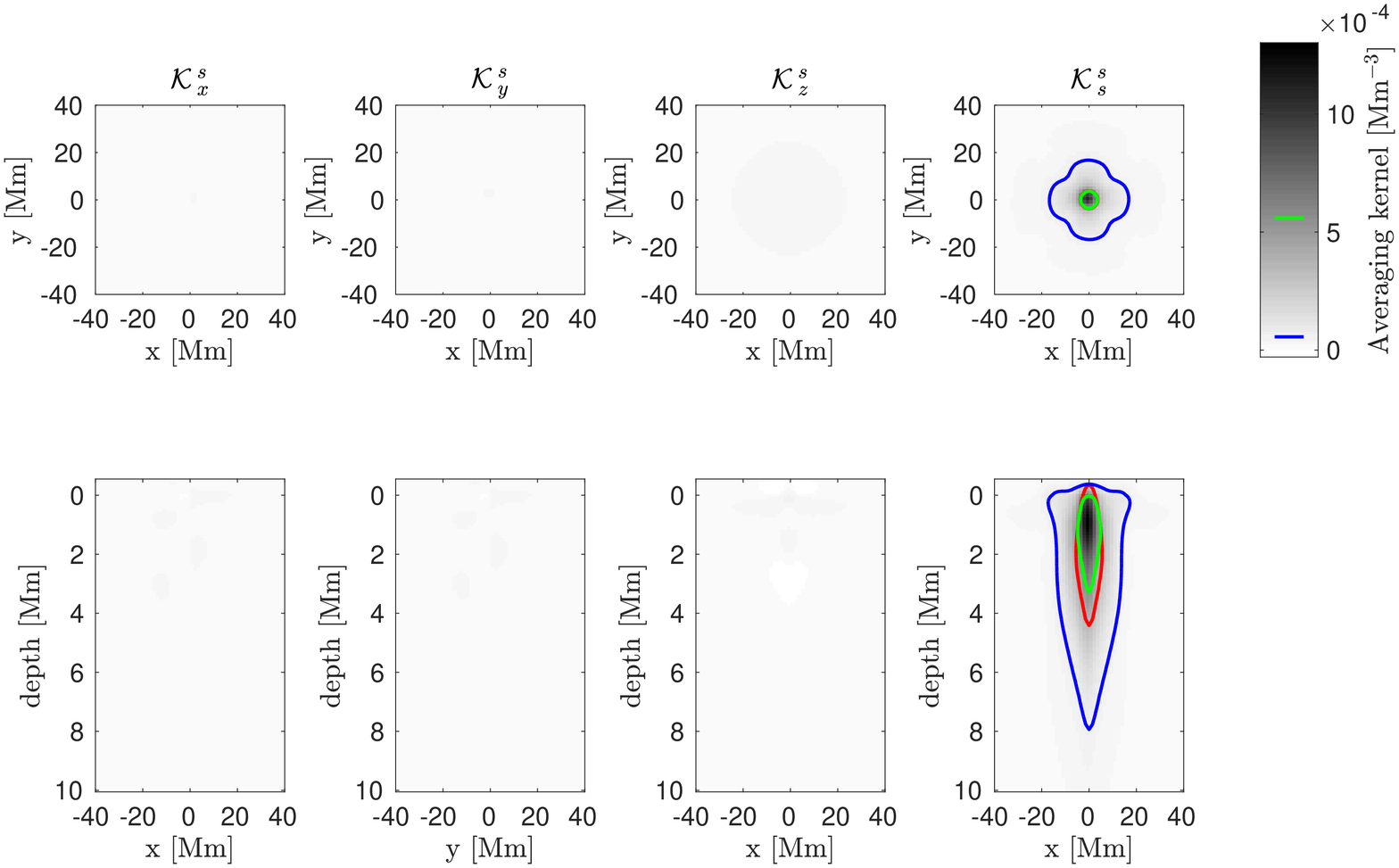}
\caption{Averaging kernels for $\delta c_s$ inversion at the depth of 0.5 Mm, JSOC-like target. See Fig. \ref{pic:RLS_rakern_x_0.5} for details.}
\end{figure*}

\begin{figure*}[!h]
\sidecaption
\includegraphics[height=8cm]{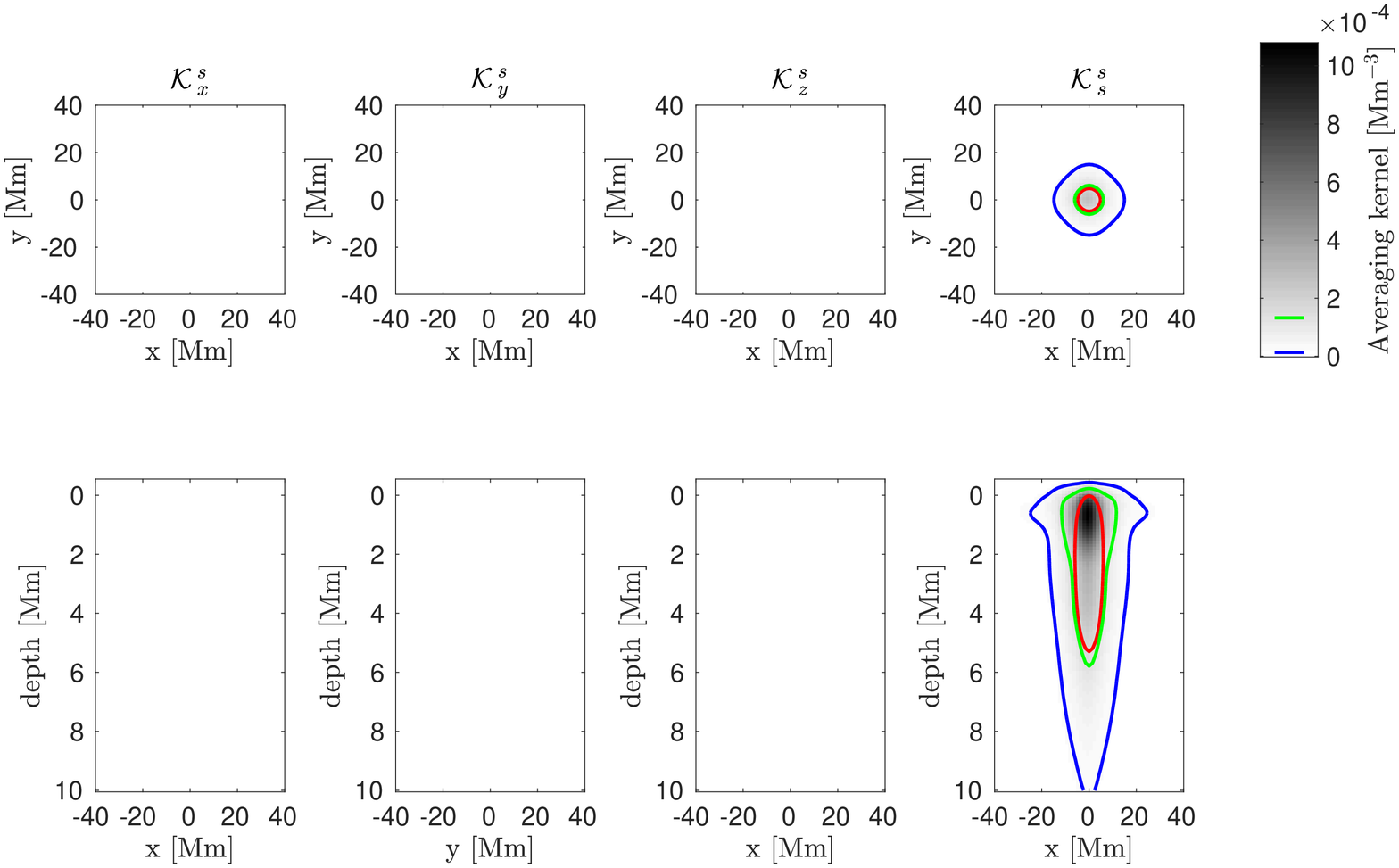}
\caption{Averaging kernels for $\delta c_s$ inversion at the depth of 4.0 Mm, JSOC-like target. See Fig. \ref{pic:RLS_rakern_x_0.5} for details.}
\end{figure*}

\begin{figure*}[!h]
\sidecaption
\includegraphics[height=8cm]{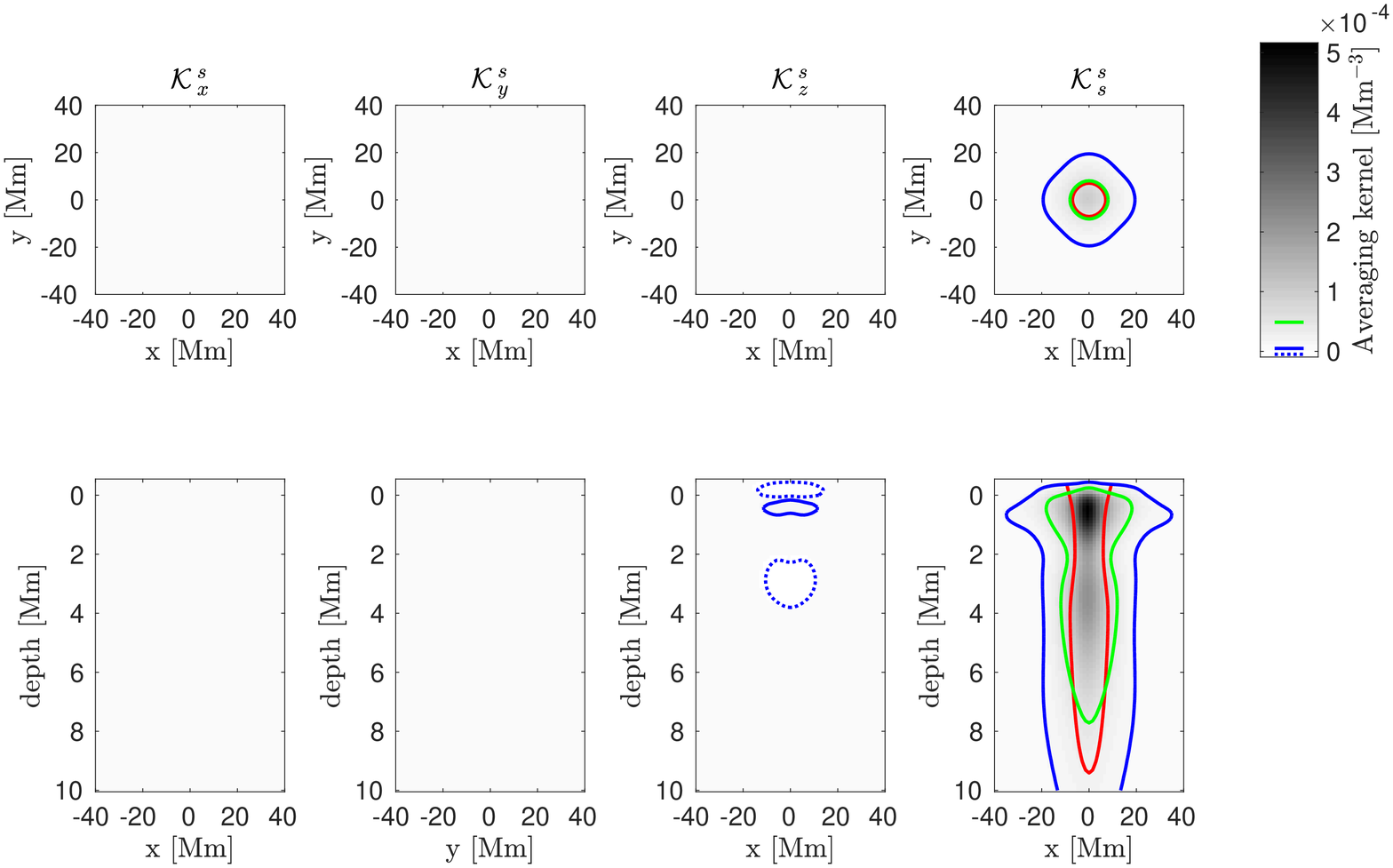}
\caption{Averaging kernels for $\delta c_s$ inversion at the depth of 6.0 Mm, JSOC-like target. See Fig. \ref{pic:RLS_rakern_x_0.5} for details.}
\end{figure*}

\begin{figure*}[!h]
\sidecaption
\includegraphics[height=8cm]{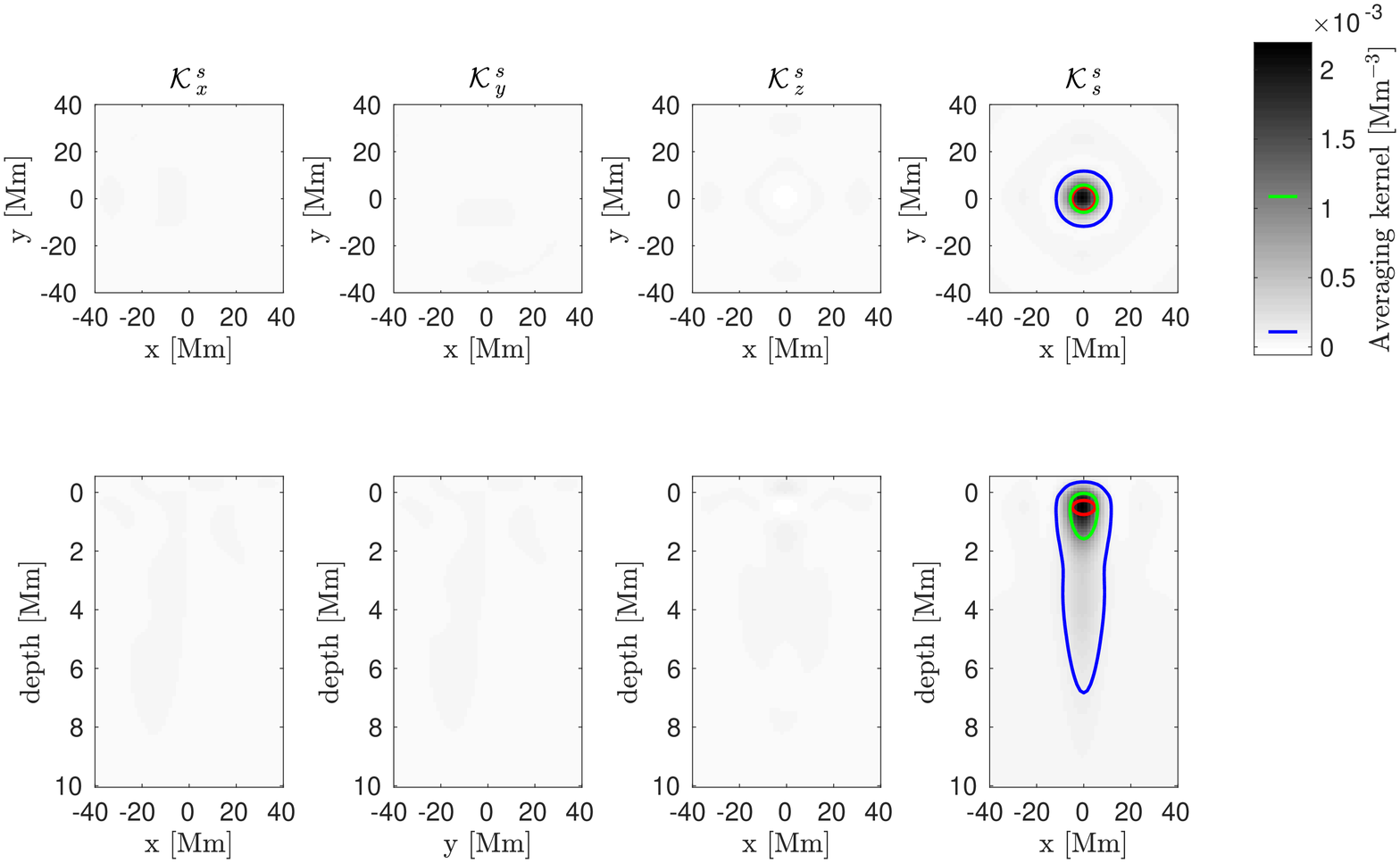}
\caption{Averaging kernels for $\delta c_s$ inversion at the depth of 0.5 Mm, JSOC-indicated target. See Fig. \ref{pic:RLS_rakern_x_0.5} for details.}
\end{figure*}

\begin{figure*}[!h]
\sidecaption
\includegraphics[height=8cm]{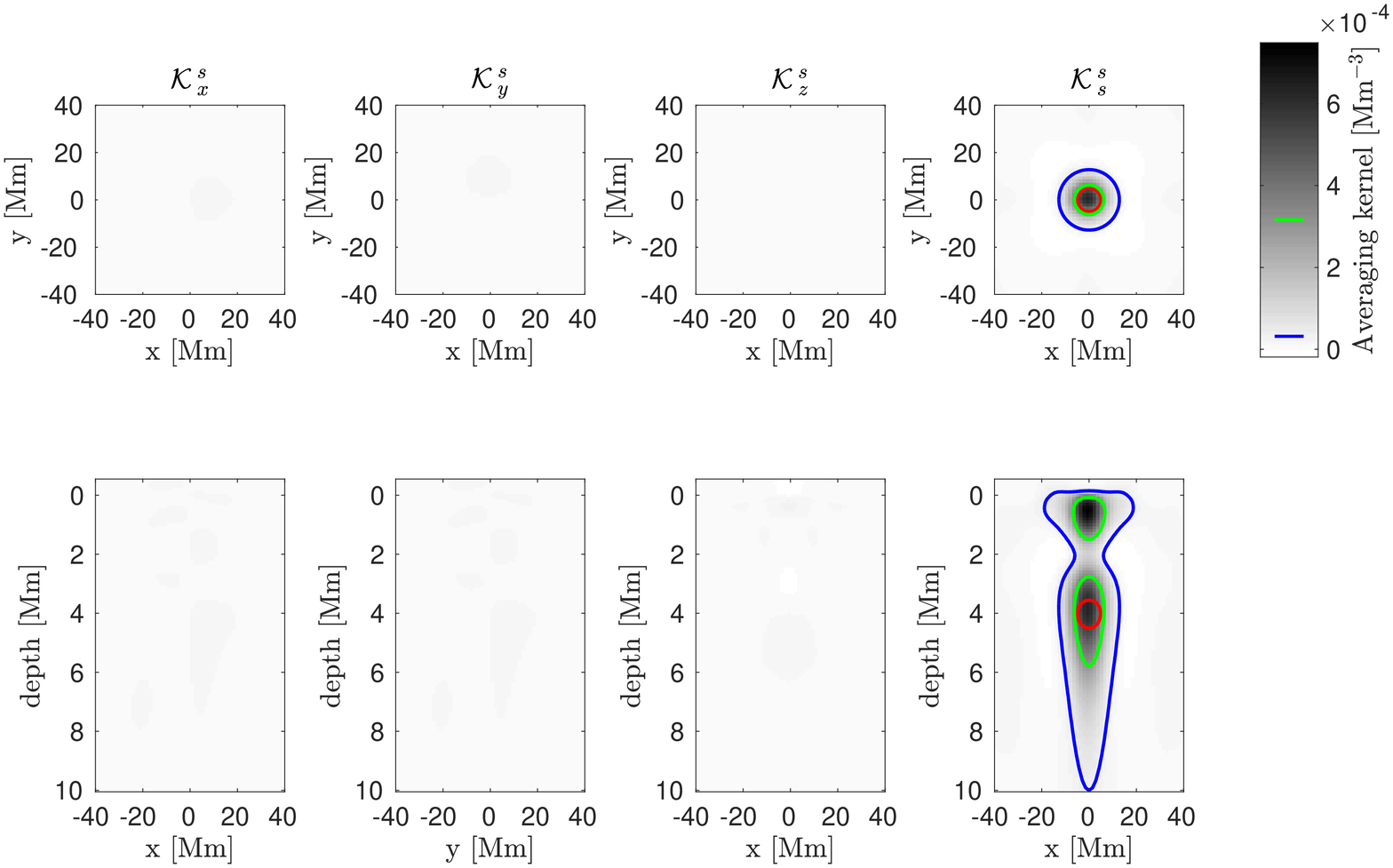}
\caption{Averaging kernel for $\delta c_s$ inversion at the depth of 4.0 Mm, JSOC-indicated target. See Fig. \ref{pic:RLS_rakern_x_0.5} for details.}
\end{figure*}

\begin{figure*}[!h]
\sidecaption
\includegraphics[height=8cm]{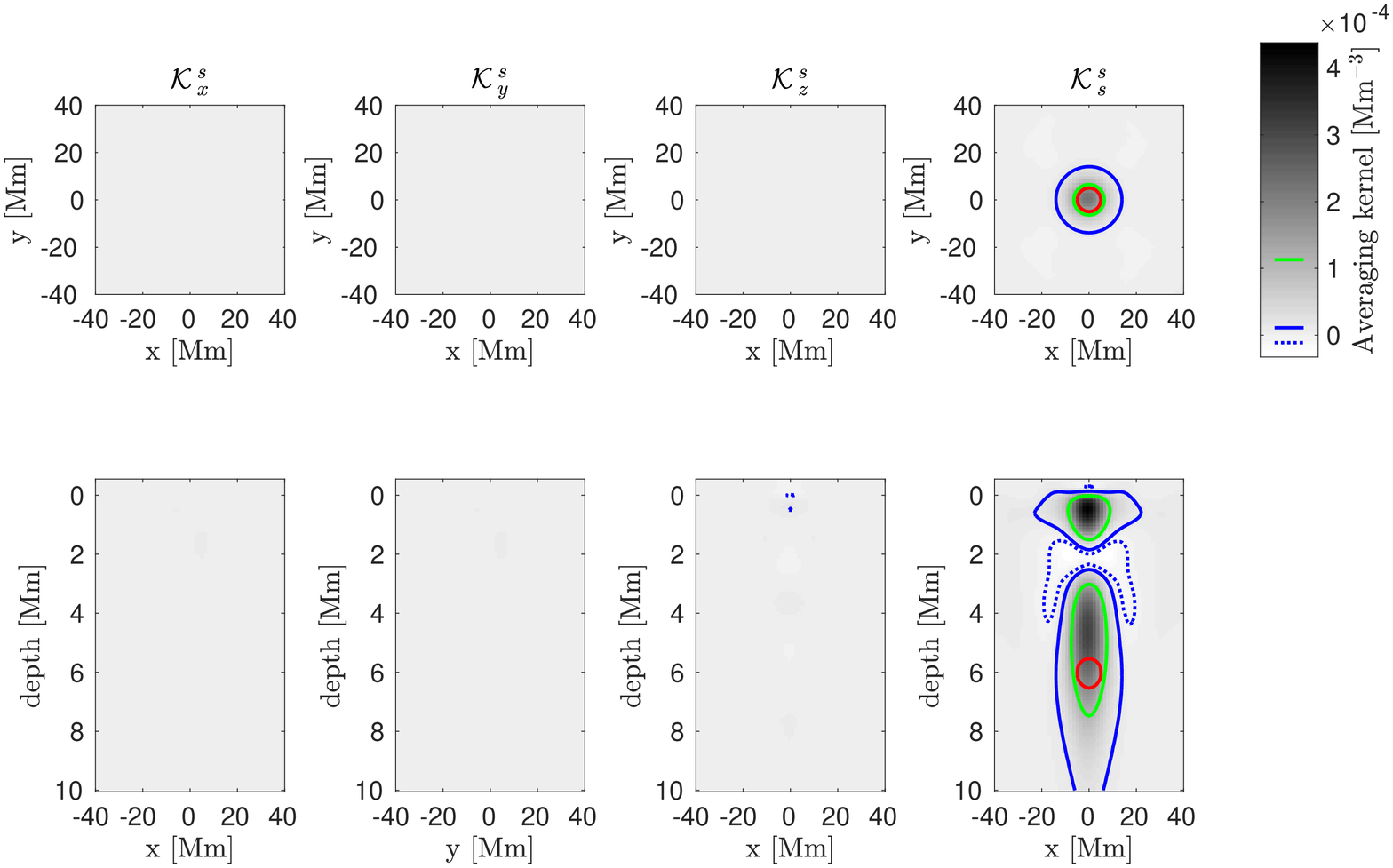}
\caption{Averaging kernels for $\delta c_s$ inversion at the depth of 6.0 Mm, JSOC-indicated target. See Fig. \ref{pic:RLS_rakern_x_0.5} for details.}
\end{figure*}

\begin{figure*}[!h]
        \includegraphics[width=0.30\textwidth]{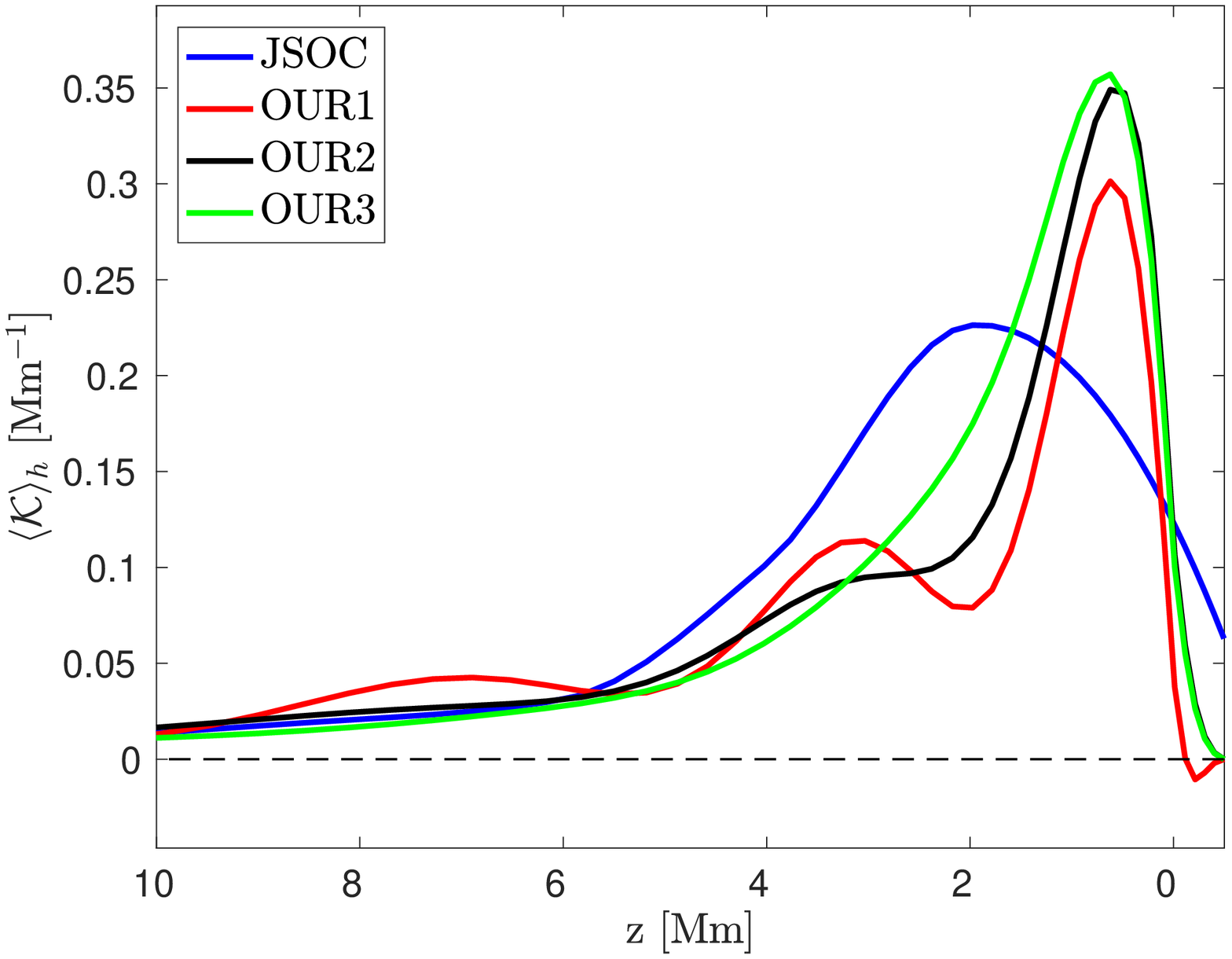}
        \includegraphics[width=0.30\textwidth]{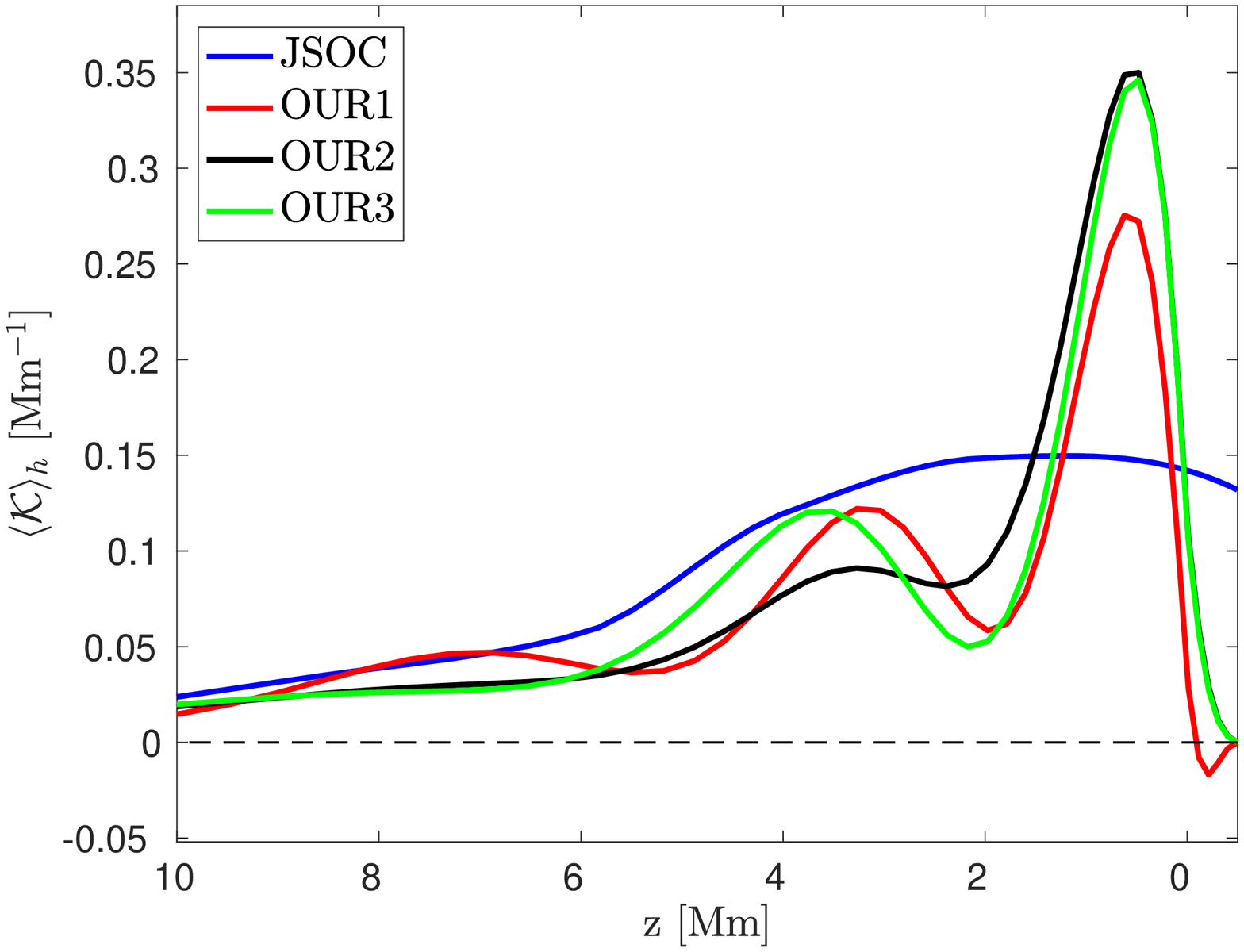}
        \caption{Horizontally averaged averaging kernels for $\delta c_s$ inversions as a function of depth. Left: kernels for the depth of 1--3~Mm (or 2.0~Mm for our inversion), Right: kernels for the depth of 3--5~Mm (or 4.0~Mm depth).}
        \label{pic:rakern_h_cs_all}
\end{figure*}

\begin{figure*}[!h]
\sidecaption
\includegraphics[height=8cm]{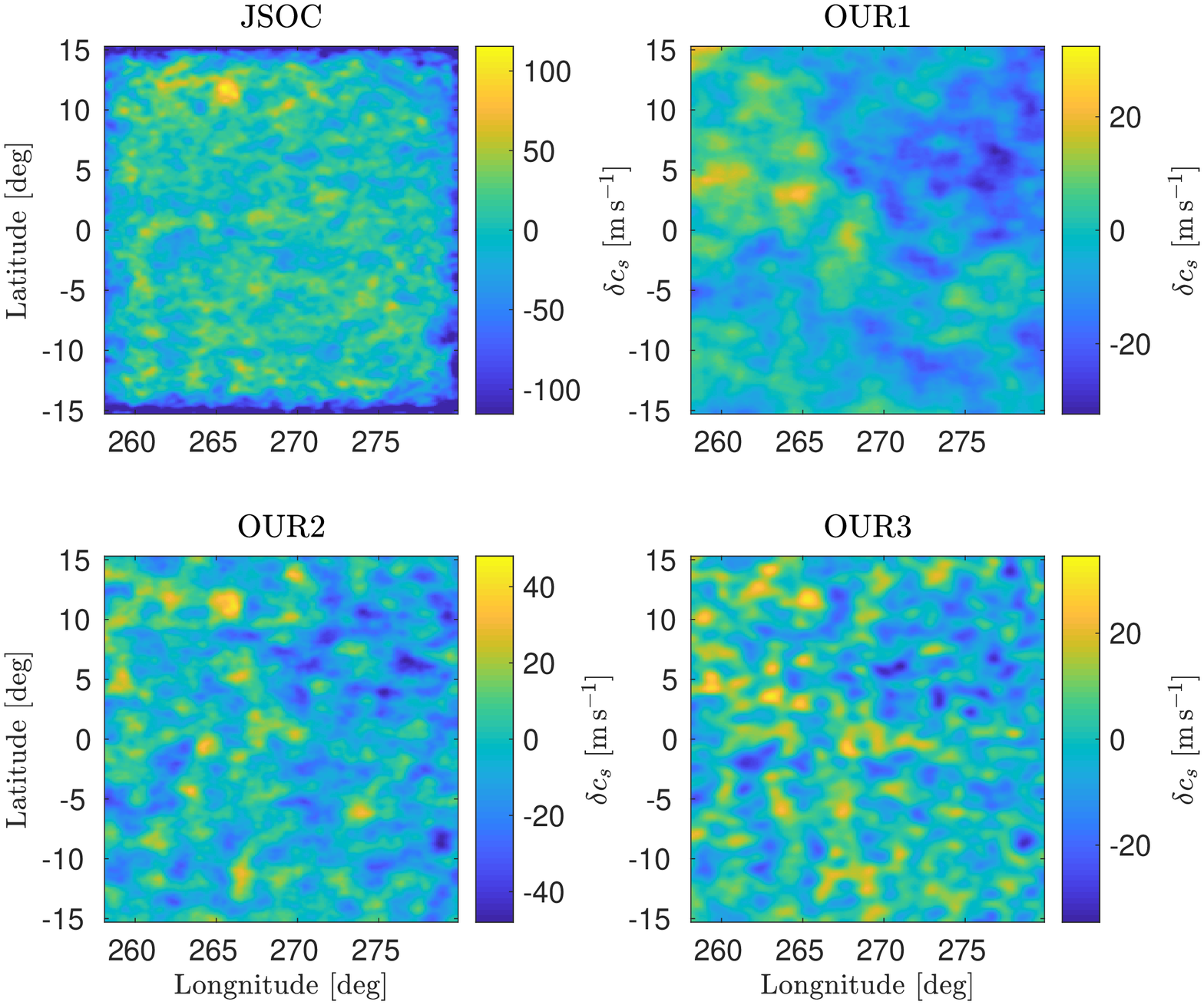}
\caption{Inversions for $\delta c_s^{{\rm inv}}$ at 0.5 Mm depth.}
\end{figure*}

\begin{figure*}[!h]
\sidecaption
\includegraphics[height=8cm]{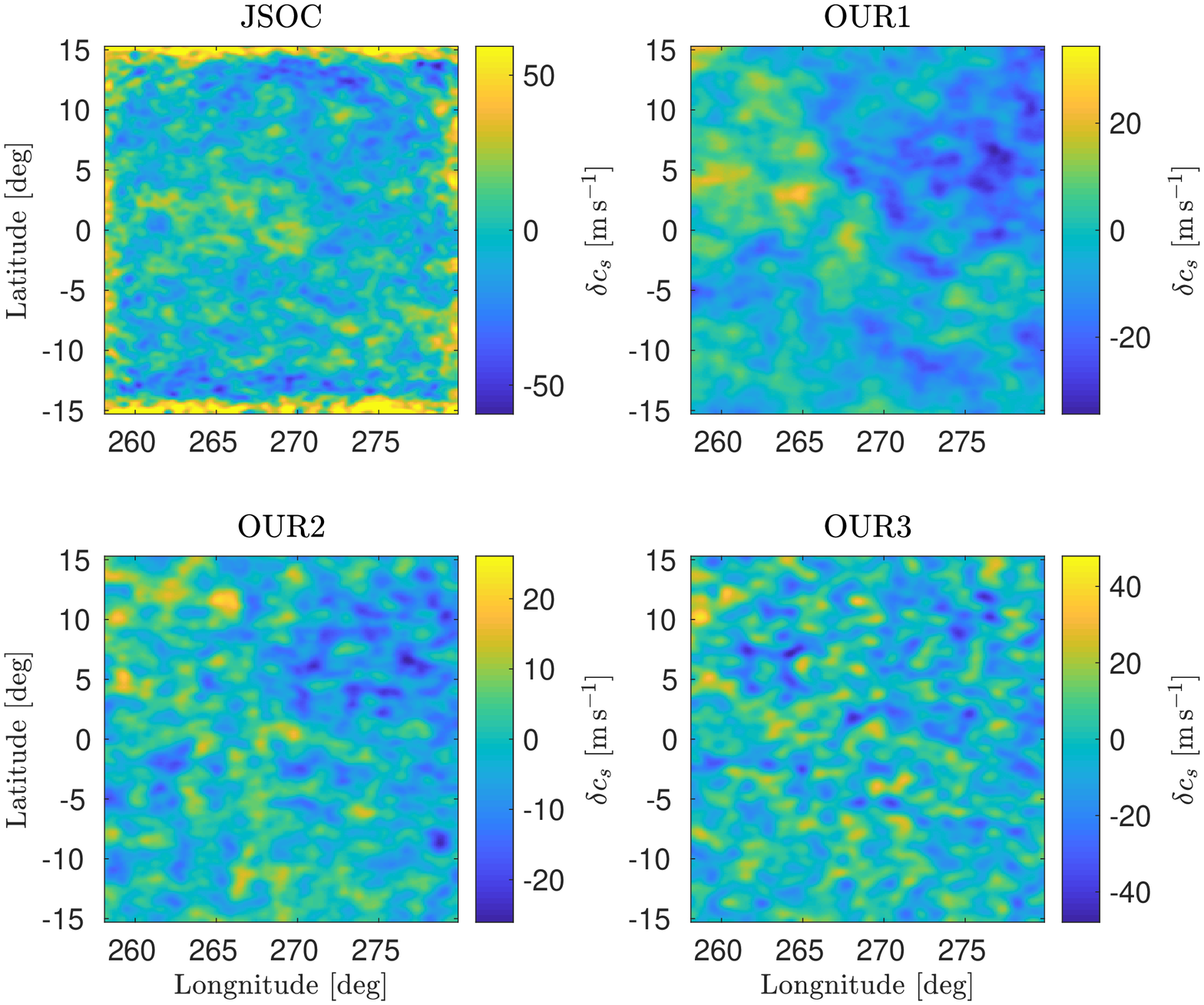}
\caption{Inversions for $\delta c_s^{{\rm inv}}$ at 4.0 Mm depth.}
\end{figure*}

\begin{figure*}[!h]
\sidecaption
\includegraphics[height=8cm]{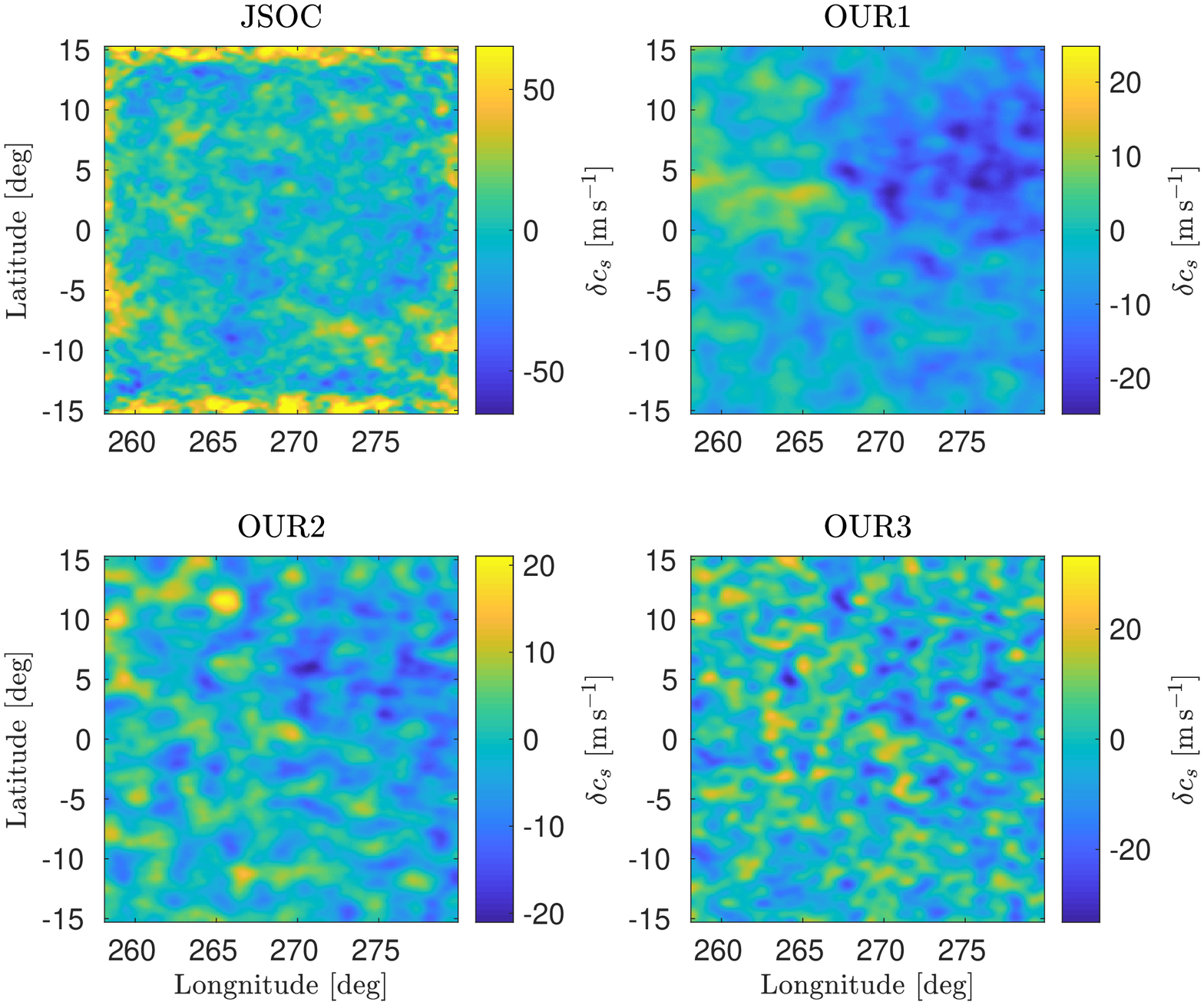}
\caption{Inversions for $\delta c_s^{{\rm inv}}$ at 6.0 Mm depth.}
\end{figure*}

\clearpage

\section{Vertical flows}
\label{app:vertical}

\begin{figure*}[!h]
\sidecaption
\includegraphics[height=8cm]{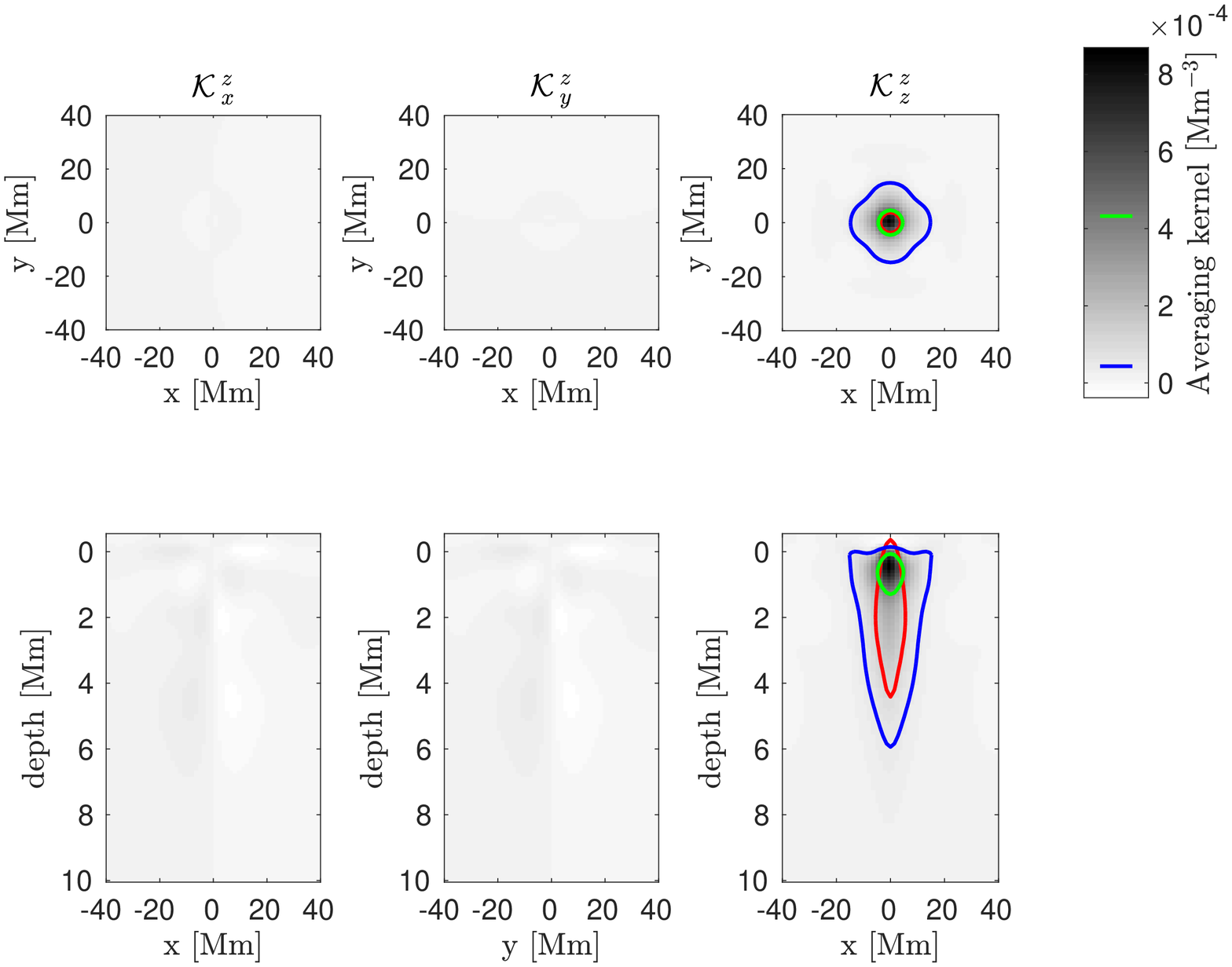}
\caption{Averaging kernels for $v_z$ inversion at the depth of 0.5 Mm, JSOC-like inversion. See Fig. \ref{pic:RLS_rakern_x_0.5} for details.}
\end{figure*}

\begin{figure*}[!h]
\sidecaption
\includegraphics[height=8cm]{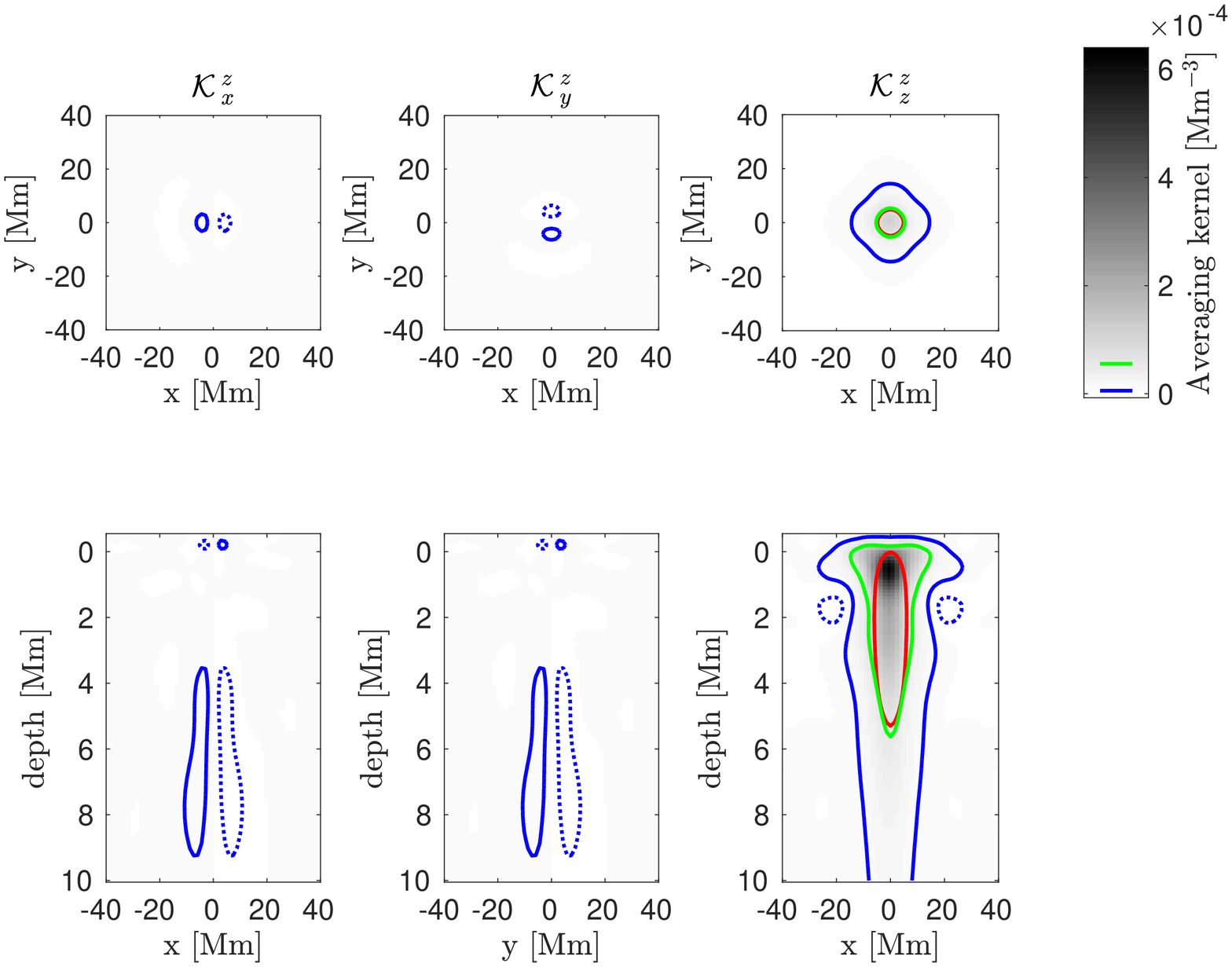}
\caption{Averaging kernels for $v_z$ inversion at the depth of 4.0 Mm, JSOC-like inversion. See Fig. \ref{pic:RLS_rakern_x_0.5} for details.}
\end{figure*}

\begin{figure*}[!h]
\sidecaption
\includegraphics[height=8cm]{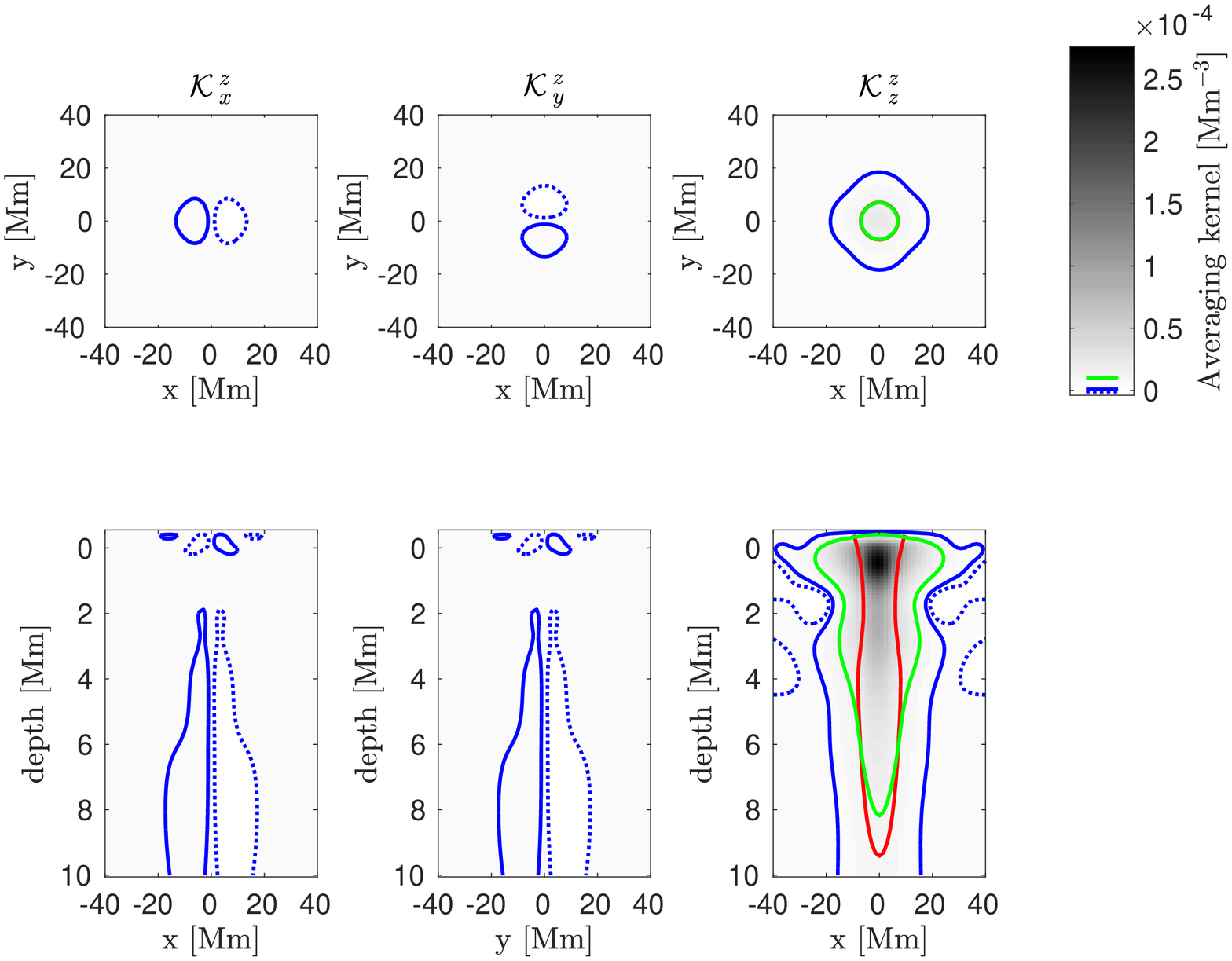}
\caption{Averaging kernels for $v_z$ inversion at the depth of 6.0 Mm, JSOC-like inversion. See Fig. \ref{pic:RLS_rakern_x_0.5} for details.}
\end{figure*}

\begin{figure*}[!h]
\sidecaption
\includegraphics[height=8cm]{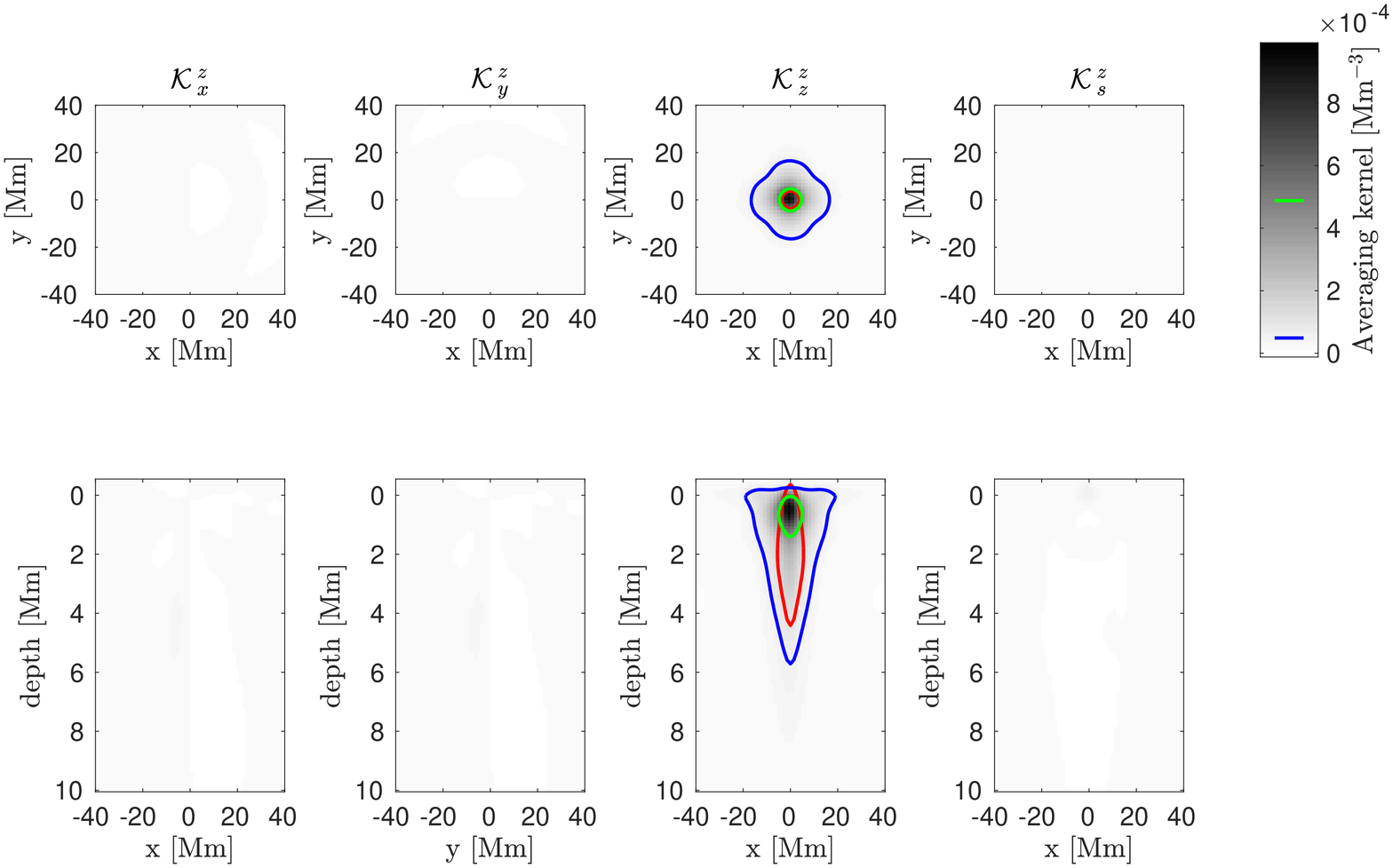}
\caption{Averaging kernels for $v_z$ inversion at the depth of 0.5 Mm, JSOC-like target. See Fig. \ref{pic:RLS_rakern_x_0.5} for details.}
\end{figure*}

\begin{figure*}[!h]
\sidecaption
\includegraphics[height=8cm]{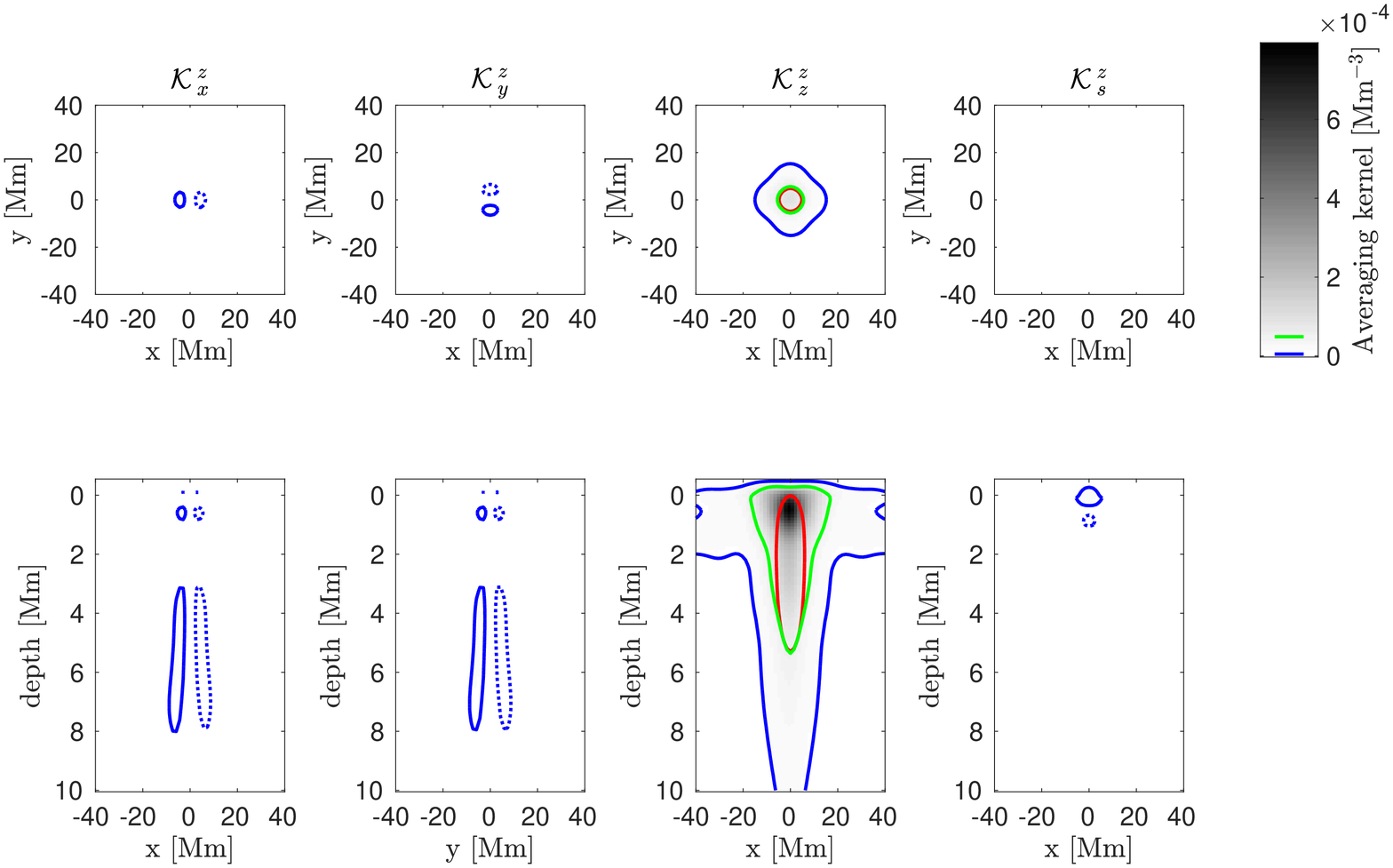}
\caption{Averaging kernels for $v_z$ inversion at the depth of 4.0 Mm, JSOC-like target. See Fig. \ref{pic:RLS_rakern_x_0.5} for details.}
\end{figure*}

\begin{figure*}[!h]
\sidecaption
\includegraphics[height=8cm]{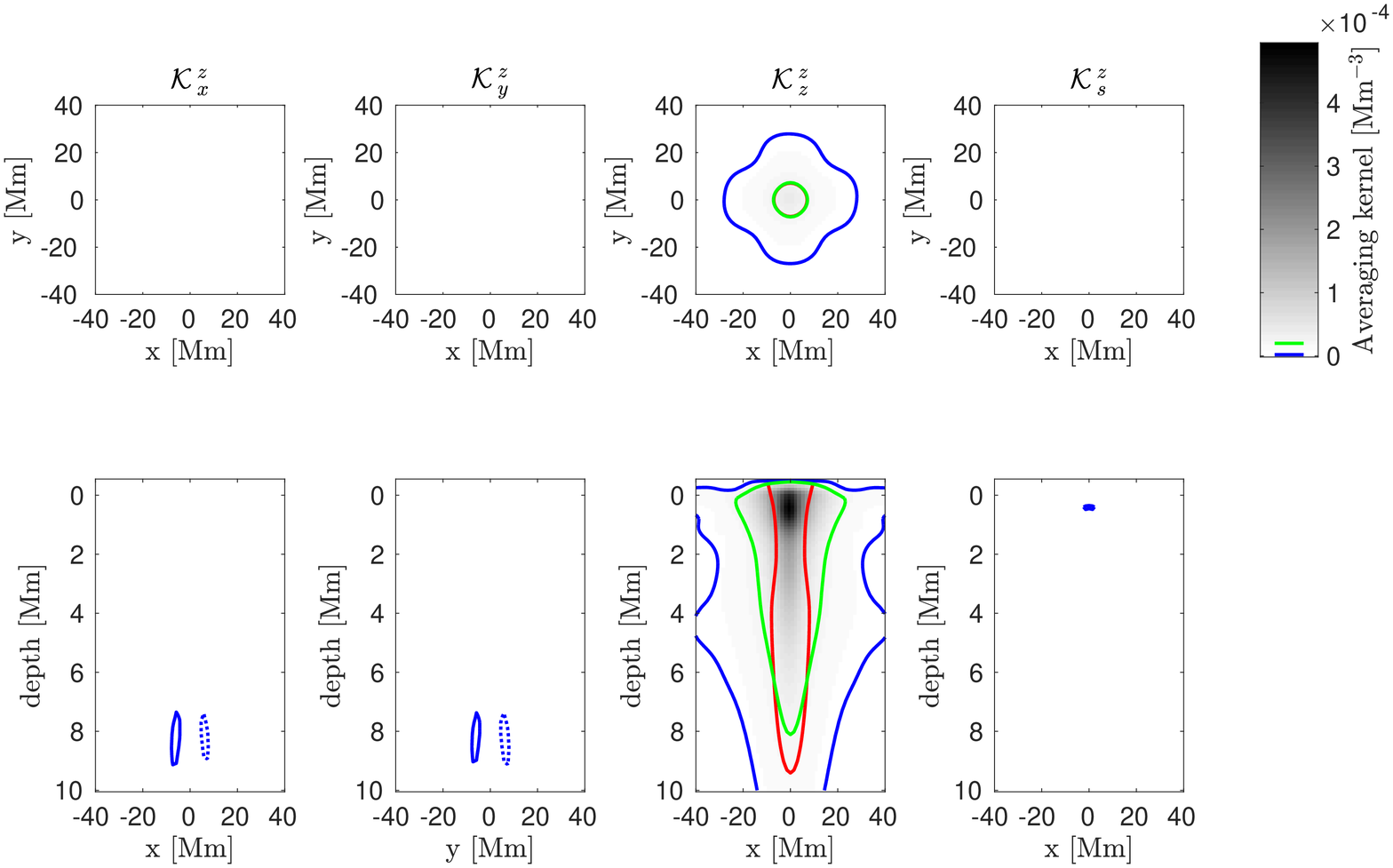}
\caption{Averaging kernels for $v_z$ inversion at the depth of 6.0 Mm, JSOC-like target. See Fig. \ref{pic:RLS_rakern_x_0.5} for details.}
\end{figure*}

\begin{figure*}[!h]
\sidecaption
\includegraphics[height=8cm]{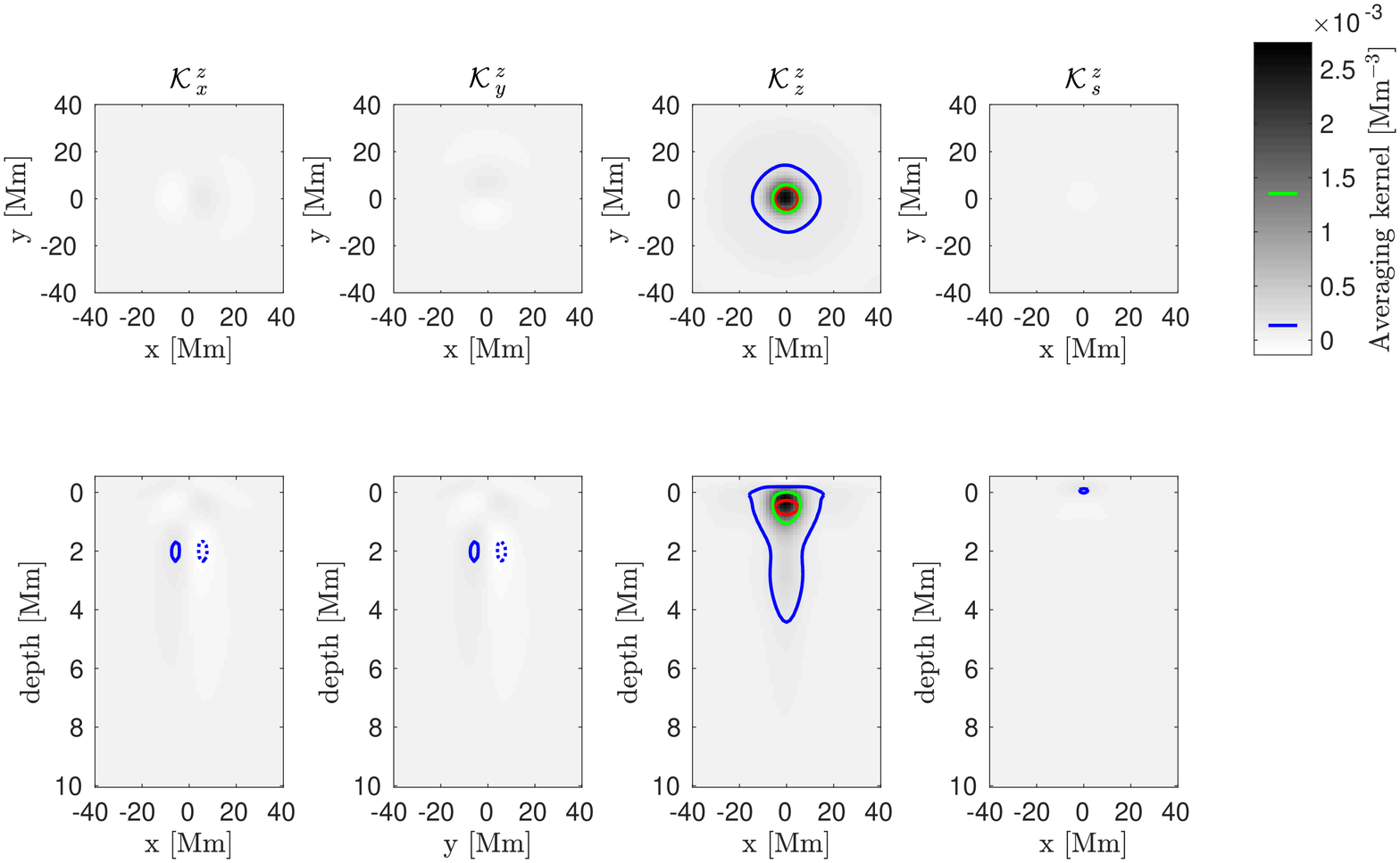}
\caption{Averaging kernels for $v_z$ inversion at the depth of 0.5 Mm, JSOC-indicated target. See Fig. \ref{pic:RLS_rakern_x_0.5} for details.}
\end{figure*}

\begin{figure*}[!h]
\sidecaption
\includegraphics[height=8cm]{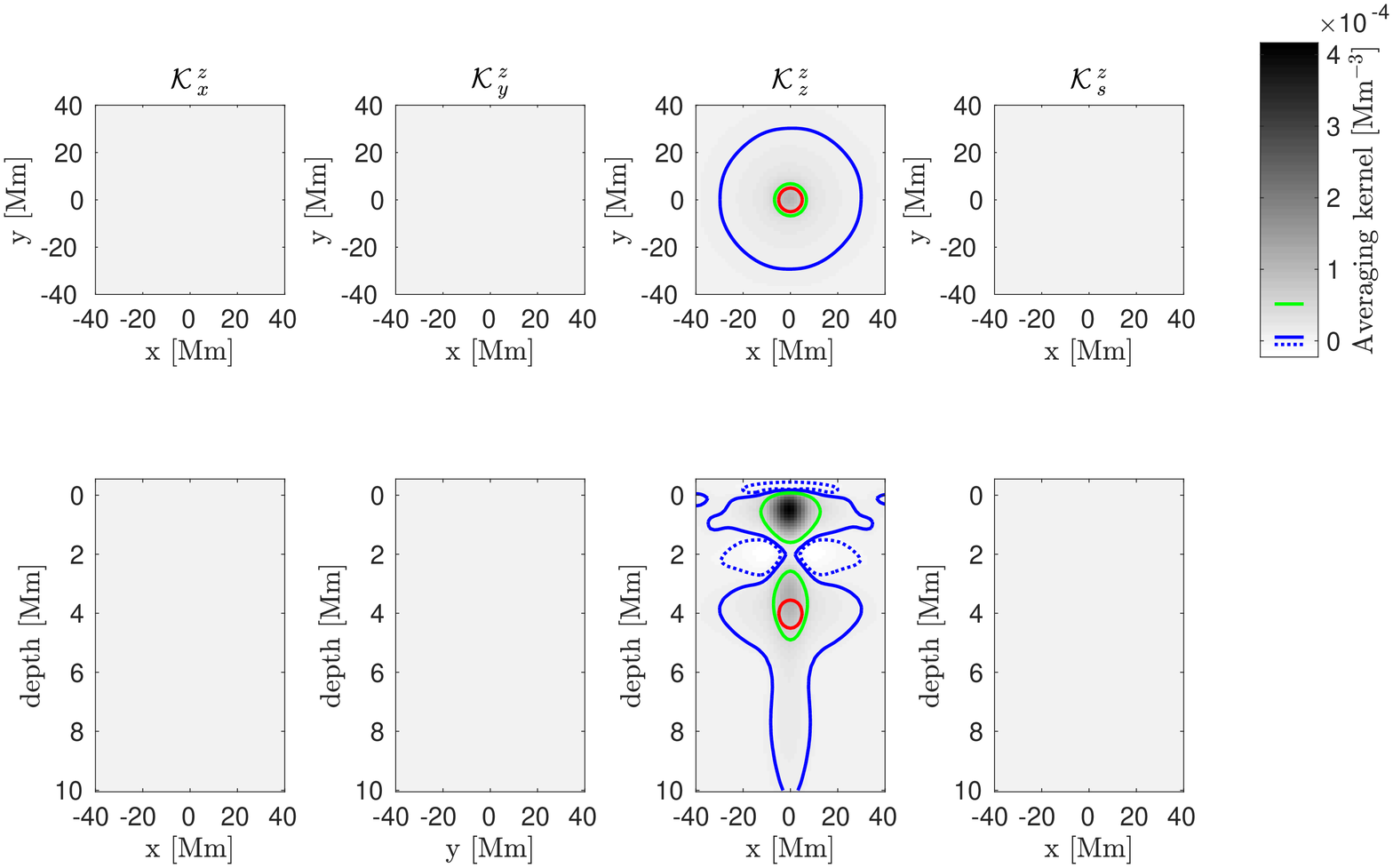}
\caption{Averaging kernels for $v_z$ inversion at the depth of 4.0 Mm, JSOC-indicated target. See Fig. \ref{pic:RLS_rakern_x_0.5} for details.}
\end{figure*}

\begin{figure*}[!h]
\sidecaption
\includegraphics[height=8cm]{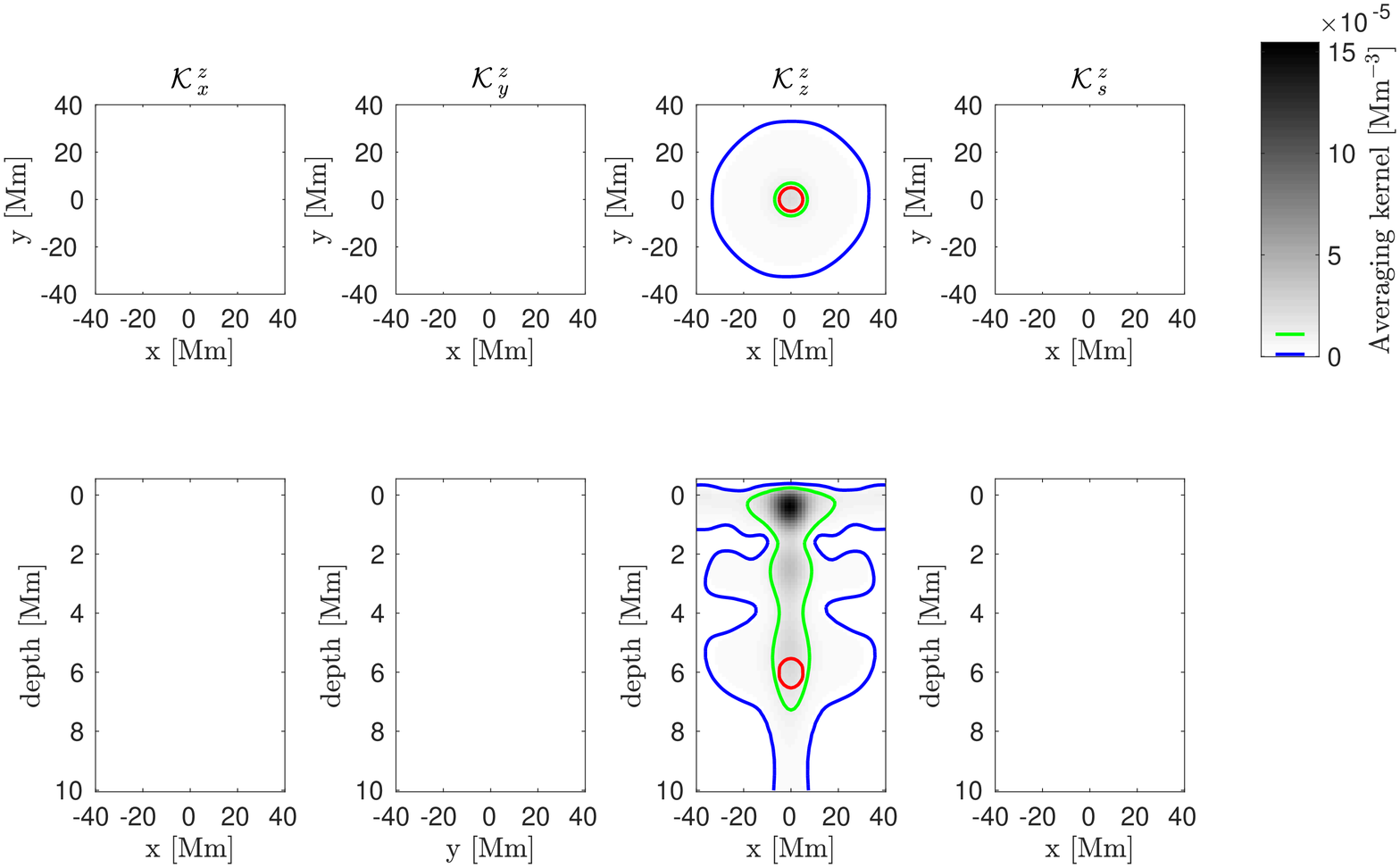}
\caption{Averaging kernels for $v_z$ inversion at the depth of 6.0 Mm, JSOC-indicated target. See Fig. \ref{pic:RLS_rakern_x_0.5} for details.}
\end{figure*}

\begin{figure*}[!h]
        \includegraphics[width=0.30\textwidth]{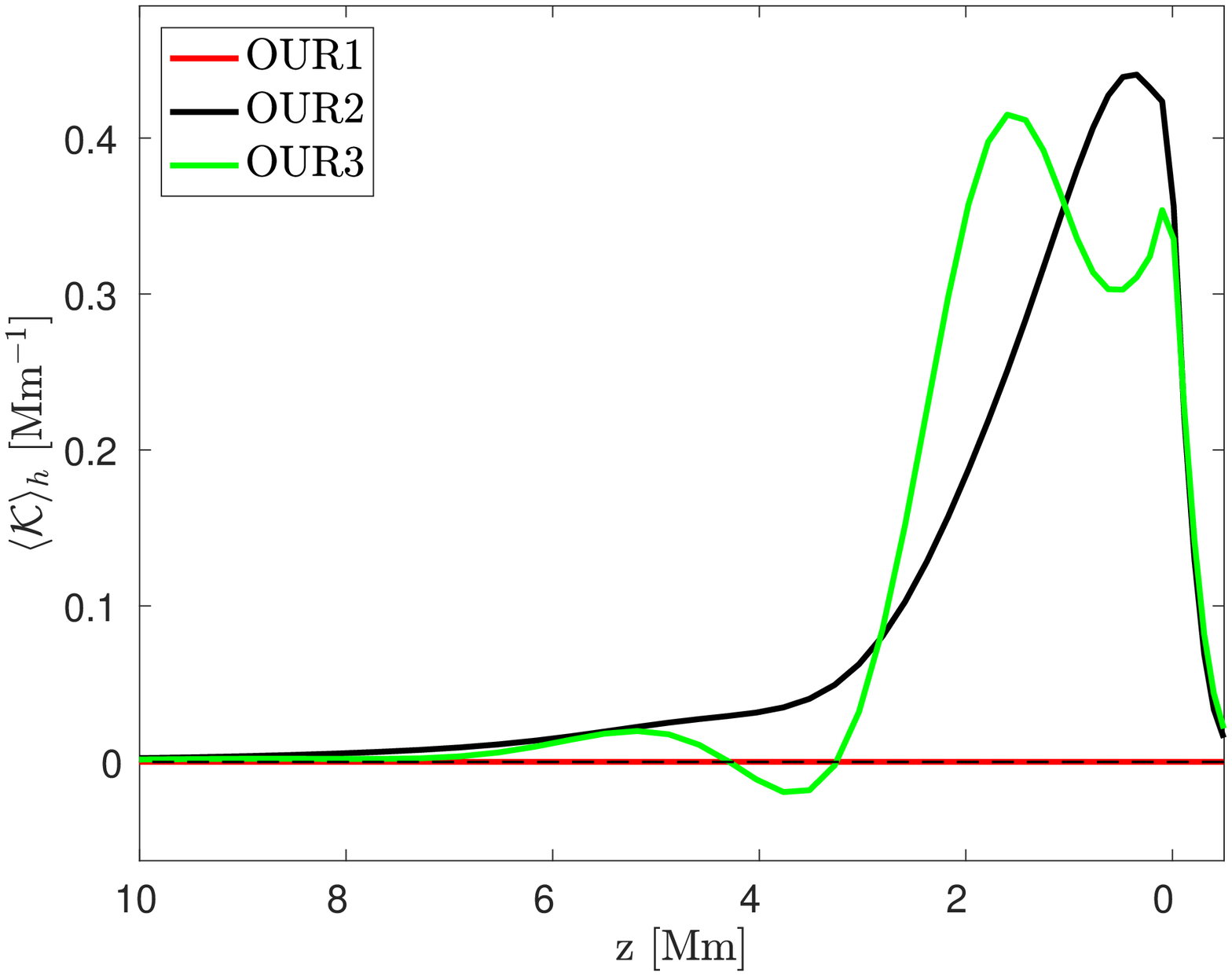}
        \includegraphics[width=0.30\textwidth]{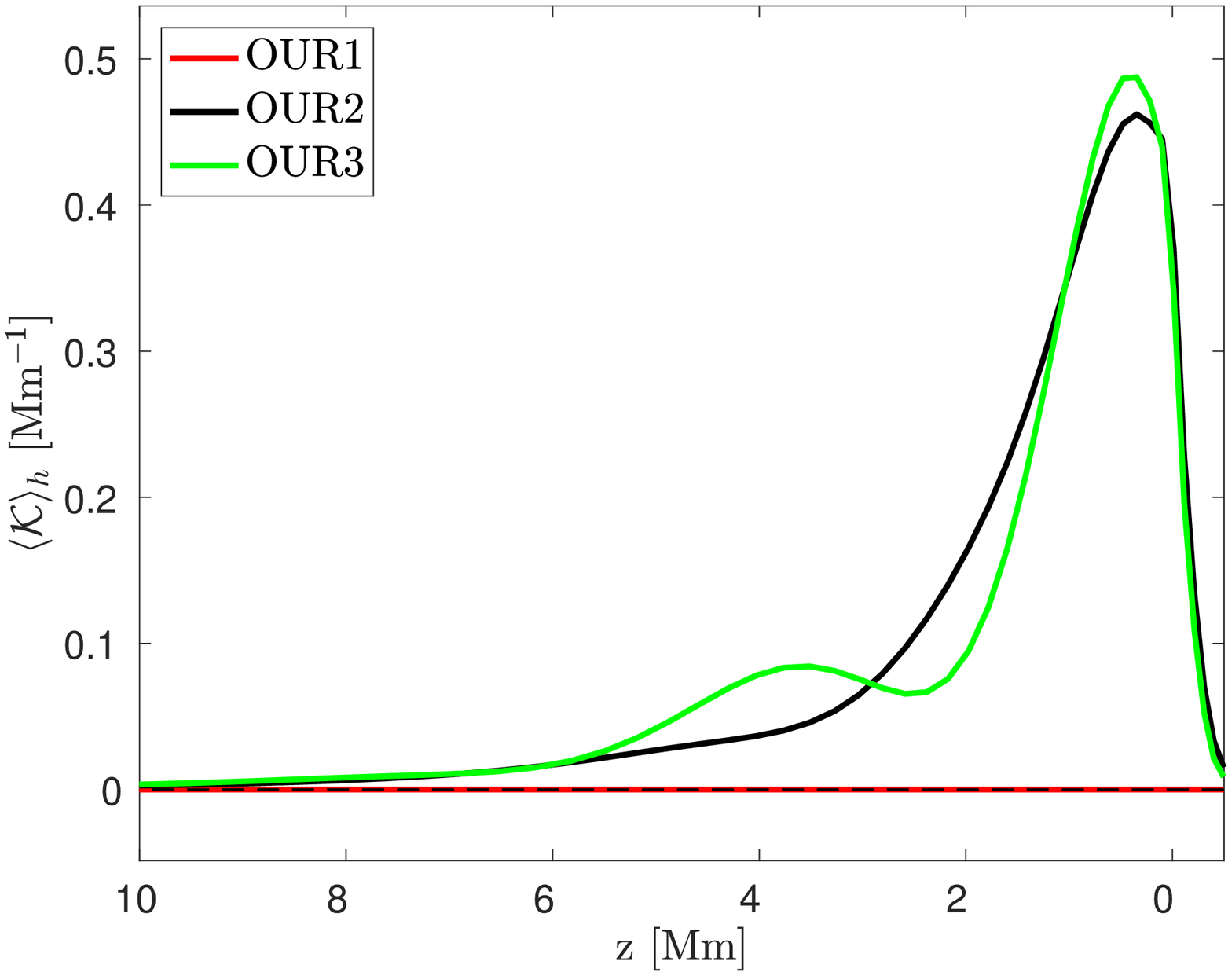}
        \caption{Horizontally averaged averaging kernels for $v_z$ inversions as a function of depth. Left: kernels for the depth of 1--3~Mm (or 2.0~Mm for our inversion), Right: kernels for the depth of 3--5~Mm (or 4.0~Mm depth).}
\end{figure*}

\begin{figure*}[!h]
\includegraphics[width=\textwidth]{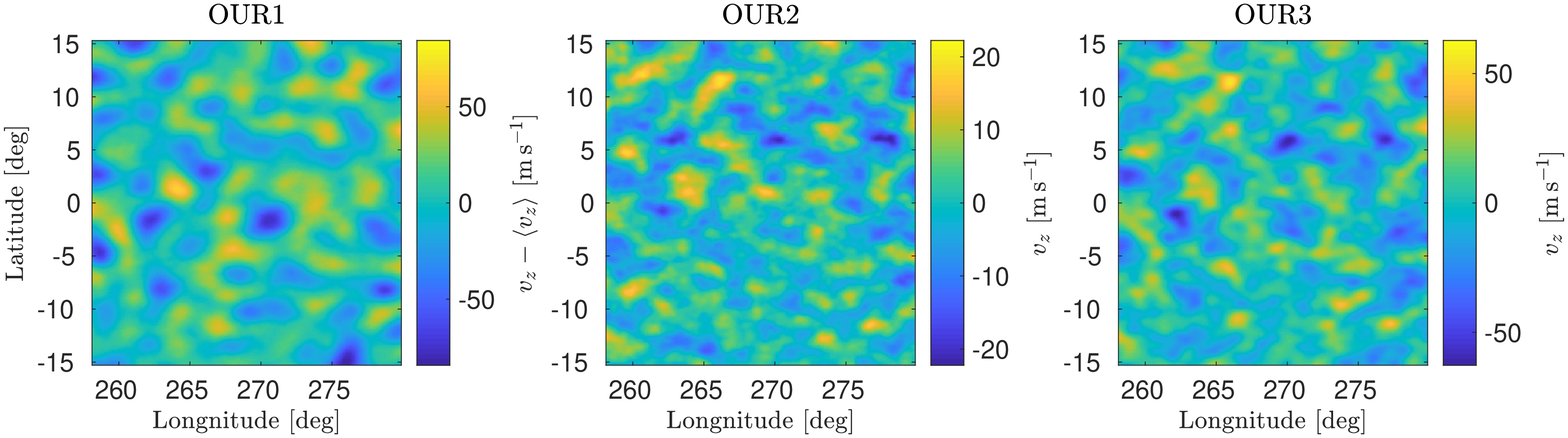}
\caption{Inversions for $v_z^{{\rm inv}}$ at 0.5 Mm depth.}
\end{figure*}

\begin{figure*}[!h]
\includegraphics[width=\textwidth]{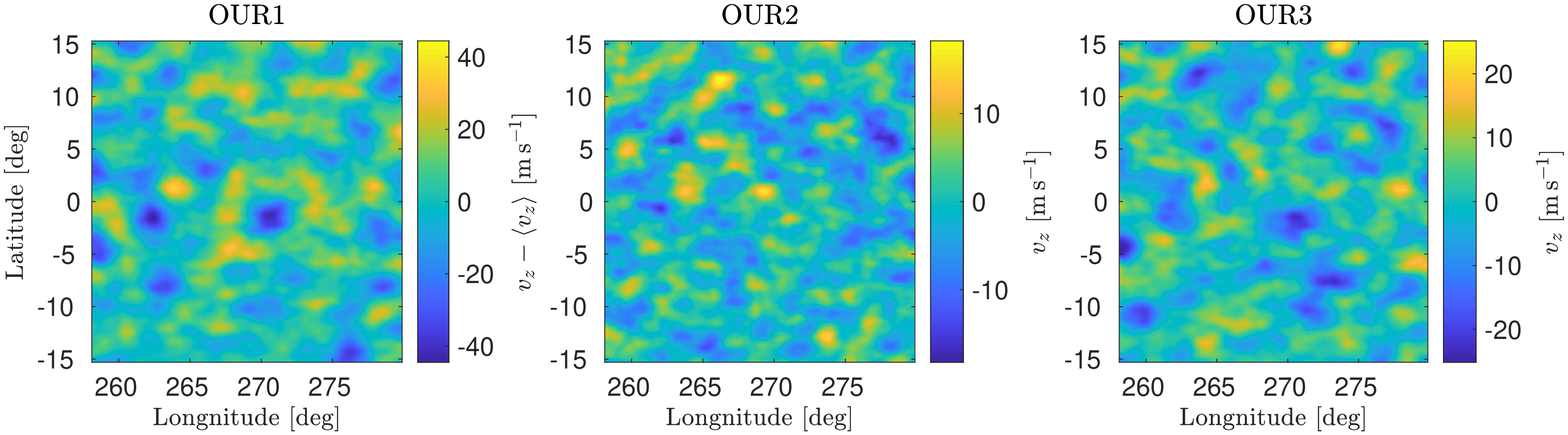}
\caption{Inversions for $v_z^{{\rm inv}}$ at 4.0 Mm depth.}
\end{figure*}

\begin{figure*}[!h]
\includegraphics[width=\textwidth]{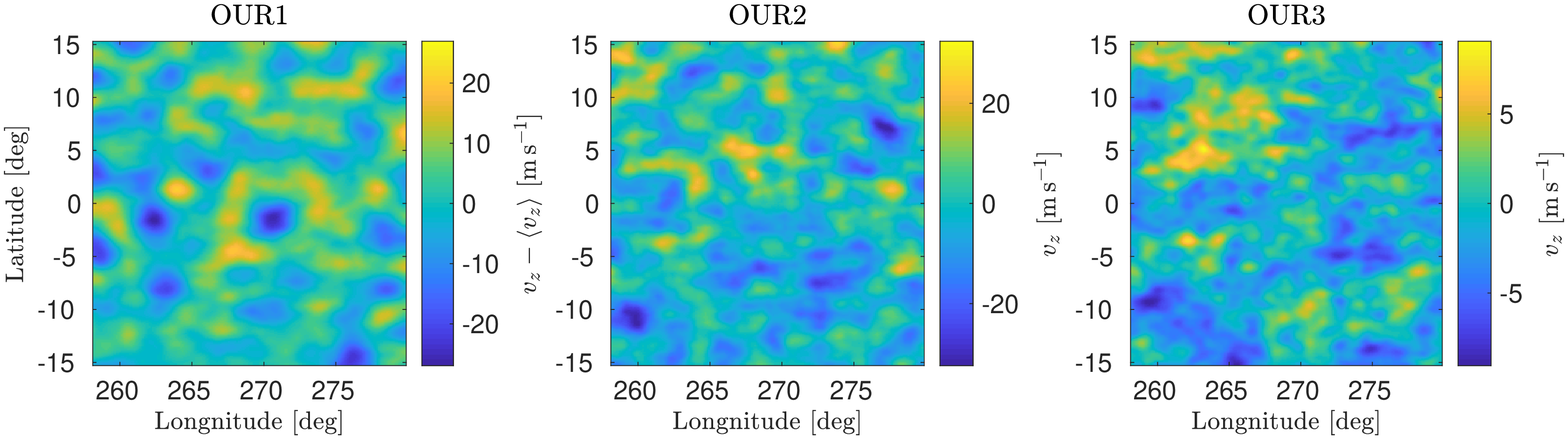}
\caption{Inversions for $v_z^{{\rm inv}}$ at 6.0 Mm depth.}
\end{figure*}

\clearpage

\end{appendix}

\end{document}